\documentclass[fleqn,usenatbib]{mnras}

\usepackage{booktabs,siunitx}
\usepackage[T1]{fontenc}
\usepackage{ae,aecompl}
\usepackage{float}

\usepackage{graphicx}	% Including figure files
\usepackage{amsmath}	% Advanced maths commands
\usepackage{nccmath}
\usepackage{amssymb}	% Extra maths symbols
\usepackage{comment}
\usepackage{arydshln}
\usepackage[table]{xcolor}
\usepackage{multirow}
\usepackage{subcaption}
\usepackage{notoccite}
\hypersetup{breaklinks=true}
\usepackage{newtxtext,newtxmath}
\usepackage{dblfloatfix}    % To enable figures at the bottom of page

\DeclareMathOperator{\sign}{sgn}
\newcommand{\angstrom}{\mbox{\normalfont\AA}}
\usepackage{units}

\usepackage{subfiles}

\title[Angular momentum of $z\sim1.5-2$ galaxies]{Multiresolution angular momentum measurements of $z\sim1.5-2$ star-forming galaxies}

\author[Juan M. Espejo Salcedo et al.]{Juan M. Espejo Salcedo$^{1}$\thanks{E-mail: jespejosalcedo@swin.edu.au}, Karl Glazebrook$^{1}$, Deanne B. Fisher$^{1}$, Sarah M. Sweet$^{3,4}$,\and Danail 
Obreschkow$^{2}$, A. M. Swinbank$^{5}$, Steven Gillman$^{5,6,7}$ and Alfred L. Tiley$^{2}$
\\
$^{1}$Centre for Astrophysics \& Supercomputing, Swinburne University of Technology, PO Box 218, Hawthorn, VIC 3122, Australia\\
$^{2}$International Centre for Radio Astronomy Research, University of Western Australia, 7 fairway, Crawley, WA 6009, Australia\\
$^{3}$School of Mathematics and Physics, University of Queensland, Brisbane, QLD 4072, Australia\\
$^{4}$ARC Centre of Excellence for All Sky Astrophysics in 3 Dimensions (ASTRO 3D)\\
$^{5}$Centre for Extra-galactic Astronomy, Durham University, South Road, Durham DH1 3LE UK\\
$^{6}$Cosmic Dawn Centre (DAWN), Copenhagen, Denmark\\
$^{7}$DTU-Space, Technical University of Denmark, Elektrovej 327, DK-2800 Kgs. Lyngby, Denmark
}

\date{Accepted 2021 September 20. Received 2021 September 6; in original form 2021 May 5}

\pubyear{2021}

\begin{document}
\label{firstpage}
\pagerange{\pageref{firstpage}--\pageref{lastpage}}
\maketitle

\begin{abstract}
We present detailed stellar specific angular momentum ($j_*$) measurements of ten star-forming galaxies at $z\sim1.5-2$ using both high and low spatial resolution integral field spectroscopic data. We developed a code that simultaneously models the adaptive optics (AO) assisted observations from OSIRIS/SINFONI along with their natural seeing (NS) counterparts from KMOS at spatial resolutions of [$0.1-0.4$] arcsec and [$0.6-1.0$] arcsec respectively. The AO data reveals 2/10 systems to be mergers and for the remaining eight the mean uncertainties $\bar \Delta j_*$ decrease from 49\% (NS), and 26.5\% (AO), to 16\% in the combined analysis. These $j_*$ measurements agree within 20\% with simple estimates ($\tilde{j_*}$) calculated from the Hubble Space Telescope photometry and NS kinematics, however higher resolution kinematics are required to first identify these disks. We find that the choice of surface mass density model and the measurement of effective radius from photometry are the key sources of systematic effects in the measurement of $j_*$ between different analyses. Fitting the $j_*$ vs $M_*$ relations (Fall, 1983) with a fixed power-law slope of $\beta=2/3$, we find a zero-point consistent with prior NS results at $z\geq1$ within $\sim 0.3$ dex. Finally, we find a $\sim 0.38$ dex scatter about that relation that remains high despite the AO data so we conclude it is intrinsic to galaxies at $z>1$. This compares to a scatter of $\leq 0.2$ dex for disks at $z\simeq0$ pointing to a settling of the Fall relation with cosmic time.
\end{abstract}


%%%%%%%%%%%%%%%%%%%%%%%%%%%%%%%%%%%%%%%%%%%%%%%%%%

%%%%%%%%%%%%%%%%% BODY OF PAPER %%%%%%%%%%%%%%%%%%

\section{Introduction}

Approximately half of all the stars in the Universe were formed during the time period when it was 3-6 Gyr old ($1<z<3$), with the peak of star formation happening at around $z\sim2$ (\citealp{Madau_SFH}). At this epoch, the majority of the galaxy population appears very different from the rapidly rotating disks that make up the bulk of the star-forming galaxies in the local Universe. At $z\sim2$, star-forming galaxies had clumpy morphologies (\citealp{Glazebrook95}; \citealp{Conselice}; \citealp{Elmegreen}; \citealp{Livermore}), their gas fractions ($f_\mathrm{gas}$) were high compared to local analogues (\citealp{Bouche_2007}; \citealp{Tacconi}; \hbox{\citealp{Decarli}}) and for a fixed stellar mass ($M_*$), their stellar specific angular momentum $j_*=J_*/M_*$ was lower than that of local disks (\citealp{Law_2}; \citealp{Obreschkow_2015}; \citealp{Teklu}).

The morphological diversity of regular galaxies at low redshift, typically classified by the Hubble sequence (\citealp{Hubble_sequence}), emerged at $z\sim2$ (\citealp{Szomoru}), where the transition from the high-redshift turbulent disks to the grand design spirals we observe in the local Universe begins to take place. Both observations and theoretical models suggest that galaxy-wide instabilities play an important role in driving the evolution of galaxies during this epoch (\hbox{\citealp{Dekel_2009}}; \hbox{\citealp{Dekel_Burkert}}; \hbox{\citealp{Tacchella}}; \hbox{\citealp{Genzel_2008}}). These strong instabilities can be produced in major merger events and gravitational interactions which are more frequent at $z>1.5$ (\hbox{\citealp{Ventou17,Ventou19}}). In the absence of mergers, a galaxy-wide instability in a rotating system can be generated by a combination of high gas surface density and low angular momentum.

Angular momentum is a fundamental physical quantity that sets the disk size and thickness and induces scaling relations such as the fundamental plane of spiral galaxies and the Tully-Fisher relation (\citealp{Tully_fisher}). It provides an alternative to the Hubble-type morphological classification (\hbox{\citealp{Fall_Efstathiou}}; \hbox{\citealp{van_den_Bosch}}; \hbox{\citealp{KarlandDanail}}) and drives the dynamical evolution of a galaxy throughout its history.

In the current framework for galaxy formation, self-bound structures of non-baryonic matter assemble due to gravitational collapse to form dark matter haloes (\citealp{Peebles}; \citealp{Blumenthal:1984bp}; \citealp{Wechsler}). These protohaloes acquire angular momentum as a result of tidal torques from the random alignment and orientation of the large-scale environment (\citealp{Mo_galaxy_formation}; \citealp{Liao}). The acquired specific angular momentum of the halo ($j_h$) has a halo mass ($M_h$) dependency of the form $j_h \propto M_h^{2/3}$ (\citealp{Catelan_Theuns_b}; \citealp{Catelan_Theuns}). As the gas cools and collapses to the centre of the dark matter halo, a rotationally supported disk is formed. Baryonic angular momentum is distributed within the disk by local torques and stellar feedback. In the case of star-forming disk galaxies (and in the absence of major mergers), the stellar specific angular momentum approximately preserves the halo dynamical properties and scales with the stellar mass as $j_* \propto M_*^{2/3}$ (\citealp{Romanowsky}; \citealp{Harrison_2017}; \citealp{Posti_2018}; \citealp{Fall_Romanowsky_2018}).

%From this relation and assuming a $z$-independent stellar-to-halo mass ratio (\citealp{Behroozi}), it is expected that for a constant stellar mass, $j_*$ increases with cosmic time as
%the relation between $J$ and $M_h$ changes with cosmic time, as predicted from $\Lambda$CDM cosmology. This implies an

Stellar specific angular momentum is expected to increase with cosmic time as $j_* \propto (1+z)^{-n}$ with $n\sim 0.5-1$ (e.g. \citealp{Obreschkow_2015} from theoretical predictions and \citealp{Swinbank} from observations). The power-law slope, $n$, may vary as angular momentum is redistributed due to star formation feedback, outflows and winds (\citealp{Naab}). Empirically constraining the angular momentum evolution is key in reproducing galaxies with realistic morphology, rotation and size in cosmological hydrodynamical simulations (\citealp{Pichon}; \citealp{Teklu}).

Galaxy angular momentum is therefore a fundamental quantity that is not directly observed but derived from kinematics and stellar mass distributions. At high-redshift this is particularly difficult, one needs to both resolve the steep rising part of the rotation curve as well as the flat outer end to derive rotation profiles.

The majority of the current kinematic measurements at high-redshift come from seeing limited observations (\citealp[SINS;][]{Forster_2008_no_AO}, \citealp[KMOS$^\mathrm{3D}$;][]{KMOS3D}, \citealp[KROSS;][]{Swinbank}, \citealp[KGES;][]{Gillman}). These measurements are heavily affected by beam smearing leading to degraded spatial resolutions of ${\sim}8$ kpc and due to the possible misclassification of mergers and disks (\hbox{\citealp{Epinat_2012}}; \citealp{Bellocchi}; \citealp{Rodrigues}; \citealp{Sweet}; \citealp{Simons2019}).

\begin{comment}
It has been shown that the morphological classification of galaxies at high-redshift using seeing limited kinematic observations is often unreliable since low spatial resolution obscures the identification of substructure and mergers can be mistakenly classified as disks \citep{Sweet}. Therefore, measurements of $j_*$ in these systems as tilted rotating disks are inadequate. Moreover,  show that by smoothing IFS data of nearby galaxies to the natural seeing resolution at $z\sim2$, kinematic components and small-scale features become absent.
\end{comment}

Adaptive optics assisted observations are required to resolve the inner part of the rotation curves since they improve resolution (at the ${\sim}1$ kpc level which is comparable to galaxy $r_\mathrm{eff}$ values) but are typically limited to a smaller field of view and have a lower throughput than seeing limited observations (\citealp{Burkert}; \citealp{Gillman_Hizels}). One way to account for this is to combine data at different resolutions as suggested in \hbox{\cite{Glazebrook_2013}} and implemented by \hbox{\cite{Obreschkow_2015}} who made a combined analysis of galaxies at $z\sim0$ with both low- and high-resolution data. Currently, despite the large amounts of seeing limited measurements at $z\sim 1-2$, only a small number overlap with adaptive optics observations so any combined measurement of $j_*$ is limited to a small number of galaxies (e.g. \hbox{\citealp{Sweet}} with two galaxies at $z\sim1.5$).

In this paper, we implement a combined measurement of the dynamics of ten star-forming galaxies to investigate the systematics in the existing measurements from seeing-limited surveys. In Section \ref{section:Observations} we present the details of our kinematic and photometric sample. We describe the methods used to measure the dynamical properties, the PSF modelling and the combination of the different datasets in Section \ref{section:Methods}. In Section \ref{section:Discussions} we discuss the enhanced angular momentum measurements and the capabilities of our modelling and in Section \ref{section:Conclusions} we present the results and conclusions of the paper. We include an appendix describing the kinematic fitting code and the different sanity checks used to avoid systematics, as well as a supplementary (online) section with an additional appendix describing each galaxy individually. Throughout the paper we assume a $\Lambda$CDM cosmology with $\Omega_\mathrm{m}$ = 0.3, $\Omega_\Lambda$ = 0.7, and $H_0$ = 70 km s$^{-1}$ Mpc$^{-1}$.

\section{Observations and sample description.}
\label{section:Observations}

\begin{figure*}
    \centering\includegraphics[width=17.5cm]{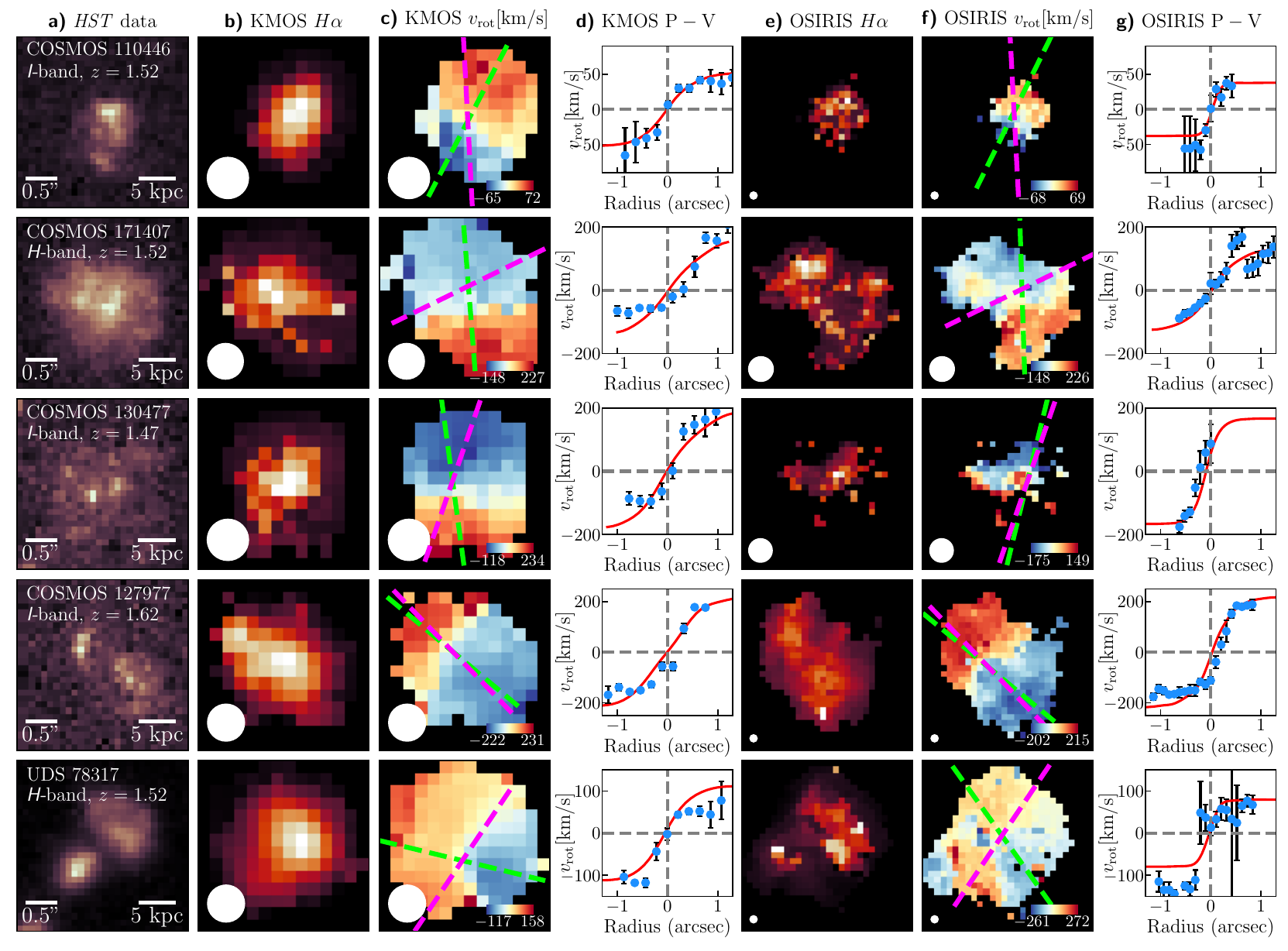}
    \caption{2D maps of the $z\sim1.5$ subsample. \textbf{a)} \textit{HST} broad-band continuum maps with the corresponding redshift and spatial scales. \textbf{b)} Natural seeing H$\alpha$ intensity maps from KMOS with the PSF FWHM indicated by the white circle. \textbf{c)} Extracted velocity fields of the natural seeing data where the green and purple dashed lines indicate the kinematic and morphological main axes respectively. \textbf{d)} Position-velocity (P-V) diagrams where blue dots indicate the velocities of the pixels that lay along the kinematic position axis as a function of the projected radius in arcsec. The error bars correspond to the 1-$\sigma$ error from the emission line Gaussian fit in each spaxel. The red line indicates the velocities along the kinematic position axis of the 2D model built from the individual data. The discrepancies between the model and datapoints are expected since the points are only along the major axis whereas the fit is built from the full 2D velocity field. Columns \textbf{e)}, \textbf{f)} and \textbf{g)} are the H$\alpha$ intensity, velocity field and P-V diagram from the adaptive optics observations from OSIRIS.}
    \label{fig:all_galaxies_1}
\end{figure*}

We analyse a sample of ten galaxies that have adaptive optics ``AO'' assisted observations and natural seeing ``NS'' limited counterparts. This sample consists of five galaxies at $z\sim1.5$ from Keck/OSIRIS (AO) + VLT/KMOS (NS) shown in Figure \ref{fig:all_galaxies_1} and five galaxies from VLT/SINFONI (AO) + VLT/KMOS (NS) at $z\sim2$ shown in Figure \ref{fig:all_galaxies_2}. We do not include galaxies at $z<1.5$ in our analysis due to low Strehl. In this Section, we describe the sample selection and the characteristics of the data from integral field spectroscopy, as well as the near-infrared photometric data from \textit{HST}. Table \ref{tab:sample_description} summarizes the instruments used in our full sample and some of the main parameters of the 10 galaxies.

\label{sec:Sample}
\subsection{Sample at \texorpdfstring{$z\sim1.5$}{z1.5}}

\subsubsection{\textbf{KGES observations (natural seeing)}}

We used the reduced datacubes of the galaxies COSMOS 110446, COSMOS 171407, COSMOS 130477, COSMOS 127977 and UDS 78317 from the KMOS Galaxy Evolution Survey (KGES; \citealp{KGES_alfie}). This survey comprises a sample of 288 mass-selected bright ($K < 22.7$) star-forming galaxies with known spectroscopic redshifts at $z\sim1.5$ in the ECDFS, UDS and COSMOS fields targeting their redshifted H$\alpha$ and [N\textsc{ii}] emission lines.

The instrument used for this survey is the $K$-band multi-object spectrograph \hbox{\citep[KMOS;][]{KMOS}} at a pixel scale of 0.2 arcsec per pixel. The mean PSF FWHM (point-spread function at full width at half maximum) of the KGES sample is ${\sim}0.6$ arcsec and the spectral resolution is $R\sim3975$ at the location of H$\alpha$ emission in the $K$-band. These galaxies were chosen based on the ordered rotation that they appeared to have in seeing limited IFS data for the follow-up observations with adaptive optics. The \textit{HST} images, velocity and H$\alpha$ intensity maps from this sample are shown in Figure \ref{fig:all_galaxies_1}.

\subsubsection{\textbf{OSIRIS observations (adaptive optics)}} 

We selected the five galaxies above based on the preliminary results of the KGES survey to do follow-up observations with adaptive optics. We prioritized the KGES targets that appeared to have ordered rotation (to more easily measure $j_*$), as well as proximity to stars that could be used for the tip-tilt correction. We observed four galaxies during March 13$^\textrm{th}$, 2019 with the OH-Suppressing Infra-Red Imaging Spectrograph (OSIRIS; \hbox{\citealp{OSIRIS}}) with the Keck II telescope using laser guide star mode. One of the targets was not useful due to the low signal to noise ratio (SNR). The other galaxies (COSMOS 110446, COSMOS 171407 and COSMOS 130477) were observed using the Hn3 filter ($H$-band at 15940-16760$\text{\AA}$) to cover the redshifted H$\alpha$ emission line using the 0.1 arcsec plate scale.

\begin{figure*}
    \centering\includegraphics[width=17.5cm]{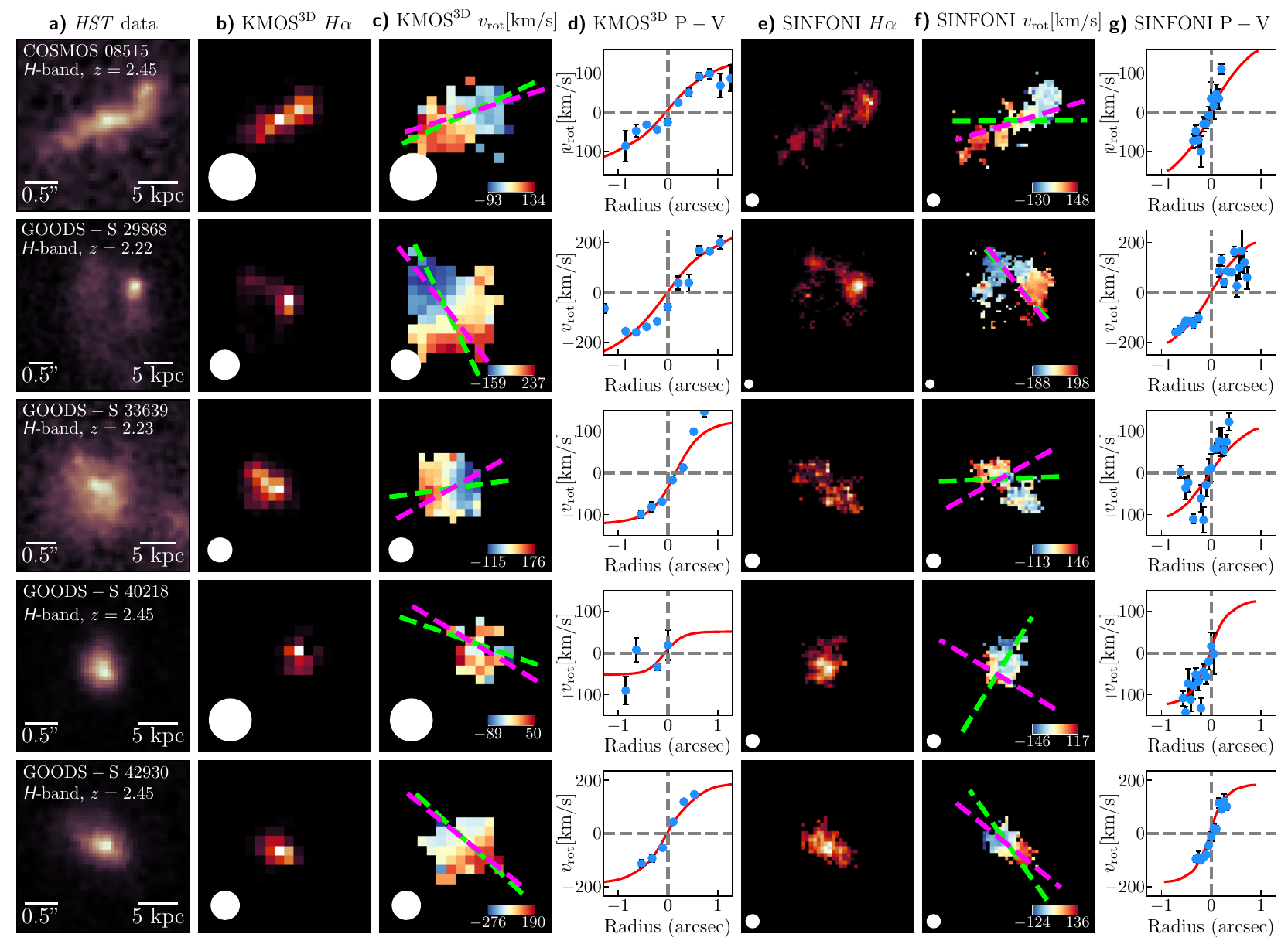}
  \caption{2D maps of the $z\sim2$ subsample. \textbf{a)} \textit{HST} broad-band continuum maps with the corresponding redshift and spatial scales. \textbf{b)} Natural seeing H$\alpha$ intensity maps from KMOS$^\mathrm{3D}$ with the size of the PSF indicated by the white circle. \textbf{c)} Extracted velocity fields of the natural seeing data where the green and purple dashed lines indicate the kinematic and morphological main axes respectively. \textbf{d)} Position-velocity (P-V) diagrams where blue dots indicate the velocities of the pixels that lay along the kinematic position axis as a function of the projected radius in arcsec. The error bars correspond to the 1-$\sigma$ error from the emission line Gaussian fit in each spaxel. The red line indicates the velocities along the kinematic position axis of the 2D model built from the individual data. The discrepancies of the model and datapoints are expected since the points are only along the major axis whereas the fit is built from the full 2D velocity field. Columns \textbf{e)}, \textbf{f)} and \textbf{g)} are the H$\alpha$ intensity, velocity field and P-V diagram from the adaptive optics sample from SINFONI.}
  \label{fig:all_galaxies_2}
\end{figure*}

\begin{table*}
	\centering
	\caption{Summary of the sample including the galaxy ID, source of the \textit{HST} photometric data, spectrograph used to acquire the data, Right Ascension and Declination in J2000 coordinates, redshift estimated from their H$\alpha$ emission (from the AO datasets), stellar mass and star formation rate (SFR) measurements from \protect\cite{Gillman} ($z\sim1.5$) and \protect\cite{SINS/zC-SINF} ($z\sim2$). We adopt typical uncertainties of the stellar properties of $0.2$ dex in $\log(M_*)$ (\citealp{Mobasher}) and $0.47$ dex in $\log(\mathrm{SFR}_\mathrm{SED})$.}
	\label{tab:sample_description}
	{\setlength{\tabcolsep}{0.55em}
	\begin{tabular}{lcccccccc}
		\hline
		Galaxy & \textit{HST} primary data & IFS Instrument & RA & Dec & $z$ & $M_*$ & SFR \\
		(ID) & & (AO+NS) & ($^\circ$) & ($^\circ$) &  & $(\log M_\odot )$ & ($M_\odot \mathrm{yr}^{-1}$) \\
		\hline
		COSMOS 110446 & $I$-band (ACS) & OSIRIS+KMOS & 149.96164 & 2.08053 & 1.5199 & 10.52 & 49\\
		COSMOS 171407 & $H$-band (DASH) & OSIRIS+KMOS & 149.89164 & 2.34849 & 1.5247 & 10.41 & 31\\
		COSMOS 130477 & $I$-band (ACS) & OSIRIS+KMOS & 150.00314 & 2.32999 & 1.4651 & 10.41 & 33\\
		COSMOS 127977 & $I$-band (ACS) & OSIRIS+KMOS & 149.90790 & 2.30060 & 1.6200 & 10.05 & 45\\
		UDS 78317 & $H$-band (CANDELS) & OSIRIS+KMOS     & 34.39170 & -5.17110 & 1.5247 & 10.49 & 45\\
		COSMOS 08515 (ZC-410041) & $H$-band (CANDELS) & SINFONI+KMOS & 150.18448 & 2.26626 & 2.4541 & 9.88 & 47\\
		GOODS-S 29868 (K20-ID7) & $H$-band (HLF) & SINFONI+KMOS & 53.12133 & 27.77461 & 2.2239 & 10.28 & 112\\
		GOODS-S 33639 (K20-ID6) & $H$-band (HLF) & SINFONI+KMOS & 53.12133 & 27.75588 & 2.2348 & 10.53 & 45\\
		GOODS-S 40218 (GMASS-2303) & $H$-band (HLF) & SINFONI+KMOS & 53.16195 & -27.72266 & 2.4507 & 9.85 & 21\\
		GOODS-S 42930 (GMASS-2363) & $H$-band (HLF) & SINFONI+KMOS & 53.16416 & -27.70990 & 2.4519 & 10.34 & 64\\
		\hline
	\end{tabular}}
\end{table*}

For each galaxy, we took eight exposures of 900 seconds for each target (seven for COSMOS 130477 due to bad weather conditions). The data were reduced using the OSIRIS data reduction pipeline DRP 4.2.0\footnote{https://github.com/Keck-DataReductionPipelines/OsirisDRP} with their appropriate rectification matrices. The OSIRIS PSF FWHM is $\sim0.1$ arcsec for the high SNR data but two galaxies (COSMOS 171407 and COSMOS 130477) had low SNR so we had to apply some spatial smoothing (discussed in Section \ref{section:Methods}) on the datacubes, leading to the degrading of their spatial resolution as shown in Table \ref{tab:sigma_all_galaxies}. The spectral resolution of the datacubes is $R\sim3800$.

Additionally, we included in our sample the two $z\sim1.5$ galaxies with similar spatial and wavelength resolutions studied in \cite{Sweet} (COSMOS 127977 and UDS 78317).

\subsection{Sample at \texorpdfstring{$z\sim2$}{z2}}

Our second set of data was obtained from the public releases of the SINS/zC-SINF \citep{SINS/zC-SINF} and KMOS$^\textrm{3D}$ \citep{KMOS3D} surveys, using the Spectrograph for Integral Field Observations in the Near Infrared \citep[SINFONI;][]{SINFONI} and KMOS. The \textit{HST} images, velocity and H$\alpha$ intensity maps are shown in Figure \ref{fig:all_galaxies_2}.

\subsubsection{\textbf{KMOS$^\textrm{3D}$ observations (natural seeing)}}

This seeing limited subsample was selected from KMOS$^\textrm{3D}$ which consists of a $0.7<z<2.7$ galaxy survey in the near-infrared with $K$-band magnitudes of $K<23$, where H$\alpha$ emission is located in the $YJ$, $H$, and $K$-band filters of KMOS. The five galaxies have a spatial pixel scale of 0.2 arcsec per pixel and the mean PSF FWHM is 0.73 arcsec. The spectral resolution is $R\sim3975$ at the location of H$\alpha$ emission.

\subsubsection{\textbf{SINS/zC observations (adaptive optics)}}

The SINS/zC-SINF AO survey consists of a sample of 35 galaxies in the $K$-band with the SINFONI instrument in adaptive optics mode presented and discussed by \cite{SINS/zC-SINF}. This is a follow-up survey to the seeing limited observations of the same galaxies and was designed to resolve the nebular emission and kinematics on scales of ${\sim}1.5$ kpc. We selected the five galaxies that had available natural seeing limited observations from KMOS$^\mathrm{3D}$ for our combination method.

The spatial pixel scale of this sample is 0.05 arcsec per pixel and the spectral resolution is $R\sim4000$ in the $K$-band. The mean PSF FWHM of this sub-sample is 0.17 arcsec.

\subsection{\textit{HST} imaging}

We use high-resolution broad-band \textit{HST} imaging of our targets from the Cosmic Evolution Survey \citep[COSMOS;][]{COSMOS}, UKIDSS UDS \citep[UDS;][]{UDS} and Great Observatories Origins Deep Survey \citep[GOODS-South;][]{Goods} extragalactic fields. 

For the galaxies in the COSMOS and UDS fields, we use the Cosmic Assembly Near-Infrared Deep Extragalactic Legacy Survey \citep[CANDELS;][]{Candels} in the F160W filter with a pixel scale of 0.06 arcsec per pixel and PSF FWHM of 0.18 arcsec. We also use COSMOS-DASH (COSMOS-Drift And SHift) which is a Wide-Field WFC3 Imaging survey in the COSMOS field \citep{DASH-HST} with the same filter and pixel scale of 0.1 arcsec per pixel and a PSF FWHM of 0.15 arcsec. For the galaxies in the GOODS-S field, we use the Hubble Legacy Field \citep[HLF;][]{HLF} also with the F160W filter and a pixel scale of 0.06 arcsec per pixel and the same PSF size.

The galaxies COSMOS 110446, COSMOS 130477 and COSMOS 127977 do not have available $H$-band images so we use $I$-band\footnote{\textit{I}-band is a bluer band so these images provide a less accurate representation of the stellar mass distribution.} \textit{HST} images with the Advanced Camera for Surveys \citep[ACS;][]{ACS} in the F814W filter and a pixel scale of 0.03 arcsec per pixel and PSF FWHM of 0.08 arcsec.

\section{Methods and analysis}
\label{section:Methods}

In this section, we describe the methods used to obtain the photometric and kinematic models and our strategy in dealing with the different resolutions and pixel scales from each dataset. We provide the details of the spatial smoothing needed in the two galaxies with low signal-to-noise. We then describe the method used to combine the different resolutions and the capabilities and limitations of our code.

\subsection{\texorpdfstring{$\mathrm{H}\alpha$}{Ha} kinematic maps}

We measure the ionized gas kinematics from each galaxy's data cube by fitting a Gaussian profile to the redshifted H$\alpha$ emission line. From the fit, we extract intensity and velocity maps with their corresponding errors. For the natural seeing data, which has higher SNR, it was possible to include the [N\textsc{ii}] doublet in the fitting routine. Given the large contamination from skylines in the near-infrared, the fitting steps must be done carefully and in a consistent manner.

First, we obtain the systemic redshift of each galaxy by fitting a Gaussian profile to the integrated spectrum in the wavelength range where we expect H$\alpha$ emission (there is good consistency of our measured redshift with respect to the literature values from the KGES and SINS surveys). Then we make a Gaussian fit along the spectral direction in each spaxel to find the redshifted line using \textsc{mpfit} (\citealp{MPFIT}) which is a least-squares curve fitting routine. In order to systematically select the real emission only, we use a SNR cut along the spectral axis by comparing the goodness of the Gaussian fit and the goodness of the fit to the continuum, as:

\begin{ceqn}
\begin{align}
   \textrm{SNR} = \sqrt{\chi^2_\textrm{cont} - \chi^2_\textrm{fit}} .
\label{eq:SNR}
\end{align}
\end{ceqn}

where $\chi^2_\mathrm{cont}$ is associated with the fit of a straight line to the continuum and $\chi^2_\mathrm{fit}$ is associated with the Gaussian profile fit to the emission line. During the Gaussian fitting, we down-weight the residual skylines in the spectrum to prevent erroneous fits and we only keep the line fits that result in velocities lower than 500 km/s and a width of at least 1.25 times the width of the skylines, similarly to the procedures employed to the KGES sample in \cite{KGES_alfie}.

\begin{figure}
	\includegraphics[width=0.45\textwidth]{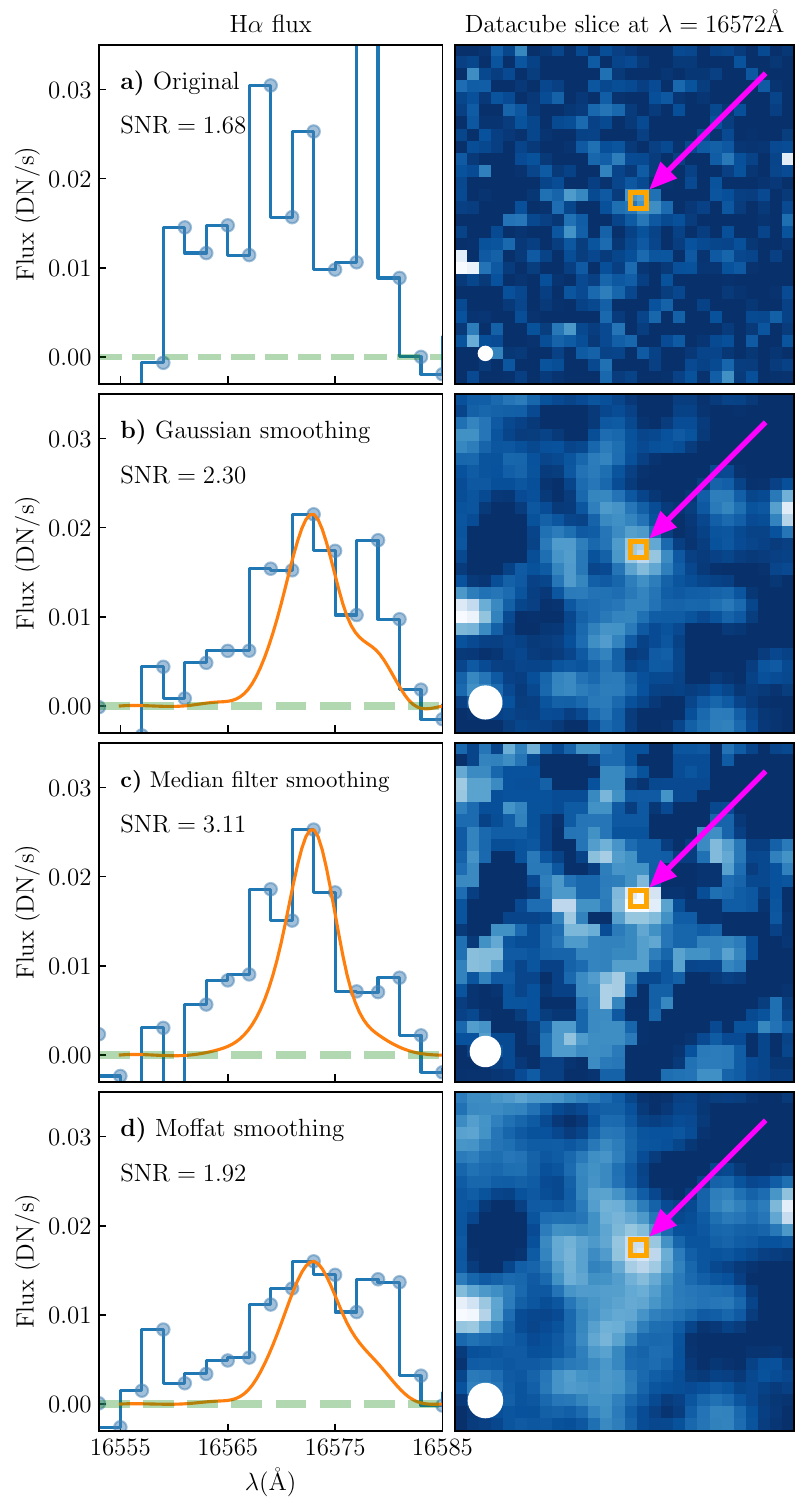}
    \caption{\textbf{Left panels:} \textbf{a)} Spectrum covering H$\alpha$ emission along the direction of a single spaxel (indicated by the orange square in the right panels) from the original data cube of COSMOS 171407. Panels \textbf{b)}, \textbf{c)} and \textbf{d)} are the spectra after applying Gaussian ($\sigma=1$ pix), median filter (size = 3 pix) and Moffat smoothing (core size = 2 pix) respectively. The green dashed line represents the continuum level. We opt for method \textbf{c)} that results in a significant improvement in the signal to noise. \textbf{Right panels:} The slice of the datacubes at $\lambda=16572 \angstrom$ to visualize the effect of the smoothing where the size of the PSF is shown at the bottom left corner. The pink arrow indicates the location of the spectrum shown in the left panel.}
    \label{fig:Smooth_all_methods_110446}
\end{figure}

The fitting routine allowed us to obtain the velocity fields for most of our galaxies. However, for the two galaxies from the adaptive optics data set (COSMOS 171407 \& COSMOS 130477) with low SNR, we had to smooth the data cubes spatially at the cost of losing spatial resolution and spreading noise through the seeing elements. We applied Gaussian, median filter and Moffat smoothing with different kernel sizes to increase SNR, as shown in Figure \ref{fig:Smooth_all_methods_110446}. The median filter with a kernel size of 3 pixels provided the largest increase in SNR (a factor of 2) and still preserved a finer resolution than the natural seeing maps so it was our chosen method.

\subsection{Spatial Resolution}

\begin{figure*}
\begin{subfigure}{16.0cm}
    \centering\includegraphics[width=0.99\textwidth]{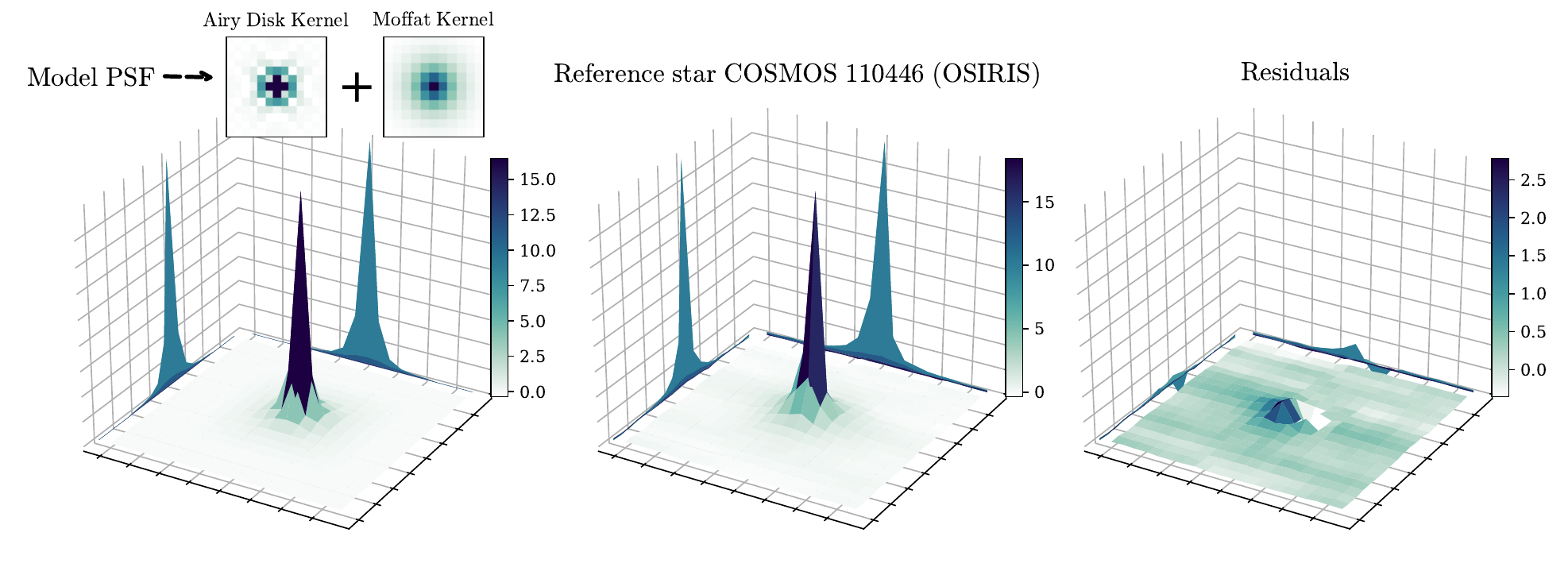}
  \end{subfigure}
  \begin{subfigure}{16.0cm}
    \centering\includegraphics[width=0.99\textwidth]{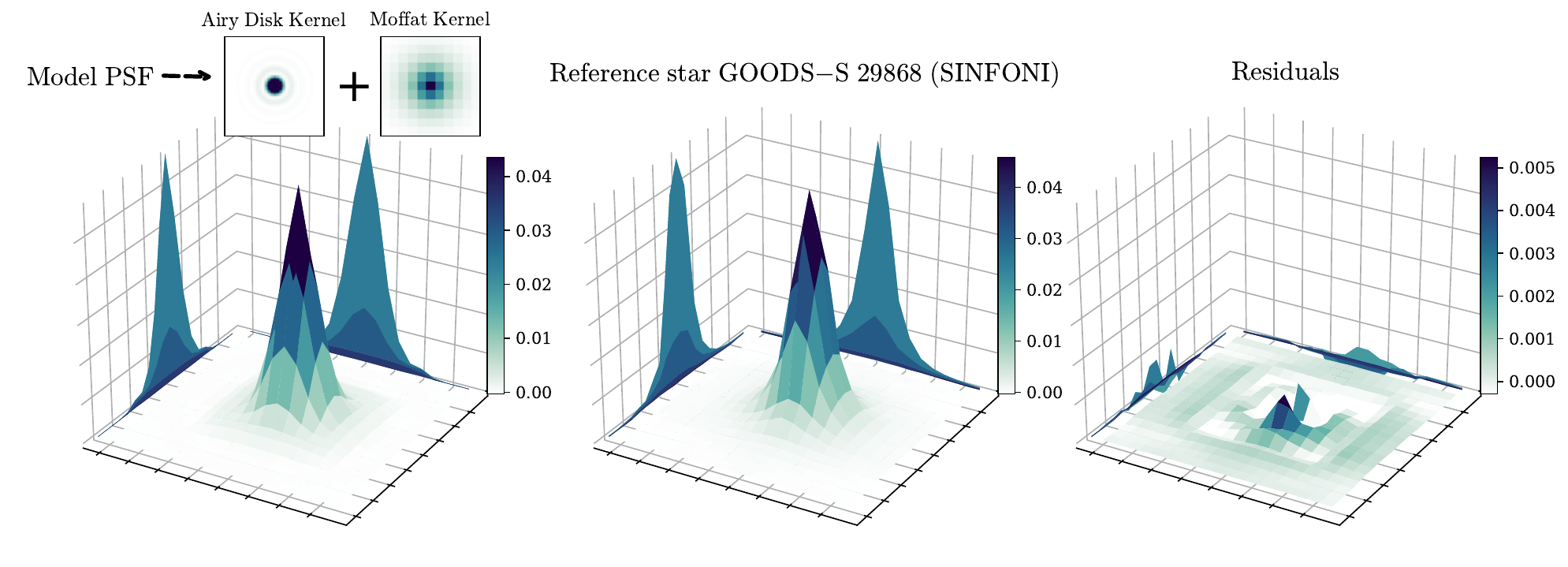}
  \end{subfigure}
  \caption{3D visualization of the PSF modelling on the acquisition stars of COSMOS 110446 (top) and GOODS-S 29868 (bottom) with the peak (Airy disk) and broad (Moffat) components. Left panels show the modelled PSF a well as the 2D visualization of the two components, middle panels show the flux from the reference stars and right panels show the residuals. The pixel scale of the instruments used in these observations (OSIRIS and SINFONI) are different as shown by the background projections of the data and models.}
    \label{fig:PSF_modelling}
\end{figure*}

Modelling the point spread functions of each dataset is crucial in order to derive intrinsic deconvolved kinematic models. Since the PSF associated with seeing limited observations is significantly different from that with adaptive optics correction, we model them separately throughout this analysis. For the natural seeing data, we fit Gaussian profiles to the acquisition stars during the time of observations (where available) to calculate the PSF size. For the adaptive optics sample, the PSF has a more complex shape with two key components (\citealp{Davies_AO}) where the peak component accounts for the adaptive optics correction and the broad component accounts for the uncorrected atmospheric blurring.

The peak component is modelled using an Airy disk profile (\citealp{Airy}) and the broad component using a Moffat profile (\citealp{Moffat}). Additionally, we quantify the efficiency of the adaptive optics correction or Strehl as the ratio of the observed peak intensity in the reference star, compared to the maximum peak intensity of the modelled star at the diffraction limit (i.e a model star with the same total flux convolved only with the measured Airy disk kernel). Figure \ref{fig:PSF_modelling} shows the PSF modelling of two of the reference stars from the adaptive optics observations and Table \ref{tab:sigma_all_galaxies} shows the PSF size of the whole sample. 

\begin{table}
\setlength\tabcolsep{3.5pt}
  \caption{Resolution parameters of the acquisition stars associated to all the IFS observations. The values marked with * are the galaxies for which we needed to apply spatial smoothing to increase SNR. Strehl in the $z\sim2$ sample are the quoted values in \protect\cite{SINS/zC-SINF}.}
%\makebox[0.5\textwidth][c]{
{\setlength{\tabcolsep}{0.35em}
    \begin{tabular}{ccccc}
\toprule
 &      & \multicolumn{1}{c}{NS} & \multicolumn{2}{c}{AO}\\
\cmidrule(lr){3-3} \cmidrule(lr){4-5}

 & Galaxy & FWHM(") &  FWHM(") & Strehl(\%) \\ \midrule

    &    COSMOS 110446 & 0.67 & 0.11 & 29\\
	&	COSMOS 171407* & 0.58 & 0.39 & 17 \\
$z\sim1.5$ &		COSMOS 130477* & 0.67 & 0.38 & 14 \\
	&	COSMOS 127977 & 0.60 & 0.11 & 29\\
	&	UDS 78317     & 0.60 & 0.11 & 29\\ \midrule
	&	COSMOS 08515 & 0.96 & 0.16 & 24\\
	&	GOODS-S 29868 & 0.60 & 0.15 & 23\\
$z\sim2$	&	GOODS-S 33639 & 0.60 & 0.20 & 13\\
	&	GOODS-S 40218 & 0.87 & 0.17 & 25\\
	&	GOODS-S 42930 & 0.62 & 0.17 & 13\\
\bottomrule
\end{tabular}}
%}
\label{tab:sigma_all_galaxies}
\end{table}

\subsection{Angular momentum}

We calculate the stellar specific angular momentum\footnote{Using $j_*$ instead of $J_*$ is a common practice in kinematic studies because it removes the stellar mass scaling $\left(j_*=\left|\frac{J_*}{M_*}\right|\right)$ whilst combining the disk size and rotation velocity profile.} $j_*$ content of the galaxies by assuming cylindrical axisymmetric rotation as in:

\begin{ceqn}
\begin{align}
j_*=\frac{\int_0^R 2\pi r^2\Sigma(r)v(r)dr}{\int_0^R 2\pi r \Sigma(r) dr}, 
\label{eq:j definition}
\end{align}
\end{ceqn}
where $\Sigma(r)$ and $v(r)$ are the one-dimensional models of the surface mass density and velocity field profiles respectively. In order to infer the resolved stellar mass profiles, we adopt the model of an exponential disk:

\begin{ceqn}
\begin{align}
   \Sigma(r) = s_d \exp
   \left(\frac{-r}{r_{d}}\right),
   \label{eq:sigma}
\end{align}
\end{ceqn}
where $r_d$ is the characteristic disk scale length and $s_d$ is the surface mass density normalization factor. The surface density model is obtained using a maximum likelihood estimation using the \textit{HST} photometric maps with $r_d$, position angle, inclination and centre coordinates as free parameters and convolving the model with the \textit{HST} PSF in each case. For the initial guess of the disk size $r_d$ we use the reported values of $r_\textrm{eff}$ from \hbox{\cite{Gillman}} and \hbox{\cite{SINS/zC-SINF}} (where $r_\textrm{eff} = 1.68 r_d$ for an exponential disk)\footnote{Note that the $r_\textrm{eff}$ values from \hbox{\cite{Gillman}} were obtained from a more general S\'ersic's profile fitting with \textsc{galfit} where $n$ is in the range $[0.2,8]$ and thus they naturally differ from our measurement.} while for the deprojection parameters we use the best fit parameters from an initial iteration of the kinematic fit. Besides defining the extent of the stellar mass density model, the measured scale length allows us to quantify the radii where the bulk of angular momentum resides in terms of a characteristic size.

There are various systematics associated with the chosen surface density modelling. First, the morphological complexity of each individual system is not accounted for, as the model does not include additional components such as a thick disk, a central bulge or bright massive clumps. The assumed exponential profile is a simplification that aids consistency in the method and facilitates a reference scale but can potentially lead to over- (or under-) estimations of the stellar mass content\footnote{The radial surface brightness profiles $\Sigma(r)$ of 6/10 galaxies in this paper follow an exponential decay to within 10\%. The rest of the sample (4/10) are within ${\sim}26$\% with GOODS-S 29868 having the largest RMS residual errors (${\sim}25.7$\%) due to its complex light distribution.}. Second, the analysis of the light distribution is limited to their \textit{H}-band (or \textit{I}-band) imaging which is not representative of all the stellar populations within the galaxies. This can lead to a systematic bias in the total stellar mass distribution and can affect the determination of the photometric centre measurement since the brightest region does not necessarily coincide with the kinematic centre of the galaxy.

Lastly, since we are interested in the \textit{specific} angular momentum, we do not need to explicitly assume a mass-to-light ratio $(M/L)_*$ as it appears on both the numerator and denominator of Equation \ref{eq:j definition}, however this assumption implies that the surface mass density profile is dependent only on one stellar population with no additional components. If the galaxies do have other components such as central bulges with a larger concentration of old stellar populations, then the galaxy mass-to-light ratio is not expected to be constant. We quantify these effects in section \ref{subsection:Effect of galaxy clumpiness}, where we use complementary \textit{J}- and \textit{H}-band imaging (available for the $z\sim2$ sample from \cite{Tacchella_sins_sizes} via private communication) to estimate the $(M/L)_*$ ratio differences in the \textit{J-H} color profiles and the effect of a central bulge and clump components in the measurement of $j_*$.

The chosen functional form of the rotation curve $v(r)$ is one characterized by an asymptotic velocity $v_\textrm{flat}$ and the distance $r_\textrm{flat}$ proposed by \cite{Boissier_2003}. This choice is based on the overwhelming evidence that rotation curves flatten at large radii for low and high-redshift disk galaxies (\citealp{Van_de_Hulst}; \citealp{Caringan}; \citealp{Catinella}; \citealp{de_Blok}; \citealp{Zasov};  \citealp{Tiley}; \citealp{Marasco2019}) and is given by:

\begin{ceqn}
\begin{align}
   v(r) = v_{\textrm{flat}} \left( 1 - \exp \left(\frac{-r}{r_{\textrm{flat}}}\right) \right).
\label{eq:v model}
\end{align}
\end{ceqn}

The velocity profile $v(r)$ is obtained from the IFS data based on the assumption that the kinematics of the ionized gas is an acceptable tracer of the motions of the stars. At $z\sim 0.1$, a direct comparative study of two star-forming galaxies considered to be analogues of high redshift clumpy disks from the DYNAMO survey (\citealp{Bassett_2014}) found that the stellar kinematics are closely coupled to the kinematics of the ionized gas. Similarly, \cite{Guerou_2017} found a similar conclusion for a sample of 17 galaxies at intermediate redshift ($0.2\leq z \leq 0.8$). However, this remains untested directly at $z>1$ given the observational challenges in measuring stellar kinematics directly (\citealp{Bezanson_2018}) and only feasible when facilities such as JWST come online. Thus, using $v(H\alpha)$ as a proxy for $v_*$ has been a common approach in multiple $z>1$ IFS studies (e.g. \citealp{Burkert}; \citealp{Swinbank}) that assume a small effect of the increased star formation activity at $1<z<3$\footnote{The increased supernovae feedback in star-forming galaxies drives outflows and winds that can affect the gas kinematics throughout the disk.}. In this work, we follow the same assumption but note that it can lead to systematic uncertainties at the $\sim0.1$ dex level as estimated with the SAMI survey at $z=0$ by \cite{Cortese2016a}.

To measure the kinematics, we use a maximum likelihood estimation where the free parameters are $r_\textrm{flat}$ and $v_\textrm{flat}$ as well as the position angle $\theta_\mathrm{PA}$\footnote{A visual inspection of the \textit{HST} maps shows that the $\theta_\mathrm{PA}$ measured from kinematics is more reliable than that of photometry, especially since it is constrained with two different datasets.} and kinematic centre coordinates [$x_{0}$,$y_{0}$]. The inclination $i$ is constrained from a fit to the surface brightness profile to break the degeneracy with the velocity field as explained in more detail in Section \ref{subsection:inclinations}.

\begin{figure*}
\begin{subfigure}{17.5cm}
  %\textbf{Adaptive Optics}
    \centering\includegraphics[width=17.5cm]{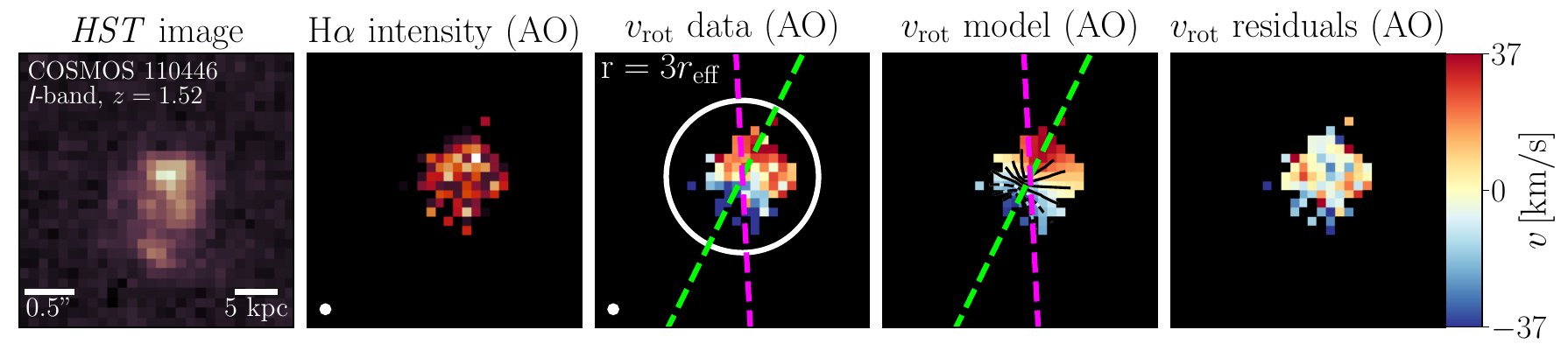}
  \end{subfigure}
\begin{subfigure}{17.5cm}
  %\textbf{Natural Seeing}
    \centering\includegraphics[width=17.5cm]{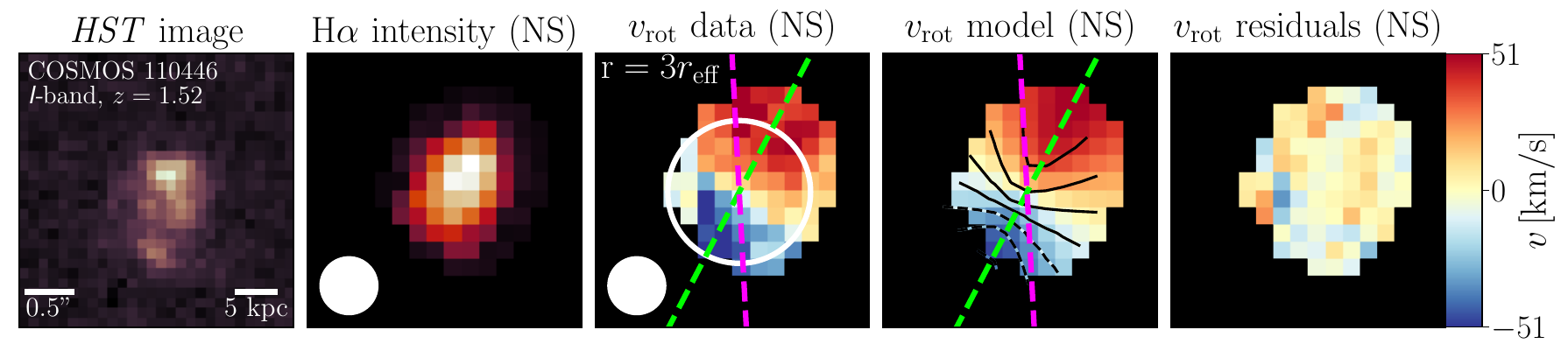}
\end{subfigure}
  \caption{Example of the kinematic fit of galaxy COSMOS 110446 at the individual resolutions (top row for AO and bottom row for NS). \textbf{From left to right:} \textit{HST} image, H$\alpha$ intensity map, velocity field, best model from the fitting routine and residuals. The purple and green dashed lines represent the morphological and kinematic main axes respectively. The color bar is the same for all of the kinematic maps. The frames are not aligned to the continuum centre but this is accounted for in the modelling.}
  \label{fig:example_fit_with_residuals}
 \end{figure*}

A two-dimensional kinematic modelling approach, where 2D velocity field models are smoothed by the PSF to resemble the extracted velocity fields (e.g. \citealp[\textsc{gipsy},][]{GIPSY}; \citealp[\textsc{ringfit},][]{RINGFIT}; \citealp[\textsc{nemo},][]{NEMO}; \citealp[\textsc{kinemetry},][]{Kinemetry}; \citealp[\textsc{diskfit},][]{Diskfit}) is only accurate if the pixel-to-pixel variations in surface brightness are negligible, which is not the case for the sample in this work. A full 3D approach, where intensity cubes are smoothed to resemble the real data cubes is computationally expensive and is limited to single resolution measurements in the publicly available codes (e.g. \citealp[\textsc{tirific},][]{Tirific}; \citealp[\textsc{dysmal},][]{Cresci2009_DYSMAL}; \citealp[\textsc{galpak$^\mathrm{3D}$},][]{Galpak};
\citealp[GalPaK$^\mathrm{3D}$,][]{Galpak};
\citealp[$^\mathrm{3D}$\textsc{Barolo},][]{DiTeodoro2015}; \citealp[\textsc{gbkfit},][]{Bekiaris2016}).

For these reasons, and in order to make a kinematic modelling that accounts for the difference in the emission line intensities, we use what we call a hybrid ``2$\nicefrac{1}{2}$D'' approach. It consists of creating model cubes using the galaxy size and kinematic parameters with an intensity profile based on the \textit{HST} imaging. We then convolve the cubes using the modelled PSF and extract the velocity fields to find those that best resemble the real $v(x,y)$ data. The accuracy of the modelling is quantified using simulations and it is explained in detail in Appendix \ref{section:appendix_code}. An example of the velocity field model and its residuals for one of the galaxies in our sample is shown in Figure \ref{fig:example_fit_with_residuals}.

We define $j_*$ as the \textit{total} stellar specific angular momentum, which represents the asymptotic value at $r\rightarrow \infty$ from Equation \ref{eq:j definition}. We find that in our sample, $j_*$ approaches the asymptotic value (at least at a ${>}95\%$ level) at 4$r_\mathrm{eff}$ as discussed in more detail in Subsection \ref{subsection:angular momentum convergence} so we take the total stellar specific angular momentum $j_\mathrm{*}$ at the arbitrary value of $r=10r_\mathrm{eff}$ which ensures convergence. Additionally we compute the commonly used approximation of $j_*$ for high-redshift galaxies from \hbox{\cite{Romanowsky}} as:
\begin{ceqn}
\begin{align}
    \tilde{j_*}\approx k_n v_{\textrm{s}} r_\textrm{eff}
\label{eq:j approx}
\end{align}
\end{ceqn}
where the factor $v_{\textrm{s}}$ is the characteristic rotation velocity evaluated at $2r_\mathrm{eff}$ and $k_n$ is a spatial weighting factor which is a function of the S\'ersic index $k(n)$ with $k_n=1.19$ for an exponential profile.

We compute angular momentum from each dataset separately but we also perform a combined measurement to improve the kinematic modelling and explore the capabilities and restrictions of the individual measurements. This combination method is explained in detail in the next section.

\subsection{The combination method}

To combine both natural seeing and adaptive optics observations of a single galaxy, we have developed a maximum likelihood estimation kinematic fitting code (see Appendix \ref{section:appendix_code}). The method combines:

\textbf{\textit{i})} Adaptive optics enabled high-resolution data (from Keck/OSIRIS and VLT/SINFONI) for the steep rising part of the rotation curve where the SNR is enough to prevail over the loss in throughput introduced by the adaptive optics correction and

\textbf{\textit{ii})} Natural seeing low-resolution data (from VLT/KMOS) to measure the flat part of the rotation curve where the bulk of angular momentum resides. At these radii, the effects of beam smearing are smaller and the SNR tends to be higher than at corresponding distances in adaptive optics assisted observations.

\begin{figure*}
	\includegraphics[width=0.98\textwidth]{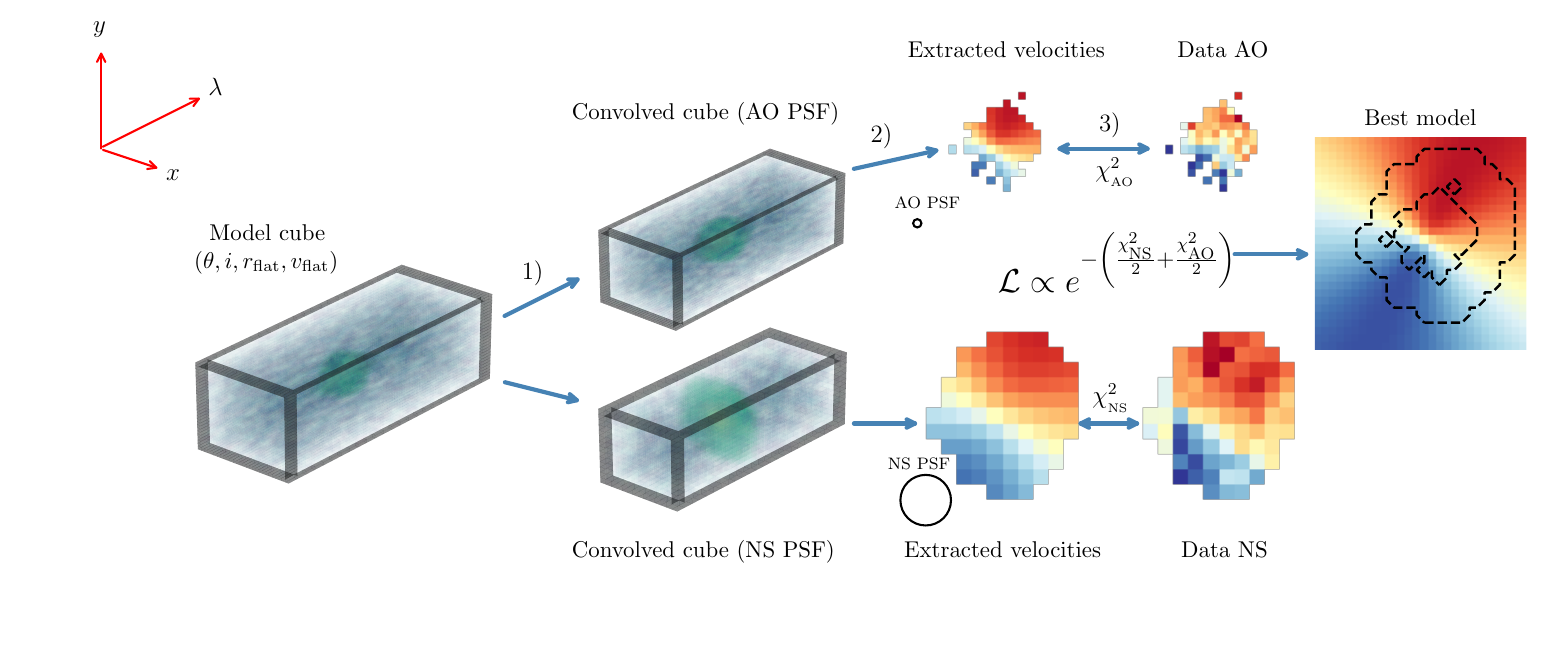}
    \caption{Representation of the kinematic modelling of our combination method. In step \textbf{1)} the model cube is convolved using the different data PSF separately. Step \textbf{2)} is the extraction of the velocity fields from the convolved model cubes. Step \textbf{3)} is the likelihood estimation based on the combination of the two $\chi^2$ values ($\chi_\mathrm{NS}^2$ and $\chi_\mathrm{AO}^2$) where the best kinematic model is obtained by maximizing that likelihood.}
    \label{fig:chi_squared_method}
\end{figure*}

The algorithm, as illustrated in Figure \ref{fig:chi_squared_method} works as follows: A model cube with the varying input parameters is created, the cube is degraded with the PSF and pixel scale of the natural seeing and adaptive optics data respectively using the corresponding convolution (single Gaussian kernel for natural seeing and Airy disk + Moffat kernel for adaptive optics), the velocity fields are extracted from the convolved cubes and the best model is the one that yields the maximum likelihood from the two datasets. The photometric maps are fitted separately in each case with a similar maximum likelihood approach.

There are multiple advantages to this method. First, it makes use of the kinematic information of the same galaxy obtained with two instruments with different sensitivities, spatial resolutions and pixel scales. Second, the convolution with the respective PSF is done separately (but simultaneously), which helps mitigate the effects of systematic errors from single modelling. Lastly, this method gives the code the freedom to centre the two models at separate spatial locations so the possible differences in the alignment of the images (of a few pixels) are accounted for. We analyse the usefulness of this combination model-fitting method by comparing it to the single model fitting as discussed in Section \ref{section:Discussions}.

\section{Results and discussions.}
\label{section:Discussions}
 
 \begin{table*}
	\centering
	\caption{Measured parameters of our kinematic and photometric modelling. Galaxies in italics are classified as mergers (see Section \ref{subsection:Kinematic state of the galaxies}) so their fit values are used for reference and comparison with the rest of the sample. \textbf{From left to right:} Galaxy ID, type of IFS data (or instrument), position angle $\theta_\mathrm{PA}$, inclination $i$, disk scalelength $r_d$, characteristic radius of the rotation profile $r_\mathrm{flat}$, asymptotic velocity $v_\mathrm{flat}$, velocity dispersion $\sigma$, total angular momentum $j_*$ and the approximation $\tilde j_*$ defined in Equation \ref{eq:j approx}. The uncertainties of the kinematic parameters as well as in $j_*$ are explained in Section \ref{Uncertainties in the angular momentum measurements}.}
	\label{tab:results}
	%\rowcolors{2}{white}{gray!20}
	{\setlength{\tabcolsep}{0.53em}
	\begin{tabular}{ccccccccccc} % four columns, alignment for each
		\hline
		Galaxy & Data & $\theta_\mathrm{PA}$ & $i$ & $r_d$ & $r_{\textrm{flat}}$ & $v_{\textrm{flat}}$ & $\sigma$ & $j_*$ & $\tilde{j_{*}}$\\% & $\Psi$ \\
		
		ID & Type & ($^\circ$) & ($^\circ$) & (kpc) & (kpc) & (km/s) & (km/s) & (kpc km/s) & (kpc km/s)\\
		\hline
		\hline
		
		\addlinespace[0.1cm]
		$\quad$         & KMOS     & $62.5\pm4.3$ & $51\pm7.1$   & $1.4\pm0.3$ & $2.35\pm1$   & $67.5\pm11.4$ & $37.5\pm6.4$  & $142.5\pm99.2$  & $164$    \\
        COSMOS 110446   & OSIRIS   & $63.8\pm7.9$ & $51.3\pm8.8$ & $1.4\pm0.2$ & $0.51\pm1$   & $48.6\pm19.5$ & $22.1\pm5.7$  & $133.5\pm62.5$  & $136$    \\
        $\quad$         & Combined & $63.3\pm2.4$ & $49.2\pm6.4$ & $1.4\pm0.1$ & $2.02\pm0.8$ & $69.3\pm7.9$  & $-$           & $154.1\pm56.4$  & $175.6$  \\
        \addlinespace[0.1cm]
		\hline
		
        $\quad$         & KMOS      & $274.5\pm8.3$ & $48.2\pm17.3$ & $2.4\pm0.2$ & $8.84\pm1.5$ & $306.9\pm35$   & $97.6\pm27.3$ & $756.5\pm251.6$  & $887.6$  \\
        COSMOS 171407   & OSIRIS    & $271.8\pm4.1$ & $46.1\pm14.6$ & $2.4\pm0.2$ & $3.31\pm1.1$ & $185.6\pm23.2$ & $81.8\pm36.8$ & $717.3\pm158.5$  & $814.9$  \\
        $\quad$         & Combined  & $274.8\pm1.3$ & $50.8\pm7.1$  & $2.4\pm0.1$ & $5.91\pm0.5$ & $226.4\pm7.2$  & $-$           & $695.8\pm112.6$  & $813.3$  \\
        \addlinespace[0.1cm]
		\hline
		
        $\quad$         & KMOS      & $277.7\pm5$    & $37.5\pm11.4$ & $2.4\pm0.2$ & $5.59\pm1.4$ & $354.1\pm19.8$ & $89\pm14.3$   & $1117.6\pm369$   & $1304.8$  \\
        COSMOS 130477   & OSIRIS    & $256.3\pm13.8$ & $39.1\pm14.1$ & $2.4\pm0.2$ & $1.35\pm1.4$ & $265\pm52.3$   & $56.3\pm10$   & $1212.7\pm281.7$ & $1268.5$  \\
        $\quad$         & Combined  & $269.2\pm1.5$  & $35.8\pm8.4$  & $2.4\pm0.1$ & $2.07\pm1.1$ & $267.7\pm7.3$  & $-$           & $1157.4\pm220.8$ & $1259.5$  \\
        \addlinespace[0.1cm]
		\hline
		
        $\quad$         & KMOS      & $140.2\pm8.2$  & $65.2\pm18.1$ & $2.1\pm0.2$ & $3.41\pm1.4$ & $250.9\pm28.1$ & $103.7\pm20.3$ & $804\pm294.5$    & $923.3$  \\
        COSMOS 127977   & OSIRIS    & $140.6\pm2.3$  & $65.1\pm5.3$  & $2.1\pm0.2$ & $2.27\pm0.5$ & $243.5\pm14.3$ & $82\pm25.1$    & $879.4\pm167.3$  & $978.3$  \\
        $\quad$         & Combined  & $137.6\pm1.2$  & $65.1\pm4.4$  & $2.1\pm0.1$ & $3.13\pm0.4$ & $252.4\pm7.4$  & $-$            & $832.8\pm132.3$  & $951.5$  \\
        \addlinespace[0.1cm]
		\hline
		
        $\quad$         & KMOS      & $166.4\pm9.7$ & $33\pm13.2$   & $1.7\pm0.2$ & $2.33\pm1.8$ & $209.6\pm28$  & $97.5\pm22.4$  & $574.9\pm200.8$  & $652.6$  \\
\textit{UDS 78317}      & OSIRIS    & $125.5\pm9.7$ & $36.9\pm17.4$ & $1.7\pm0.2$ & $0.62\pm1.6$ & $133\pm38.7$  & $105.3\pm31$   & $443.6\pm161.8$  & $452$    \\
        $\quad$         & Combined  & $143.3\pm1.4$ & $37.2\pm5.1$  & $1.7\pm0.1$ & $1.12\pm0.6$ & $128.8\pm8.5$ & $-$            & $410.5\pm112.1$  & $435.2$  \\
        \addlinespace[0.1cm]
		\hline
		
        $\quad$         & KMOS      & $204.9\pm8.5$  & $81.7\pm4$   & $2.9\pm0.2$ & $6.37\pm1.1$ & $154.1\pm20.2$ & $75.9\pm8.5$   & $603.8\pm360.7$  & $703.2$   \\
        COSMOS 08515    & SINFONI   & $180.4\pm5.7$  & $81.7\pm7.7$ & $2.9\pm0.2$ & $6.92\pm1.1$ & $245.2\pm23.5$ & $81.7\pm13.8$  & $924.5\pm197.8$  & $1072.7$  \\
        $\quad$         & Combined  & $189.3\pm3.7$  & $81.6\pm2.3$ & $2.9\pm0.1$ & $6.98\pm0.4$ & $262.9\pm7.8$  & $-$            & $987.2\pm95.9$   & $1145.6$  \\
        \addlinespace[0.1cm]
		\hline
		
        $\quad$         & KMOS      & $296\pm15$     & $64.6\pm35.9$ & $2.8\pm0.2$ & $8.2\pm1.2$  & $361.3\pm31.4$ & $88.3\pm17$    & $1185.1\pm529.6$ & $1389.3$ \\
        GOODS-S 29868   & SINFONI   & $311.7\pm2.9$  & $62.1\pm11$   & $2.8\pm0.2$ & $6.83\pm0.8$ & $346.8\pm23.6$ & $101.2\pm28.5$ & $1249.2\pm205.4$ & $1459.5$ \\
        $\quad$         & Combined  & $302.2\pm1.8$  & $62.1\pm5.9$  & $2.8\pm0.1$ & $6.89\pm0.4$ & $343.9\pm11.5$ & $-$            & $1233.5\pm103.7$ & $1441.5$ \\
        \addlinespace[0.1cm]
		\hline
		
        $\quad$         & KMOS      & $187.9\pm22.9$ & $32.1\pm9.9$ & $3\pm0.2$  & $2.39\pm2.1$ & $230\pm33$     & $78\pm10.8$    & $1259.7\pm688.8$ & $1360.2$ \\
        GOODS-S 33639   & SINFONI   & $182.3\pm18.8$ & $33.1\pm22$  & $3\pm0.2$  & $6.51\pm2.4$ & $287.3\pm32.7$ & $89.8\pm19.6$  & $1170.8\pm412.7$ & $1205.5$ \\
        $\quad$         & Combined  & $187.9\pm4.3$  & $34.3\pm6$   & $3\pm0.1$  & $6.11\pm1.1$ & $297.8\pm12.1$ & $-$            & $1247.7\pm143.8$ & $1290.1$ \\
        \addlinespace[0.1cm]
		\hline
		
        $\quad$         & KMOS      & $160.9\pm16.8$ & $52.4\pm16.6$ & $1.1\pm0.2$ & $0.51\pm1.6$ & $66.3\pm30.4$  & $109.7\pm40.2$ & $141.2\pm153.7$ & $145.7$  \\
\textit{GOODS-S 40218}  & SINFONI   & $238.8\pm17.1$ & $52\pm18.1$   & $1.1\pm0.2$ & $2.52\pm1.3$ & $166.1\pm40.6$ & $123.3\pm29$   & $242.1\pm103.6$ & $282.5$  \\
        $\quad$         & Combined  & $232.4\pm6.4$ & $49\pm11.6$   & $1.1\pm0.1$ & $1.45\pm0.7$ & $112.1\pm10.6$ & $-$            & $201.3\pm44.1$  & $227.8$  \\
        \addlinespace[0.1cm]
		\hline
		
        $\quad$         & KMOS      & $316.8\pm16$  & $58.8\pm28.3$ & $1.6\pm0.2$ & $2.97\pm1.7$ & $229.4\pm29.7$ & $70.1\pm11.2$ & $532.6\pm321.8$  & $616.5$  \\
        GOODS-S 42930   & SINFONI   & $305.1\pm5.9$ & $58.8\pm11.4$ & $1.6\pm0.2$ & $1.93\pm0.5$ & $219.4\pm24.1$ & $64.5\pm11.6$ & $587.3\pm162.5$  & $659.9$  \\
        $\quad$         & Combined  & $311.4\pm2.8$ & $57.7\pm16.6$ & $1.6\pm0.2$ & $2.38\pm0.5$ & $229.4\pm11.3$ & $-$           & $577.1\pm62.2$   & $659.1$\\
		\hline

	\end{tabular}}
\end{table*}

The high-redshift sample studied in this paper is to date the largest sample with angular momentum measurements from the combination of high- and low-spatial resolution data. Consequently, we can address the relevance of a combination method. Throughout this section, we discuss the effect of combining the different resolutions in the determination of rotation velocity profiles, kinematic state as well as the convergence of $j_*(r)$ to the asymptotic value $j_*$. Table \ref{tab:results} shows the main results of the modelling and additional figures are included in the supplementary (online) section.

\begin{figure*}
	\includegraphics[width=0.98\textwidth]{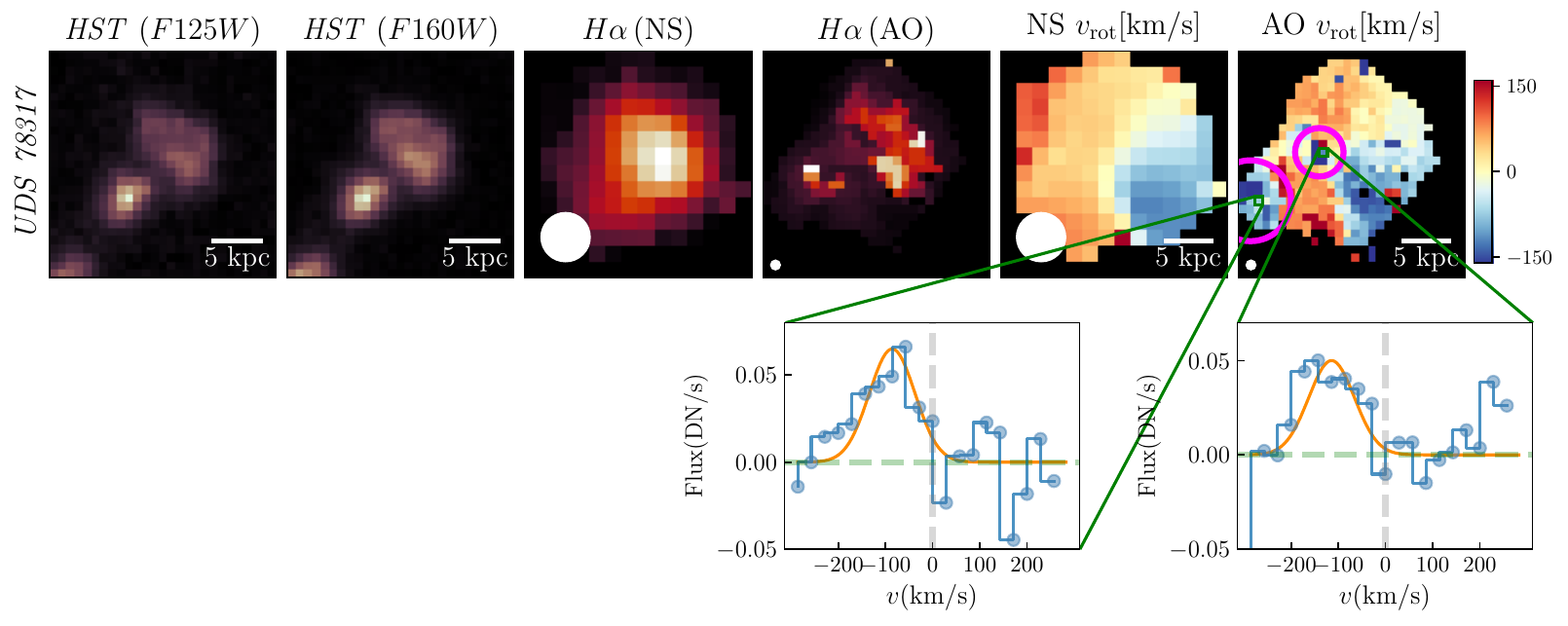}
    \caption{Various maps of galaxy \textit{UDS 78317} which appears to be a rotating disk from natural seeing kinematics but shows a complex kinematic state at high-resolution as well as from photometry. \textbf{From left to right:} \textit{J} and \textit{H}-band $HST$ imaging, H$\alpha$ intensity maps at low and high resolution and velocity fields at the two resolutions sharing the same color bar. Purple circles in the adaptive optics map indicate the kinematic components that are absent in the natural seeing counterpart, with the bottom panels showing the detected H$\alpha$ emission as a function of radial velocity in the zoomed-in pixels. The maps have different pixel scales as well as different PSF sizes as shown by the white circles at the bottom left of each panel.}
    \label{fig:maps with bulges}
\end{figure*}

\subsection{Classification as rotating disks}
\label{subsection:Kinematic state of the galaxies}

High-spatial resolution observations, to sub-kpc scales, are necessary to accurately determine the kinematic state of a galaxy and resolve large star-forming clumps. (\hbox{\citealp{Law}}; \hbox{\citealp{Genzel_2008}}; \hbox{\citealp{Epinat_2012}}; \hbox{\citealp{Livermore}}; \hbox{\citealp{Mieda}}; \hbox{\citealp{KMOS3D}}; \protect\citealp{Fisher}; \protect\citealp{SINS/zC-SINF}; \protect\citealp{Sweet}). In our full sample, the mean PSF FWHM of the adaptive optics sample (${\sim}0.15$ arcsec) corresponds to ${\sim}1.3$ kpc whereas the seeing-limited sample (${\sim}0.69$ arcsec) corresponds to ${\sim}5.7$ kpc. In the high-resolution data, additional components with proper line of sight velocities become visible as indicated in Figure \ref{fig:maps with bulges}. 

\begin{figure}
	\includegraphics[width=0.47\textwidth]{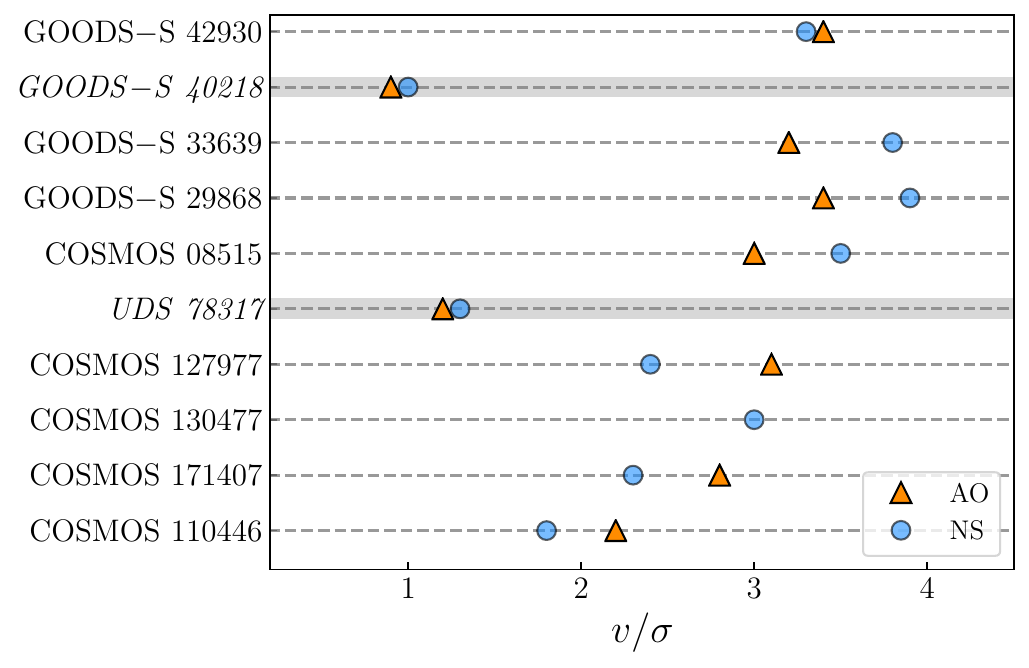}
    \caption{$v/\sigma$ ratios for the ten galaxies from the NS and AO datasets used for the disk/merger classification. Blue points indicate the measurements from low resolution, and orange triangles indicate the measurements from the adaptive optics data. Grey horizontal lines indicate the mergers \textit{UDS 78317} and \textit{GOODS-S 40218}, which have the lowest values in $v/\sigma$ ($<1.5$) at both resolutions.}
    \label{fig:v_over_sigma}
\end{figure}

Another way to assess the kinematic state of the sample is by determining the dynamical support using the $v/\sigma$ ratio. Following \cite{Genzel_2006}, a system is considered dispersion supported for $v/\sigma<1$, and rotation supported for $v/\sigma>1$. However, as discussed by \cite{Tiley_2019}, a galaxy is truly supported by its rotation when $v/\sigma>3$. In our sample, we use $v_\mathrm{flat}$ for the velocity $v$ and for the velocity dispersion $\sigma$ we take the median of the annulus at $r=1.5r_\mathrm{eff}$ in the moment two maps, where the effects of beam smearing are lower. The $v/\sigma$ ratios at both resolutions are shown in Figure \ref{fig:v_over_sigma}.

A commonly used approach to classify high-redshift galaxies as disks from their 2D kinematic maps (e.g \hbox{\citealp{SINS/zC-SINF}}; \hbox{\citealp{Wisnioski_2019}}) consists of a set of criteria based on their rotation and dispersion profiles. These criteria (as proposed by \hbox{\citealp{KMOS3D}}) are: 

\textbf{1.} a smooth monotonic velocity gradient across the galaxy, defining the kinematic axis;

\textbf{2.} a centrally peaked velocity dispersion distribution with maximum at the position of steepest velocity gradient, defining the kinematic centre;

\textbf{3.} dominant rotational support, quantified by the $v/\sigma$ ratio; 

\textbf{4.} co-aligned morphological and kinematic major axes (a.k.a. kinematic misalignment); 

\textbf{5.} spatial coincidence of the kinematic and morphological centres.

We follow these criteria, the results from running the kinematic code and the visual inspection of the photometric maps to classify our sample: Galaxy UDS 78317 has four kinematic poles in the high-resolution map, the morphological and kinematic centres do not align, there is a large misalignment of the kinematic and morphological positions angles and it has a bright companion as seen in the \textit{HST} images (see Figure \ref{fig:maps with bulges}), thus we classify it as a major merger in agreement with \hbox{\cite{Sweet}}. GOODS-S 40218 is very compact and only a few pixels surpass the SNR threshold in the Gaussian fit at the two resolutions. Given these limitations, and based on the disordered velocities from both datasets and low $v/\sigma$ ratios (1 for NS and 0.9 for AO), we classify this galaxy as a merger noting the large uncertainties. For the remainder of the analysis, we label these two galaxies in italics (\textit{UDS 78317} and \textit{GOODS-S 40218}) throughout the paper to distinguish them from the rotating disks.

We treat COSMOS 171407 as a disk due to its ordered rotation but we note its complex structure in the AO H$\alpha$ intensity map and a kinematic anomaly that is completely absent in its natural seeing counterpart, which is consistent with a minor merger. Galaxy GOODS-S 33639 has only a few pixels with enough SNR for the H$\alpha$ Gaussian fit as well as poorly constrained inclinations from photometry and kinematics. However, it fulfils the conditions to be classified as a disk. The rest of the galaxies (COSMOS 110446, COSMOS 130477, COSMOS 127977, COSMOS 08515, GOODS-S 29868 and GOODS-S 42930) are classified as disks. In summary, 2/10 galaxies in our sample are classified as mergers (\textit{UDS 78317} and \textit{GOODS-S 40218}), 8/10 are classified as disks with two of them (COSMOS 171407 and GOODS-S 33639) being largely uncertain. We discuss each individual case in further detail in the supplementary (online) section.

\begin{figure*}
    \centering
    \includegraphics[width=17.5cm]{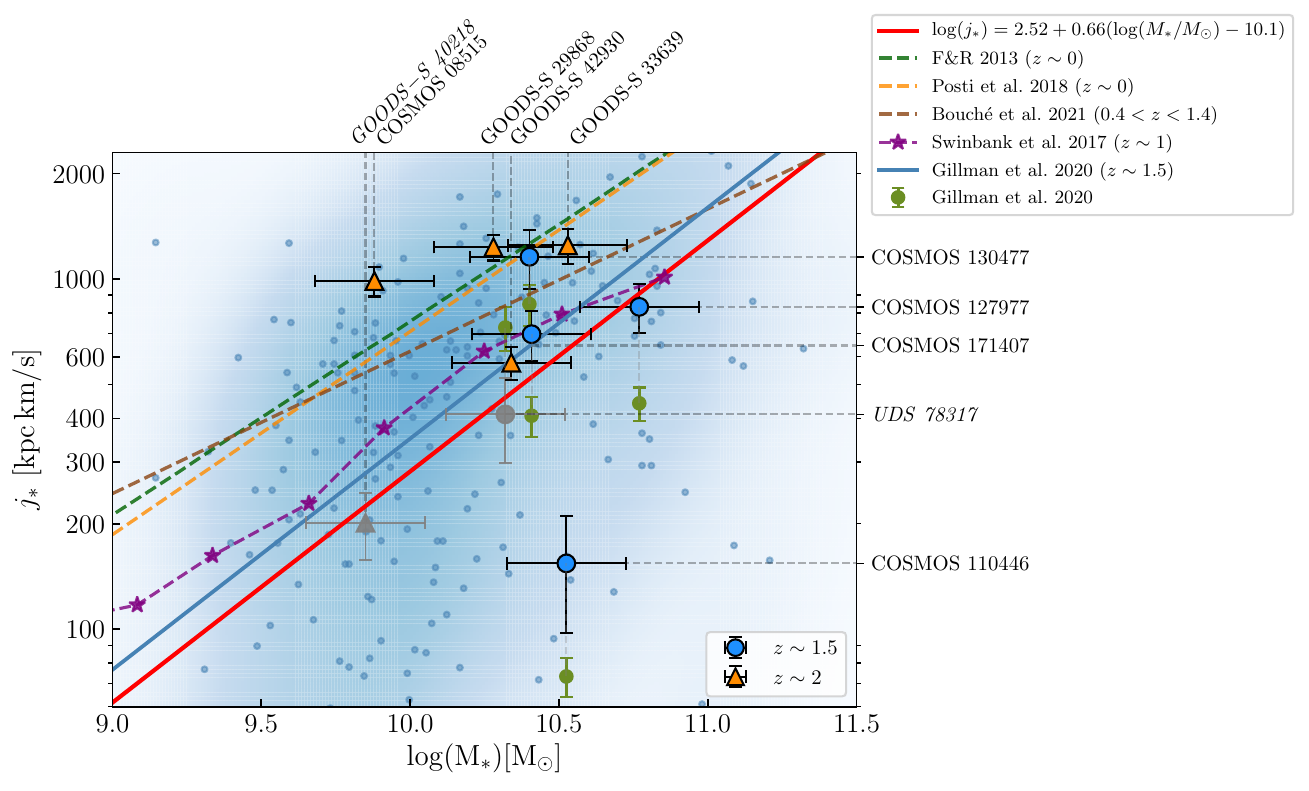}
    \caption{Stellar specific angular momentum $j_*$ vs stellar mass $M_*$ with the results from the combined analysis indicated by the blue circles ($z\sim1.5$) and orange triangles ($z\sim2$). The two grey data points correspond to the galaxies classified as mergers so they are only shown for reference. Solid red line indicates the fit to the combined data of the eight disk galaxies in the form $\log(j_*)=\alpha + \beta (\log(M_*/M_\odot)-10.1)$ for a fixed $\beta=0.66$. Green and orange dashed lines correspond to the relations at $z\sim0$ found for disks by \protect\cite{F_and_R_2013} and \protect\cite{Posti_2018} respectively. Purple stars correspond to the binned measurements from \protect\cite{Swinbank} at $z\sim1$ with MUSE and KMOS and brown dashed line corresponds to the intermediate redshift measurements ($0.4<z<1.4$) from \protect\cite{Bouche_2021}. The blue solid line corresponds to the relation found for a large sample of seeing limited data (including non-disks) by \protect\cite{Gillman} at $z\sim1.5$ with the large scatter shown by the background blue dots and blue shading is the background density determined using a Gaussian kernel density estimation. The green points represent the $j_*$ measurements of the five galaxies in our sample that overlap with the \protect\cite{Gillman} measurements.}
    \label{fig:j_vs_M}
\end{figure*}

From similar disk/merger classifications, recent kinematic surveys have measured the fraction of disks ($f_\mathrm{disk}$) at cosmic noon and at later times. At an intermediate redshift range of $0.9\leq z \leq 1.6$, \cite{Epinat_2012} found $f_\mathrm{disk} \sim44\%$ with the MASSIV survey for 50 star-forming galaxies, most of which were observed at low spatial resolution. Some larger (seeing-limited only) surveys at $z>1$ have found disk fractions as large as $70\%-80\%$ (e.g \citealp{KMOS3D, Wisnioski_2019}; \citealp{Stott_2016}). However, that range decreases to $f_\mathrm{disk}=50\%-60\%$ for the 35 galaxies assisted by adaptive optics studied in \cite{SINS/zC-SINF}. Our results are consistent with those of the $z>1$ seeing limited surveys. However, it is worth noting that the five galaxies chosen from the KGES sub-sample ($z\sim1.5$) were galaxies that at low resolution appeared to be rotating, had flat rotation curves and well-behaved dispersion profiles. This introduces a selection bias, so the 1:5 ratio of galaxies misclassified as disks is a lower bound on this systematic uncertainty. For larger samples, with a wider range in parameter space, this ratio is likely to be higher.

\subsection{Angular momentum measurements}
\label{subsection:angular momentum}

We can use our angular momentum measurements to assess their consistency with known relations. One of these relations is the scaling of stellar specific angular momentum and stellar mass typically known as the ``Fall relation'' (\citealp{fall83}). The power law that relates these quantities is well established at $z\sim0$ in the form $j_* \propto M_*^{\beta}$ with $\beta\sim 0.58$ for disks and $\beta\sim0.83$ for bulges (\citealp{Fall_Romanowsky_2018}).

In Figure \ref{fig:j_vs_M} we plot the eight disks in the $j_*$ vs $M_*$ map as well as the results from other studies that make predictions on the Fall relation at various redshift ranges (\citealp{F_and_R_2013}; \citealp{Swinbank}; \citealp{Posti_2018}; \citealp{Gillman}; \citealp{Bouche_2021}). Since $j_*$ is measured more precisely using both resolutions (as shown in Appendix \ref{section:appendix_code}), we use those values to find the scaling law that best describes the 8 disk galaxies (we exclude the mergers \textit{UDS 78317} and \textit{GOODS-S 40218}). We use the relation $\log(j_*)=\alpha + \beta (\log(M_*/M_\odot)-10.1)$ where $\beta$ is the slope from the commonly used $j_*\propto M_*^\beta$ relation and $\alpha$ is the normalization. In Table \ref{tab:results_j_vs_M} we show the best-fit parameters to the unconstrained model (free $\alpha$ and $\beta$) as well as a model with a fixed power-law slope $\beta=2/3=0.66$.

From our small sample, we find a scatter of 0.38 dex, which occupies a similar parameter space in the $j_*$ vs $M_*$ relation as the majority of the measurements from \cite{Gillman} where they used 201 spatially resolved (seeing limited) galaxies at $z\sim1.5$ with a scatter of 0.56 dex. The constrained fit ($\beta=2/3=0.66$) to their sample has a small positive offset in the vertical axis ($\Delta \alpha=0.08\pm 0.11$) as compared to our constrained fit.

\begin{table}
	\centering
	\caption{Best fit parameters for our eight disk galaxies of the form $\log(j_*)=\alpha + \beta (\log(M_*/M_\odot)-10.1)$. We show the results from the KGES sample (\citealp{Gillman}) in the third and fourth row for a direct comparison.}
	\label{tab:results_j_vs_M}
	\begin{tabular}{llc}
		\hline
		Data & $\beta$ & $\alpha$\\
		\hline
		Constrained fit ($\beta=2/3$) & $0.66$ & $2.52\pm0.11$\\
		Unconstrained fit & $-0.49\pm0.31$ & $2.92\pm0.11$\\
		KGES ($\beta=2/3$) & $0.66$ & $2.6\pm0.03$\\
		KGES unconstrained & $0.53\pm0.1$ & $2.63\pm0.04$\\
		\hline
	\end{tabular}
\end{table}

The small difference could be explained by the different mass distribution estimates from their photometric models which are more general (S\'ersic models with $n=0.2-8$) than our assumed exponential disks ($n=1$). There are also systematic differences associated to the chosen velocity profile, which in their case is measured at the kinematic position angle instead of the full velocity field. In their measurement, there could be objects that show ordered rotation at low resolution but that could be mergers when seen with high spatial resolution. We note that the limited number of objects in our sample and the small mass range provide little diagnostic ability on the slope of the $j_*$ vs $M_*$ relation. 

According to the Fall relation, low-$M_*$ ($\leq 10^{10}M_\odot$) galaxies have a low $j_*$ content so they are more susceptible to fragmentation due to a higher prevalence of galaxy-wide instabilities. Thus, they are better candidates for addressing the effect of angular momentum in driving clumpiness and are critical for constraining the low-$j_*$ regime in the $j_*$ vs $M_*$ relation. Constraints in the low-mass regime ($8\leq \log (M_*/M_\odot) \leq 10.5$) of the Fall relation have been found at an intermediate redshift range ($0.2 \leq z \leq 1.4$) by \hbox{\cite{Contini_2016}} and \hbox{\cite{Bouche_2021}} using seeing-limited observations with the Multi Unit Spectroscopic Explorer (MUSE; \hbox{\citealp{Bacon_2010_MUSE}}). In our sample, the only galaxy with trustworthy results in the low-$M_*$ regime is COSMOS 08515. Part of the reason is that the $z\sim1.5$ sample is biased towards high-$j_*$, as the targets chosen to be observed with adaptive optics with OSIRIS were the ones that appeared to be large disks with ordered rotation from the natural seeing observations in the KGES survey. A combined (NS+AO) analysis of galaxies in the low-mass regime is thus necessary to impose better constraints on the $j_*$ vs $M_*$ relation at the redshift of interest. 

\begin{figure}
	\includegraphics[width=0.47\textwidth]{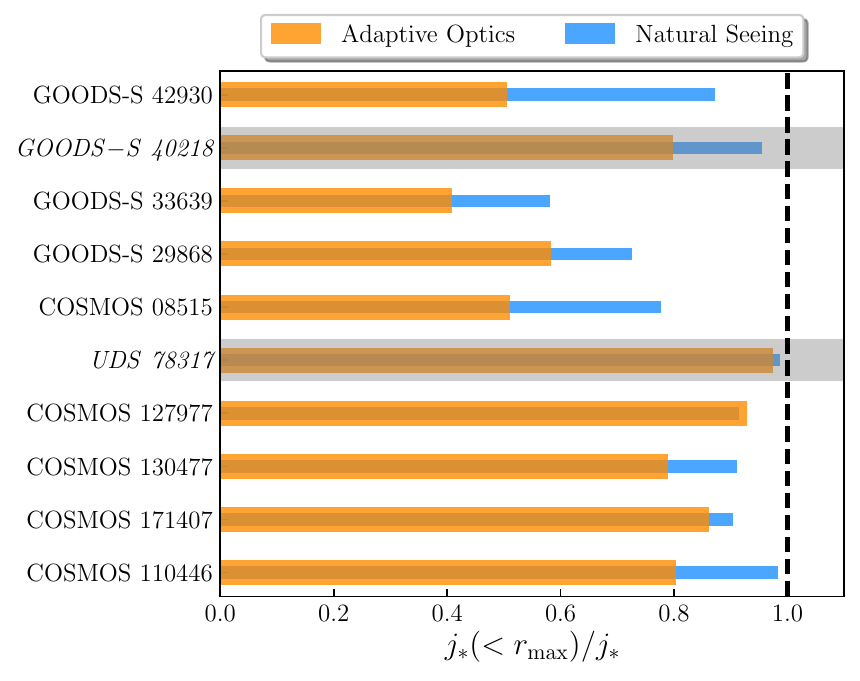}
    \caption{Angular momentum contained within the boundaries of the different data (extent of the velocity maps) in the combined analysis. The dashed vertical line represents the total stellar specific angular momentum $j_*$. The adaptive optics assisted data (orange) accounts only for about 72\% of $j_\mathrm{*}$ whereas the natural seeing data accounts for about 86\% of $j_\mathrm{*}$. Grey shades indicate the galaxies that cannot be treated as rotating disks.}
    \label{fig:j_contributions}
\end{figure}

Another important measurement is the contribution of each data type in the total measurement of $j_*$ in terms of their spatial extent. In the combined analysis, we measure the contribution of the individual data sets by measuring the amount of angular momentum in the model enclosed within the boundaries of the high- and low-resolution maps respectively as shown in Figure \ref{fig:j_contributions}. The adaptive optics data, with its sensitivity limitations, contributes to a mean value of ${\sim}72\%$ of total $j_*$ whereas the natural seeing maps reach ${\sim}86\%$. The combination is thus ideal since the adaptive optics data aids a more precise measurement of the rotation profile in the inner radius, and natural seeing aids the measurement of the bulk of $j_*$ which is built at large radii, where the rotation curve is expected to flatten and the effects of beam smearing are smaller.

\subsection{Uncertainties in the angular momentum measurements}
\label{Uncertainties in the angular momentum measurements}

In order to quantify the uncertainties in the measurement of $j_*$ associated with the individual and combined fits, we resample each galaxy $100$ times and run the same analysis for each iteration. To do this, we create a perfect model cube from the best-fit parameters using the pixel scale of the data and masking out empty pixels. We then convolve with the PSF and extract the velocity field from the convolved cube. Next, in order to recreate the observational errors, we take the residual velocity field (the difference between the best kinematic model and the real velocity field) and resample it with a normal distribution in each pixel. This new error map is then added to the model velocity field. This empirical approach better captures the contribution of artifacts and substructure that are not covered by the kinematic model, since they manifest in the residual velocity fields. If we used the Gaussian line fit errors instead, we would get unrealistically low uncertainties since the residuals are typically 0-70 km/s compared to the formal errors of $\sim15$ km/s from the Gaussian line fits. We also resample the photometric model, where the noise is estimated from the RMS of the background regions in the \textit{HST} images.

For each realization of the model, we recalculate $j_*$ in the individual and combined cases with the advantage of knowing the actual value of $j_*$ from the input model $j_\mathrm{real}$. The scatter in the deviations from this real value will give us an estimation of how good the measurement is in each case. This exercise quantifies the random error that would be obtained if a galaxy corresponding to the best fit model was observed with the same data mask, substructure and similar noise levels. It also allows us to assess the systematic error levels in the methodology.

\begin{figure}
	\includegraphics[width=0.47\textwidth]{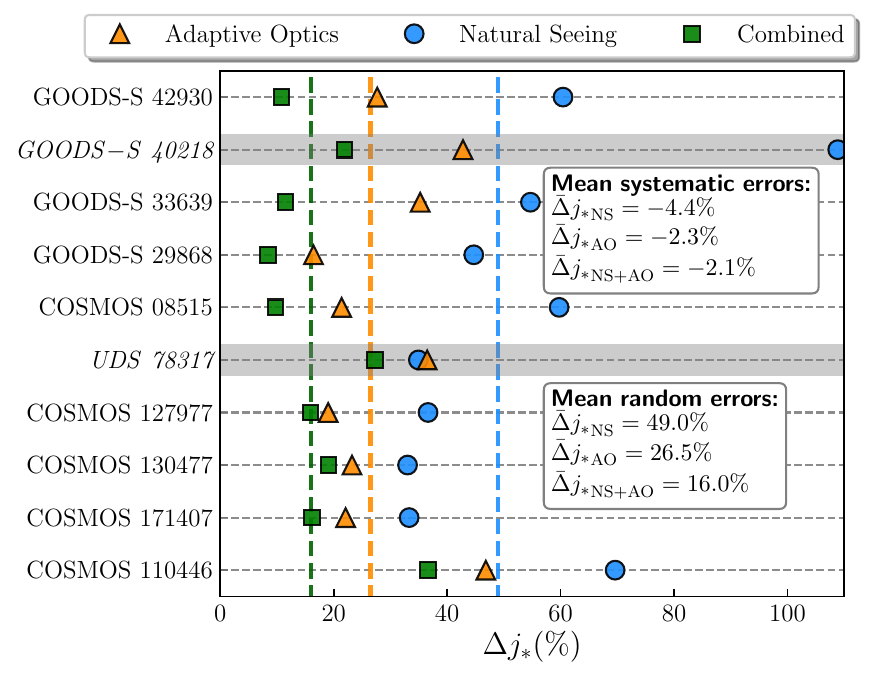}
    \caption{Uncertainties in $j_*$ (\%) obtained from resampling the best models in the individual and combined cases. The mean errors of the sample are indicated by the vertical dashed lines and are shown at the bottom right corner while the mean systematic errors are shown in the top right.}
    \label{fig:j_errors}
\end{figure}

The systematic error in $j_*$ for each galaxy is calculated from the median of the errors from the 100 realizations of the model. For the full sample, the average systematic errors are $-4.4\%$ for the natural seeing data, $-2.3\%$ for the adaptive optics data and $-2.1\%$ in the combination. More importantly, by taking the standard deviation of the individual errors from the $100$ realizations we obtain the error $\Delta j_*$ for each galaxy as well as the errors in all parameters reported in Table \ref{tab:results}. The mean percentage error for the measurements using natural seeing data is 49\%, in the case of adaptive optics 26.5\% and as for the combined resolutions it is 16\% as shown in Figure \ref{fig:j_errors}. This clear improvement can be explained by the better constraints imposed in the inner part of the rotation curves where the adaptive optics data is essential as well as the larger spatial extent of the combined data.

\subsection{Geometrical parameters}
\label{subsection:inclinations}

We pay special attention in determining the position angle and inclination as they are critical quantities in the deprojection of the velocity field. In the case of the inclination, there is a well-known degeneracy between rotation velocity and inclination that arises from kinematic modelling alone (e.g. \citealp{Begeman}; \citealp{Epinat2010}; \citealp{Kamphuis}; \citealp{Bekiaris2016}). In order to break this degeneracy, and noting that the uncertainty in the inclinations is one of the largest caveats in our analysis, we adopt a consistent approach to that of \citealp{Swinbank} for relative comparison where we constrain the inclination $i$ from the galaxies' surface brightness profiles as done by other kinematic codes (e.g. \citealp[GalPaK$^\mathrm{3D}$,][]{Galpak}). For axisymmetric disk-like galaxies, the axis ratio can be used to estimate the inclination (\citealp{Holmberg_inclination_axis_ratio}) as:

\begin{ceqn}
\begin{align}
   \cos ^2 i = \frac{q^2-q_0^2}{1-q_0^2},
\label{eq:inclination from axis ratio}
\end{align}
\end{ceqn}

where $q=b/a$ is the observed axis ratio and $q_0=c/a$ is the true axis ratio of an edge on system that depends on morphological type (\citealp{Heidmann}; \citealp{Fouque}). In the thin-disk approximation $q_0\approx0$, so this relation reduces to $\cos i \approx b/a$, however, star-forming galaxies at high redshift show vertical dispersion (\citealp{Law}; \citealp{Forster_2008_no_AO}) that corresponds to thick disks with $q_0\approx0.2$ so we adopt this value for consistency with previous studies at high redshift (e.g. \citealp{Emily}; \citealp{Harrison_2017}; \citealp{Gillman}).

\begin{figure}
	\includegraphics[width=0.47\textwidth]{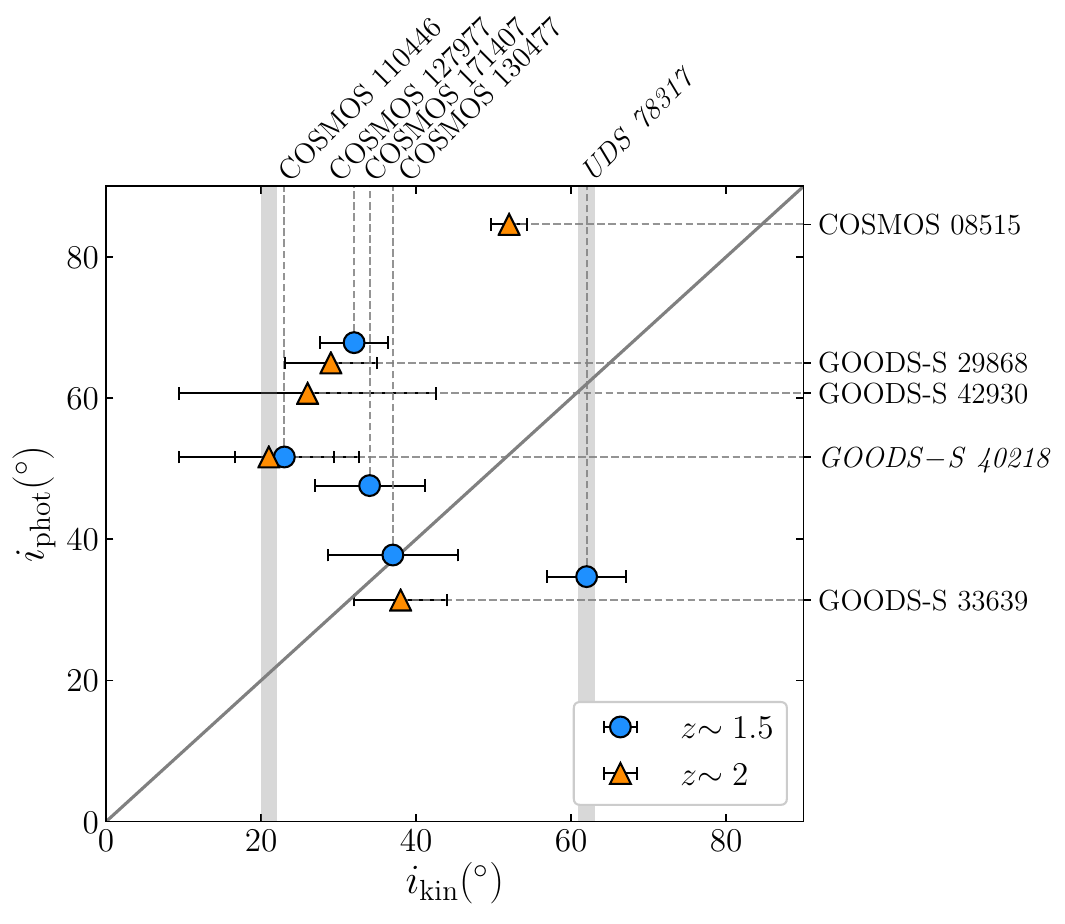}
    \caption{Direct comparison between the inclinations measured from the surface brightness profiles ($i_\mathrm{phot}$) vs the measurements from the kinematic modelling alone ($i_\mathrm{kin}$). Diagonal line represents the one-on-one correspondence.}
    \label{fig:inclinations}
\end{figure}

We compare the inclinations measured individually from photometry ($i_\mathrm{phot}$) and from the kinematic modelling alone ($i_\mathrm{kin}$) where we find that the kinematic modelling underestimates the inclinations for 8/10 of the galaxies in the sample as seen in Figure \ref{fig:inclinations}. The values of $i_\mathrm{kin}$ of those eight galaxies are low $\leq40^\circ$, where the degeneracy with the velocity field (characterized by $v_\mathrm{flat}$) escalates more rapidly as discussed by \cite{Bekiaris2016}. This degeneracy leads to an increase on the $v_\mathrm{flat}$ parameter and thus an overall increase in the measurements of $j_*$. Besides the known degeneracy, such low inclinations deviate significantly from the mean expected inclination of disk galaxies given their random orientations in space. Following geometrical arguments, this mean expected value is $<\sin i>=0.79$ (see Appendix of \citealp{Law_2}) which is consistent with the mean value of the photometric inclinations (and bootstrap errors) of our sample $<\sin i_\mathrm{phot}>=0.77\pm0.06$. On the other hand, the mean kinematic inclinations largely disagree with $<\sin i_\mathrm{kin}>=0.56\pm0.05$.
    
Moreover, the predictions on the Fall relation are significantly different when using the kinematic or photometric inclinations. If we used best fit parameters from the kinematic model alone (without constraining the inclination from photometry), we find a zero-point of 2.75 for the fixed power slope $\beta=2/3$, which has a positive 0.23 offset with the with respect to the fit with the photometric inclinations. This experiment confirms the necessity of imposing constraints in the inclinations from the surface brightness profiles.

In the case of the position angle $\theta_\mathrm{PA}$, we used the values measured from the kinematic modelling since the equivelocity contours of the ``spider diagram'' (e.g. fourth column in Figure \ref{fig:example_fit_with_residuals}) are indicative of a clear main axis. On the other hand, the position angle estimated from photometry depends on the light distribution which is more susceptible to global non-axisymmetric features within the disk (\citealp{Palunas_Williams}). The $\theta_\mathrm{PA}$ from kinematics is thus better constrained, especially in the combined case.

\begin{comment}
We calculate the misalignment between the position angle obtained from the kinematic fit and the fit to the photometric maps of the galaxies as defined \cite{Franx} as

\begin{ceqn}
\begin{align}
    \sin \Psi = \left|\sin\left( {\theta_\mathrm{PA}^\mathrm{kin}}-{\theta_\mathrm{PA}^\mathrm{phot}}\right)\right|
\end{align}
\end{ceqn}

\end{comment}

\begin{figure}
	\includegraphics[width=0.48\textwidth]{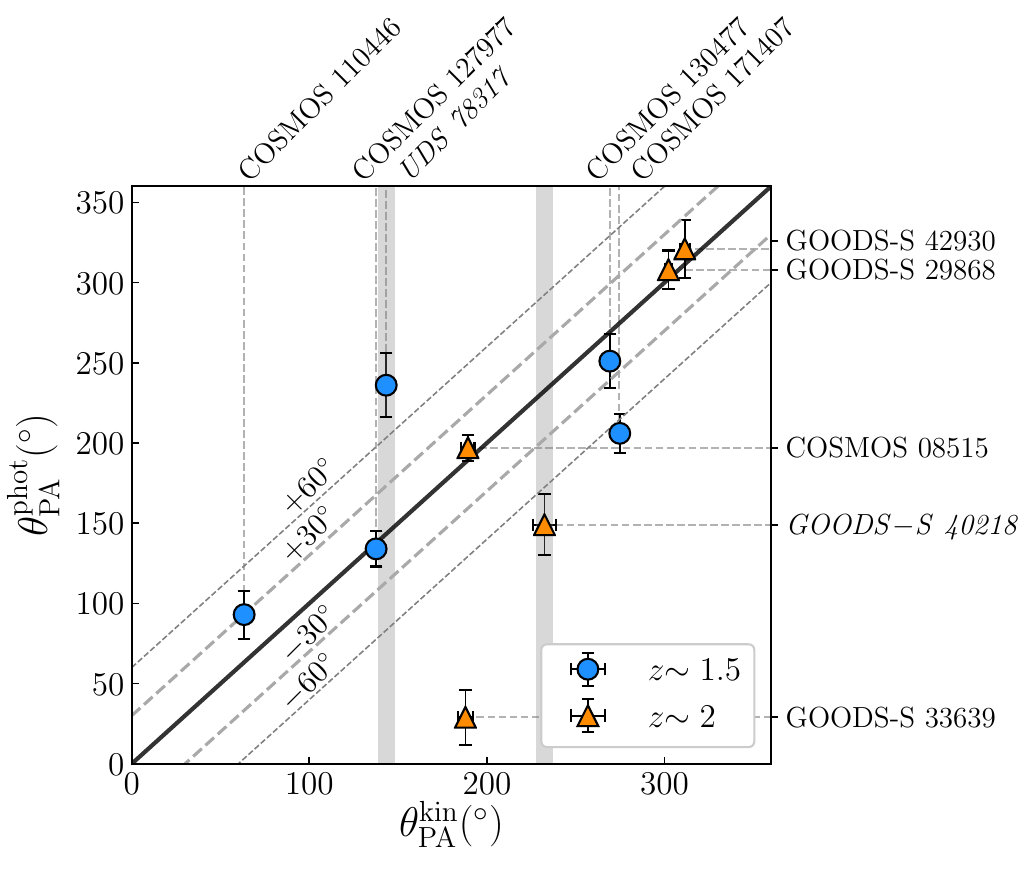}
    \caption{Comparison between the position angle obtained from photometry ($y$-axis) and kinematics ($x$-axis) where the measurements from kinematics are from the combined analysis (NS+AO). Dashed lines represent the regions with $\Delta \theta_\mathrm{PA}=\pm30^\circ,\pm60^\circ$. The galaxies with large discrepancies are ones classified as mergers (\textit{UDS 78317} and \textit{GOODS-S 40218} indicated by the grey vertical lines) as well as GOODS-S 33639 which is highly uncertain in the photometric and kinematic fits.}
    \label{fig:position angles}
\end{figure}

The misalignment between $\theta^\mathrm{kin}_\mathrm{PA}$ and $\theta^\mathrm{phot}_\mathrm{PA}$ can be a product of physical differences such as bright clumps in the continuum, the asymmetry between the stellar and gas distributions as well as complex morphologies due to mergers (\citealp{Rodrigues}). However, it can also be a product of the lack of precision in the deprojection of high redshift maps. We find that the galaxies that show the largest misalignment are the two mergers and the galaxy with large uncertainties (GOODS-S 33639) as indicated in Figure \ref{fig:position angles}, which is consistent with the disk-merger criteria discussed in Section \ref{subsection:Kinematic state of the galaxies}. For the rest of the sample, there is an overall agreement in the position angles which supports the choice of using the photometric inclinations.

\subsection{Angular momentum convergence}
\label{subsection:angular momentum convergence}

\begin{figure}
	\includegraphics[width=0.47\textwidth]{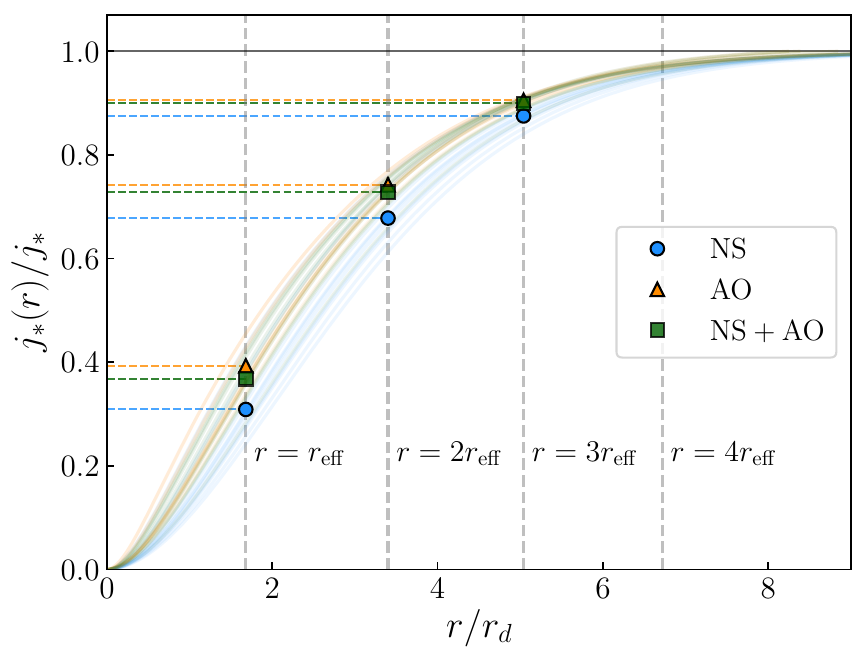}
    \caption{Cumulative profiles of $j_*(r)$ as a function of deprojected radius $r_d$ for the full sample in both single and combined resolutions. Dashed vertical lines indicate the $r_\mathrm{eff}$, $2r_\mathrm{eff}$, $3r_\mathrm{eff}$, $4r_\mathrm{eff}$ distances from the galactic centre. The dots, triangles and squares on top of these lines correspond to the mean $j_*(r)$ of the full sample to indicate the difference in the convergence from the different data types, which is significant at low radii.}
    \label{fig:j_convergence_comparison}
\end{figure}

Stellar specific angular momentum converges to an asymptotic value at large radii, with some notable differences between the different resolutions. A direct comparison of the convergence at different radii (namely $r=1,2\,\,\mathrm{and}\,\,3r_\mathrm{eff}$) between the different data types shows that the measured $j_*$ converges faster in the case of the high-resolution data as shown in Figure \ref{fig:j_convergence_comparison}. This result is expected since the rapidly growing rotation curves are better determined with the high-resolution data in the centre of the disks, whereas the flat part of the rotation curve is better determined with the seeing limited data where the effects of beam smearing are smaller.

\begin{figure}
	\includegraphics[width=0.47\textwidth]{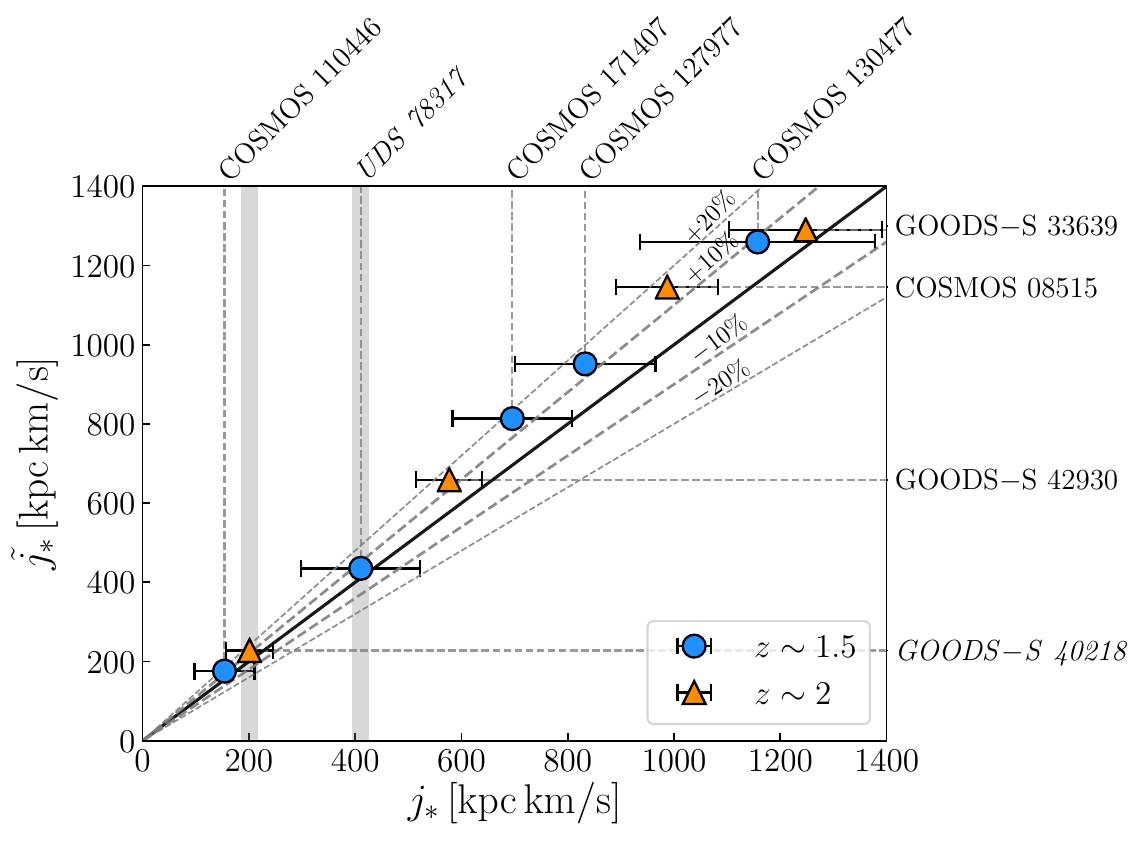}
    \caption{Stellar specific angular momentum from the approximation $\tilde{j_*} = k_n v_s r_\mathrm{eff}$ with respect to the measurements from the combined analysis ``NS+AO'' using the analytical expression of $j_*$ in Equation \ref{eq:j definition}. The diagonal line indicates the perfect agreement $\tilde{j_*}=j_*$ which is followed by the whole sample at the 20\% level.}
    \label{fig:j_approx}
\end{figure}

Another consistent result that we find throughout the whole sample is that the approximation $\tilde j_*$ (Equation \ref{eq:j approx}), agrees within $\sim 20\%$ with the total angular momentum value $j_*$ measured from the best fit models in the combined analysis as in Figure \ref{fig:j_approx}. This approximation relies on global variables that can be measured from low-resolution data such as the galaxy size and asymptotic velocity, hence natural seeing observations can provide a good estimation of the total angular momentum content of galaxies that are indeed disks as suggested by \cite{Romanowsky}.

\subsection{Comparison with Natural-seeing based results}
\label{subsection:comparison with NS based results}

\begin{figure}
	\includegraphics[width=0.47\textwidth]{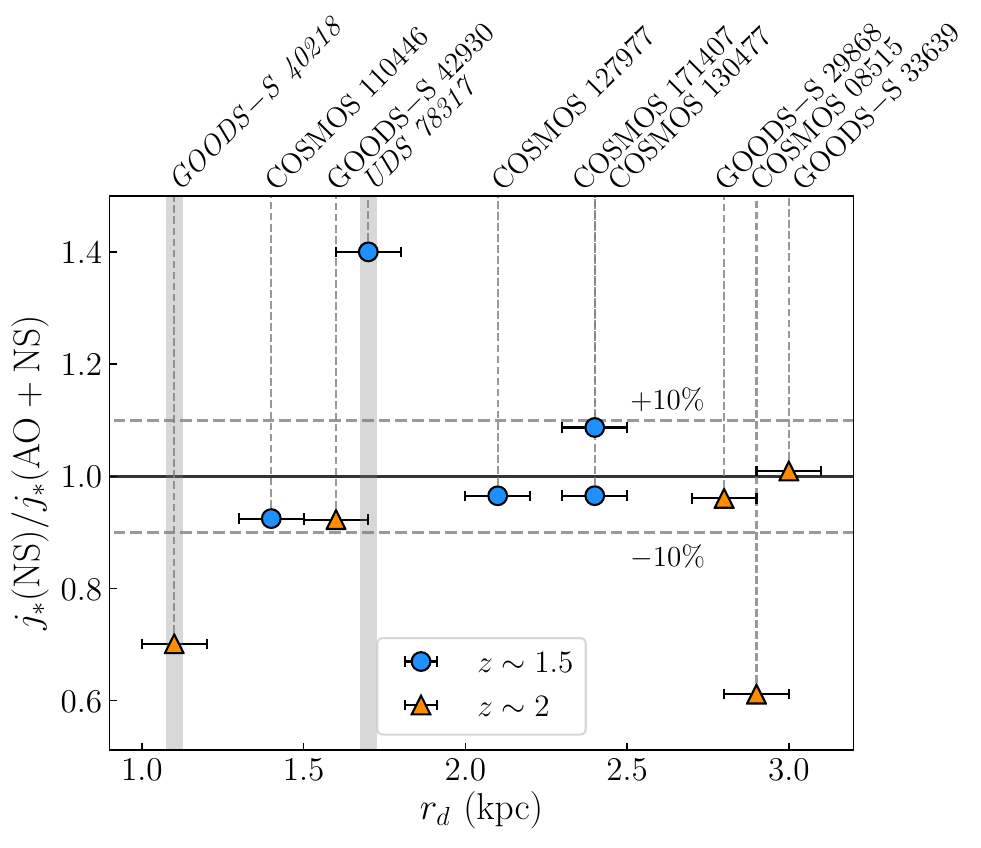}
    \caption{Ratio of the stellar specific angular momentum from the natural seeing (NS) sample and the value from the combined (NS+AO) analysis as a function of disk scale length $r_d$. Blue circles represent the sample at $z\sim1.5$ and orange triangles the sample at $z\sim2$. With the exception of COSMOS 08515, disk galaxies are in agreement within 10\% as shown by the dashed lines whereas the two galaxies classified as mergers (\textit{GOODS-S 40218} and \textit{UDS 78317}) show a large disagreement. Additionally, there is no evidence for a strong dependence on the disk scale length or redshift.}
    \label{fig:ratio_combined_ns_vs_rd}
\end{figure}

We test the reliability of natural-seeing based estimations of $j_*$ by comparing them with those from the combined analysis. In order to do that, we measure the ratio of both quantities as a function of disk scale length $r_d$ as shown in Figure \ref{fig:ratio_combined_ns_vs_rd}, where no dependence on redshift or galaxy size was found.

We find that the NS-only measurements of $j_*$ for the disk galaxies are within 10\% from the combined ones (with the exception of COSMOS 08515\footnote{The large discrepancy is due to the dramatic difference between the values of $v_\mathrm{flat}$, which is more prominently affected by beam smearing in the case of the NS fit (it has the lowest spatial resolution in the full sample with PSF FWHM = 0.966 arcsec).}), suggesting that if the galaxies are truly disks, then the natural seeing based estimations of $j_*$ are reliable. However, in the case of the two galaxies classified as mergers, the discrepancy in the measurement is large ($>30\%$), which shows how the poor determination of the real kinematic state of the galaxy (from low resolution) can lead to under- or over-estimations of $j_*$. Even if the estimation of $j_*$ from the combined analysis of those galaxies is also wrong (as the calculation is still done assuming an exponential disk with ordered rotation), the large difference between the two measurements could be used as an indicator of a complex kinematic state for future large samples.

\begin{figure}
	\includegraphics[width=0.47\textwidth]{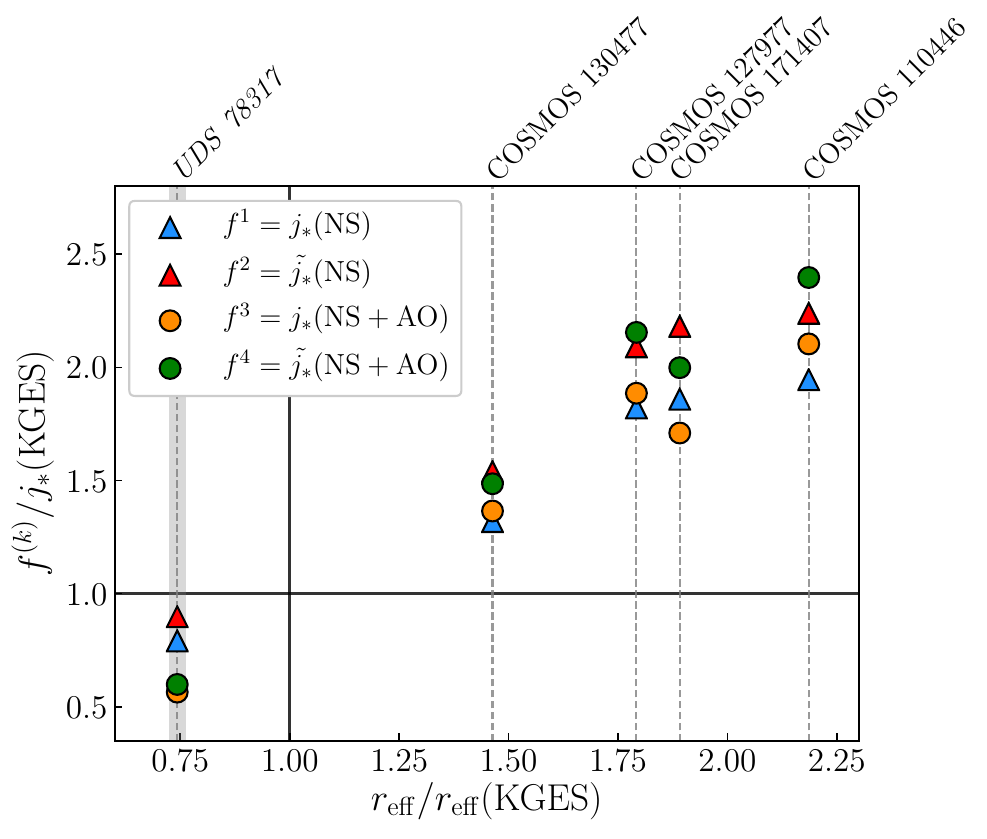}
    \caption{Comparison of our measurements of $j_*$ with those from \protect\cite{Gillman} for the five galaxies at $z\sim1.5$ that overlap with our sample. We show the ratio $f^{(k)}/j_*(\mathrm{KGES})$ (where $f^{(k)}$ is used to label the different types of measurements) with respect to the ratio of the effective radii measurements $r_\mathrm{eff}/r_\mathrm{eff}(\mathrm{KGES})$ from Table \ref{tab:reff_vs_Gillman_reff}. Blue (and red) triangles represent the total $j_*$ (and approximated $\tilde{j_*}$) from the natural seeing ``NS'' sample. Orange and green circles represent the combined measurements ``NS+AO''. Black solid lines indicate the regions where the ratios are equal. The discrepancy in the measurement of $r_\mathrm{eff}$ translates directly into the discrepancy in the measurements of $j_*$.}
    \label{fig:comparison_Gillman}
\end{figure}

In order to compare our results with other existing low-resolution studies and directly address the possible systematics associated with our chosen method, we compare our individual measurements to those obtained for the KGES sample by \protect\cite{Gillman} for the five galaxies at $z\sim1.5$ that overlap with our sample. We compare their measurements with our natural-seeing only measurements (as well as the combined resolutions) as shown in Figure \ref{fig:comparison_Gillman}. This is not a fully direct comparison since the functional form they used to measure $j_*$ is different from ours but it allows to identify the systematic differences between the two approaches. The $j_*$ values in \protect\cite{Gillman} are obtained with the approximation $j_* \approx k_n v_{\textrm{s}} r_\textrm{eff}$ (Equation \ref{eq:j approx}) where $r_\mathrm{eff}$ is measured from a S\'ersic fit with \textsc{GALFIT} and $v_s$ is obtained from a parametric model $v(r)$ that can rise or decline at large radii\footnote{This parametric model of the rotation velocity is based on an exponential light profile so it is characterized by the scale length of an exponential disk $r_d$ (Equation 10 in \citealp{Freeman}).}. Since the KGES values of $v_s$ are measured at $r=2r_\mathrm{eff}$, where beam smearing effects are low, they are similar to our $v(r=2r_\mathrm{eff})$ measurement at the $25\%$ level.

We find that the large differences in the galaxy size measurement (shown in Table \ref{tab:reff_vs_Gillman_reff}) directly translate into the large discrepancies of our measurements with the $j_*(\mathrm{KGES})$ values. For the galaxies where our measurement of $r_\mathrm{eff}$ is significantly larger ($\Delta r_\mathrm{eff} >$ 75\%) (COSMOS 110446, COSMOS 127977 and COSMOS 171407), the ratios in $j_*$ are also significantly larger, whereas for the galaxies where the measurement of $r_\mathrm{eff}$ is smaller ($\Delta r_\mathrm{eff} <$ 50\%) (COSMOS 130477 and \textit{UDS 78317}), the $j_*$ value is smaller as well.

The large difference in galaxy sizes can be attributed to the different surface brightness profiles $I(r)$ used in each case (S\'ersic profile with varying $n=0.2-8$ in \citealp{Gillman} and exponential disk in our analysis) because the radius enclosing half of the light depends strongly on the power-law slope. This implies that an erroneous fit of the S\'ersic index $n$, which could be caused by the low resolution of the \textit{HST} maps and clumpy nature of these galaxies, can lead to over or underestimations of the light profile as well as unrealistic values of $r_\mathrm{eff}$. For example, the steep gradient in SNR could bias the fit to smaller values of $n$ (steep profile), and thus bias the fit to small values of $r_\mathrm{eff}$. In fact, some of the S\'ersic fits, particularly those with $n\leq0.2$, are at the limit of the fitting range and thus the estimations of the galaxy size are unreliable. In our analysis, we chose the $n=1$ exponential profile, which has a shallower slope and thus contains a large fraction of the galaxy light in the outskirts, where the SNR of the data is low. For 6/10 galaxies, the (average) radial surface brightness profile $\Sigma(r)$ has lower RMS residuals when using the exponential disk assumption as compared to the S\'ersic fit as shown for each individual galaxy in Figures \ref{fig:COSMOS 110446} to \ref{fig:GS4_42930} in the supplementary section.

The comparison to the results of the KGES sample also allows us to determine the intrinsic scatter in the $j_*$ vs $M_*$ relation. The RMS scatter of the KGES seeing-limited sample is $\sim 0.56$ dex whereas for our eight disks we get $\sim 0.38$ dex. Even though a sample of eight disks is not enough to make statistically significant predictions, we note that this is still a large scatter, suggesting that adding the high-resolution kinematic data does have a significant effect on it. This could indicate that the observed scatter found from low resolution studies is not related to systematic effects caused by poor resolution. Instead, this intrinsic scatter could be attributed to different mechanisms of retention and redistribution of angular momentum due to inflows and outflows, morphological differences, varied bulge-to-total fractions, and complex environmental conditions such as higher incidence of minor mergers and large gas accretion from cosmic filaments. In the local Universe, the RMS scatter of the parametric relation of disks decreases to $\leq0.2$ dex (e.g. \citealp{Romanowsky}; \citealp{Obreschkow_2014}; \citealp{Posti_2018}), pointing to the emerging of rotational support that assembles massive disks in the local Universe (\citealp{Kassin_disk_assembly}).

\begin{table}
	\centering
	\caption{Effective (half-light) radius $r_\mathrm{eff}$ measurements from different studies. From left to right: Redshift, galaxy ID, $r_\mathrm{eff}$ obtained from the S\'ersic's profile fit with the corresponding S\'ersic index $n$, and our measurement from the exponential disk assumption which corresponds to $n=1$. The five top rows correspond to the measurements from \protect\cite{Gillman} of the KGES galaxies where they follow the procedures of \protect\cite{van_der_Wel} and the bottom five rows correspond to the measurements from \protect\cite{Tacchella_sins_sizes} of the SINS galaxies using the corresponding \textit{HST} \textit{H}-band images.}
	\label{tab:reff_vs_Gillman_reff}
	{\setlength{\tabcolsep}{0.5em}
	\begin{tabular}{lcccc}
		\hline
		& Galaxy & $r_\mathrm{eff}$ (kpc) & $n$ & $r_\mathrm{eff}$ (kpc)\\
		\cmidrule(lr){3-4}
		& ID &  \multicolumn{2}{c}{\protect\citealp{Gillman}} & (this work)\\
		%\hline
		\midrule
		& COSMOS 110446 & 1.08 & 0.5 & 2.35\\
		& COSMOS 171407 & 2.13 & 0.2 & 4.03\\
		$z\sim1.5$& COSMOS 130477 & 2.75 & 1.1 & 4.03\\
		& COSMOS 127977 & 2.06 & 0.2 & 3.69\\
		& \textit{UDS 78317} & 4.07 & 0.4 & 3.02\\ 
		%\hline
		\midrule
		%&  & $r_\mathrm{eff}$ (kpc) & $n$ & $r_\mathrm{eff}$ (kpc)\\
        %\cmidrule(lr){3-4} %\cmidrule(lr){4-5}
        %& Galaxy & $r_\mathrm{eff}$ (kpc) & $n$ & $r_\mathrm{eff}$ (kpc)\\
		
		&  &  \multicolumn{2}{c}{\protect\citealp{Tacchella_sins_sizes}} & (this work)\\
		\cmidrule(lr){3-4}
		%\hline
		& COSMOS 08515 & 2.2 & 0.5 & 4.9\\
		& GOODS-S 29868 & 5.7 & 0.2 & 4.7\\
		$z\sim2$& GOODS-S 33639 & 3.6 & 0.6 & 5.0\\
		& \textit{GOODS-S 40218} & 1.3 & 0.9 & 1.8\\
		& GOODS-S 42930 & 1.7 & 1.3 & 2.7\\
		\hline
	\end{tabular}}
\end{table}

\subsection{Effects of galaxy clumpiness}
\label{subsection:Effect of galaxy clumpiness}

The precision of a direct pixel-by-pixel measurement of $j_*$ from the actual data is limited by the number of pixels with useful information in $v(x,y)$ and $I(x,y)$ as well as their difference in pixel scales and spatial resolutions (this motivated the use of the analytical calculation from Equation \ref{eq:j definition}). However, it can be useful when trying to avoid bias from assumed models. In the case of the photometry, extensive star-formation activity in bright clumps (not necessarily massive) within the disks and a central bulge can affect the observed morphology from the \textit{HST} maps which is not accounted for in the exponential disk assumption. We calculate the pixel-by-pixel stellar specific angular momentum of all galaxies as

\begin{figure}
	\includegraphics[width=0.47\textwidth]{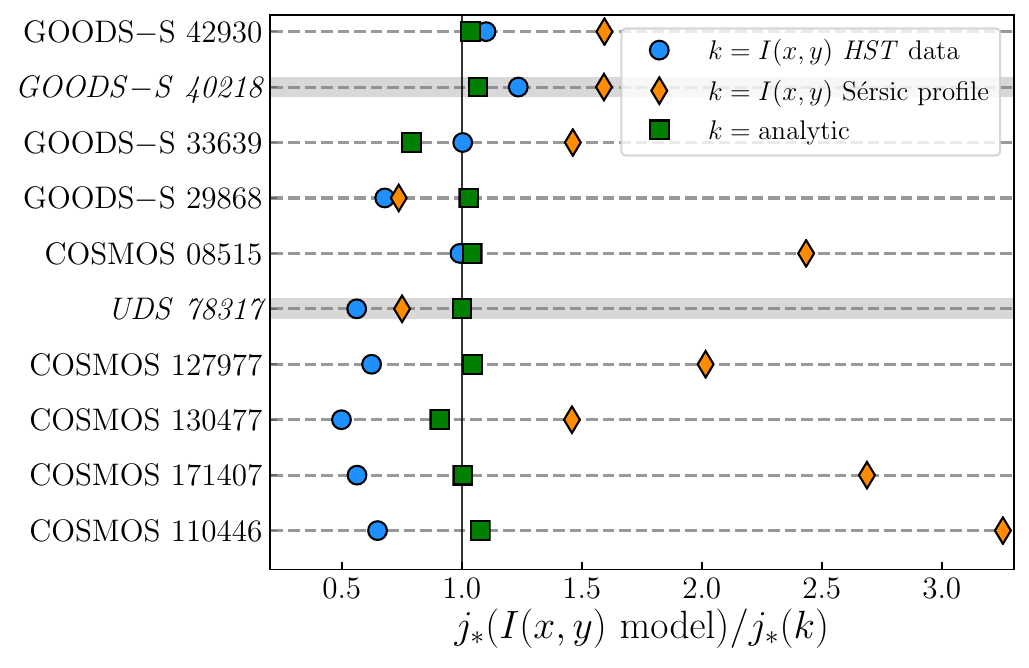}
    \caption{Fractional difference of the pixel-by-pixel measurement of $j_*$ (Equation \ref{eq:j pixelwise}) with respect to other measurements. \textbf{\textit{i)}} In blue circles, the difference with the estimation using the \textit{HST} data instead of a model intensity profile $I(x,y)$ from where the $z\sim1.5$ sample shows larger differences ($>30\%$) due to clumpiness. \textbf{\textit{ii)}} Orange diamonds indicate the ratio with the estimation using a S\'ersic profile with the fit parameters from \protect\cite{Gillman} and \protect\cite{Tacchella_sins_sizes}. The large ratios are directly related to the large difference in $r_\mathrm{eff}$ from the different surface brightness profiles shown in Table \ref{tab:reff_vs_Gillman_reff}. \textbf{\textit{iii)}} In green squares, the difference with the analytical calculation (Equation \ref{eq:j definition}) which is always below the $30\%$ level.}
    \label{fig:difference using HST}
\end{figure}

\begin{ceqn}
\begin{align}
j_*= \frac{\sum\limits_{i,j} v_{i,j}\Sigma_{i,j}r_{i,j}}{\sum\limits_{i,j} \Sigma_{i,j}},
\label{eq:j pixelwise}
\end{align}
\end{ceqn}

where $i,j$ go through all the spatially-matched pixels in velocity and photometry. In order to quantify the effects of clumpiness as well as the chosen form of $I(x,y)$, we perform four separate experiments (the first three are shown in Figure \ref{fig:difference using HST}) that we explain in the following paragraphs.

\textbf{\textit{i)} Using \textit{HST} data instead of the I(x,y) model:} We take the measurement of $j_*$ where $v(x,y)$ and $I(x,y)$ come from the best-fit models and compare it to the measurement where $I(x,y)$ comes from the \textit{HST} data directly. Galaxies COSMOS 110446, COSMOS 130477, COSMOS 127977, \textit{UDS 78317} and GOODS-S 29868 show a large difference at the $>40\%$ difference level (with $j_*$ using the \textit{HST} being higher), likely due to the multiple bright components located away from the centre in their \textit{HST} maps. The difference in the rest of the galaxies is at the $<30\%$ level. The majority of the light coming from the bright regions comes from young stellar populations that ionize the surrounding gas, so it only represents a small fraction of the full mass content. Hence if these clumps are young, then the assumption that light traces mass provides an upper bound in the systematic error which is at the $28\%$ level for the whole sample.

\textbf{\textit{ii)} Using a S\'ersic profile (in 2D) instead of the exponential profile (in 2D):} Another way to test the systematics associated with the choice of surface brightness profile is by using a S\'ersic profile instead of the exponential disk. We use the best-fit parameters from \cite{Gillman} and \cite{Tacchella_sins_sizes} to create the $I(x,y)$ models from where we find large differences throughout the whole sample with the majority of the ratios being larger than unity (model $I(x,y)$ yields larger ratios than S\'ersic profile $I(x,y)$) except for the merger UDS 78317 and GOODS-S 29868 which are the only galaxies where $r_\mathrm{eff}$ from the exponential disk fit is lower as seen in Table \ref{tab:reff_vs_Gillman_reff}. In particular, we find dramatic differences at the >50\% level for six galaxies with COSMOS 110446 and COSMOS 171407 being the most extreme cases with ratios of 3.2 and 2.5 respectively. As mentioned in the previous subsection, despite the generalities that a S\'ersic profile allows, steep gradients in the poorly-constrained light profiles can lead to small values of $n$ which translate to lower values of $r_\mathrm{eff}$ and thus underestimate the total value of $j_*$.

\textbf{\textit{iii)} 2D approach vs 1D analytical calculation:} In addition to testing the systematics associated with the assumed $I(x,y)$ model, we compare the pixel-by-pixel measurement with the estimation from the one-dimensional analytical expression in Equation \ref{eq:j definition}. We find a strong dependence in the disk scale length as the analytic calculation deviates significantly ($>10\%$) in the case of COSMOS 08515 and GOODS-S 33639 which have the largest $r_d$ (2.9 and 3 kpc respectively). On the other hand, the galaxies where the difference in $j_*$ is $<5\%$ (COSMOS 110446, \textit{UDS 78317}, \textit{GOODS-S 40218} and GOODS-S 42930) are the galaxies with the lowest $r_d$ with $r_d<2$ kpc. This scale dependence can be attributed to the error introduced by the discrete change in intensities (and velocities) from the pixel-by-pixel approach. This effect is stronger when $r_d$ is large since more pixels will have values that deviate more from the analytical expression which is more reliable since it depends only on $r_d$, $r_\mathrm{flat}$ and $v_\mathrm{flat}$.

\textbf{\textit{iv)} Bulge and clumps contribution:} We quantify the effect of adding a central bulge in the modelling that follows a S\'ersic profile with $n=4$ and a size of $r_\mathrm{eff}\sim 1$ kpc in our modelling. The choice of this size is motivated by the findings of \hbox{\cite{Fisher_Drory}} who estimate typical bulge sizes in the local Universe and the bulge-disk decomposition performed in \cite{Tacchella_sins_sizes} on the SINS galaxies where they find typical bulge sizes of $\sim 1$ kpc. We recalculate $j_*$ with the modified mass distribution model while keeping the same velocity field information (since it is constrained by the multiple resolutions). We add the extra co-rotating component by varying the bulge-to-total ratio $(B/T)_*$ in the range [0.1-0.5], noting that in the $(B/T)_*$ decomposition on \textit{J}- and \textit{H}-band \textit{HST} images performed by \cite{Tacchella_sins_sizes} on COSMOS 08515, GOODS-S 33639, \textit{GOODS-S 40218} and GOODS-S 42930, the bulge-to-total ratios were all in the range [0-0.3].

For a constant mass-to-light ratio $(M/L)_*$, we find that the effect of the bulge is a decrease in the total measurement of $j_*$, in consistency with similar studies (e.g. \citealp{Romanowsky}; \citealp{Obreschkow_2014}). The effect is minimal at $(B/T)_*=0.1$ with a small median decrease in $j_*$ of $\sim2\%$, and becomes more prominent at higher $(B/T)_*$, having the maximum effect at of $(B/T)_*=0.5$ with a median decrease of 10\% in $j_*$. However, since the stellar populations in the bulge are different from those in the disk, then the $(M/L)_*$ in each component is different as well. In order to quantify this effect, we make use of the \textit{J}- and \textit{H}- band \textit{HST} data at $z\sim2$ used in \cite{Tacchella_sins_sizes} (via private communication) to perform a color analysis. Since \textit{J-H} colors at $z\sim2.2$ correspond to rest-frame \textit{u-g}, we can use them as a proxy for stellar mass-to-light ratios $(M/L)_*$ based on color-$(M/L)_*$ relations (\citealp{Bell_2001}). From the radial color profiles and color maps we find color differences in the bulge and disk of $\Delta m \leq 0.3$ which translate to bulges that are $\leq40\%$ more massive than the disk using the relations in \cite{Bell_colors_M_L_ratio}. This difference sets an upper bound of $\Delta j_* \lesssim 15\%$, and shows that galaxies with massive central bulges (higher mass-to-light ratios) have a lower content of stellar specific angular momentum.

A similar color analysis of the galaxies with the most prominent clumps and high quality \textit{HST} multi-wavelength data (namely COSMOS 08515 and GOODS-S 29868), sets an upper bound on the effect of their bright clumps on the overall measurement of $j_*$. From the color maps, we identify clumps that are bluer (younger) than the disk with a maximum difference of $\Delta (J-H)\sim -0.6$ which translates to a difference in mass-to-light ratio of $\Delta \log (M/L_*)\sim-0.3$. Thus, these clumps are 50\% less massive than their host disks so they contribute less to the mass profile of the galaxy. In parallel to the effect of the bulge, the effect of these young clumps in $j_*$ is a slight decrease of 8\% for COSMOS 08515 (with two clumps) and 5\% for GOODS-S 29868 (one clump), where we have used clump sizes of $r\sim1$ kpc from the \textit{HST} broad-band imaging at their distance from the galaxy centre. The measurement of this effect is limited to the difference in mass profile as we did not include any extra kinematic components in the model. Besides its consistency with similar studies, this assumption is justified from the inspection of the velocity fields which show no significant differences in the regions where these clumps reside (they appear to co-rotate with the disks). For the rest of the sample, this contribution ($\Delta j_*\lesssim10\%$) is expected to be of similar order so it does not have a dramatic effect on the stellar specific angular momentum measurements in this work.

\section{Conclusions}
\label{section:Conclusions}

We have gathered and analysed a sample of 10 high-redshift ($z\sim1.5$ and $z\sim2$) galaxies that have IFS observations at high and low spatial resolution as well as near-infrared \textit{HST} observations to measure their stellar specific angular momentum $j_*$ content. This is the largest sample to date with angular momentum measurements coming from the combination of adaptive optics assisted observations and their seeing limited counterparts. We summarize our findings as follows:

\begin{itemize}

    \item We developed a code to combine and take advantage of the different capabilities of each data type (adaptive optics provides high resolution but low sensitivity and natural seeing provides low resolution but high sensitivity). In the ``2$\nicefrac{1}{2}$D'' modelling we carefully apply the right convolution to each model cube and find the best model using a maximum likelihood technique. We successfully tested the kinematic code by simulating a sample of $10^3$ mock galaxies with similar characteristics and running the code at different resolutions. Besides providing a sanity check for the code, it also confirmed that the combination of the different resolutions reduces the uncertainty in $j_*$ as explained in Appendix \ref{section:appendix_code}.
    
    \item In order to increase SNR in two of the adaptive optics assisted datacubes, we employed different spatial smoothing methods and the median filter provided the best results. The signal was improved sufficiently so that H$\alpha$ emission could be used to measure its emission line kinematics and the spatial resolution was kept high enough to improve that of seeing limited data.
    
    \item We have measured the improvement in the uncertainties of $j_*$ by resampling the models with a Monte Carlo approach. The mean uncertainties in the sample are 49\% in the natural seeing case, 26.5\% for adaptive optics and 16\% when combining the datasets. The improvement comes from a reduction in the effects of beam smearing and the accurate modelling of the rotation curves, which are expected to rise rapidly in the inner regions of disk galaxies (better determined with adaptive optics) and then flatten at large radii (better determined with the deeper seeing limited counterpart). The combination can thus improve the measurements of $j_*$ that come from large seeing-limited surveys and will be more useful once large samples of adaptive optics assisted measurements are made with upcoming facilities such as \textit{JWST} and next generation ground-based IFUs.
    
    \item Estimates of $j_*$ from the natural seeing observations are reliable (within $10\%$ of the combined analysis) when the galaxies are confidently classified as disks. The only exception was galaxy COSMOS 08515 due to the very poor resolution in NS. This suggests that the reliability of the findings from large seeing limited surveys depends strongly on the accurate kinematic classification of the samples.
    
    \item Besides the measurement from the best-fit models, we calculate the stellar specific angular momentum of all galaxies with a widely used approximation in terms of the rotation profile and size of the disk. This value labeled $\tilde{j_*}$ consistently agrees with the asymptotic angular momentum measurement obtained from the full modelling within 20\%.
    
    \item A pixel-by-pixel measurement is subject to large systematic uncertainties due to resolution effects and pixel scales. A smooth profile assumption from an analytical calculation of $j_*(r)$ can reduce the effect of these systematics and its reliability depends on the choice of surface brightness and rotation velocity profiles. In particular, the size $r_\mathrm{eff}$ is a critical quantity in measuring $j_*$ and depends strongly on the choice of $I(r)$.
    
    \item We classify \textit{GOODS-S 40218} and \textit{UDS 78317} as mergers. The latter was previously defined as a rotating disk from natural seeing only studies but turned out to be a merger from the high-resolution observations, in accordance with \cite{Sweet}. We find that two of the other systems (COSMOS 171407 and GOODS-S 33639) have complex H$\alpha$ intensity and kinematic structures but insufficient to catalog them as mergers so we treat them as disks. The other six galaxies are classified as rotating disks. 
    
    \item We used the eight galaxies identified as disks to make a prediction on the Fall relation ($j_*$ vs $M_*$). We find that our best fit with a constrained slope agrees with the findings of previous high-redshift studies within $\sim0.3$ dex. This is limited by the small number of galaxies in our current sample. A direct comparison to the relation found from a seeing limited sample at a similar redshift (\citealp{Gillman} at $z\sim1.5$) shows a minor offset, despite the differences in the methods used to measure $j_*$ (different mass distributions and radial extents).
    
    \item The scatter of our measurements in the Fall relation remains large ($\sim 0.38$ dex) even in the combined analysis in similarity with the ``NS-only'' analysis of \cite{Gillman}. The scatter is thus intrinsic and reflects diversity in galaxy assembly and complex morphology where stellar bulge-to-total ratios may play a more important role. In the local Universe, the scatter decreases to $\sim0.2$ pointing to the assembly of disks through rotational support.
    
\end{itemize}

\begin{comment}% Extra conlusion of the B/T ratio and color analisis
    \textbf{We find that the effect of a $\sim 1$ kpc central bulge within the disk is a decrease in the $j_*$ measurements and becomes more important for bulges with large mass-to-light ratios. Simple analysis of the color profiles from the $z\sim2$ \textit{HST} data shows small $(M/L)_*$ differences between the centre (where a bulge would reside) and the disk setting an upper limit of $\Delta j_* \lesssim 15\%$. Similarly, the effect of blue (young) clumps in the galaxies with more prominent ones is $\Delta j_* \lesssim 10\%$, and expected to be of similar order for the whole sample.}
\end{comment}

In future work, we will apply the method introduced here on a larger sample from the seeing limited survey \cite{Forster_2008_no_AO} with its adaptive optics counterparts from \cite{SINS/zC-SINF}. This will expand the current combined analysis to a low-mass regime and will provide further constraints in the $j_*$ vs $M_*$ relation from multi-resolution measurements of $j_*$.

\section*{Acknowledgements}

We thank the anonymous referee for helpful comments and suggestions that greatly improved this manuscript. JE is funded by the Swinburne University Postgraduate Research Award (SUPRA). JE, DBF, KG, DO and SMS acknowledge support from Australian Research Council (ARC) DP grant DP160102235. DBF acknowledges support from ARC Future Fellowship FT170100376. DO is a recipient of an ARC Future Fellowship (FT190100083) funded by the Australian Government. ALT acknowledges support from a Forrest Research Foundation Fellowship. SG acknowledges the support of the Science and Technology Facilities Council through grant ST/N50404X/1 and ST/L00075X/1 for support and the Cosmic Dawn Centre of Excellence funded by the Danish National Research Foundation under grant No. 140. Some of the data presented herein were obtained at the W. M. Keck Observatory, which is operated as a scientific partnership among the California Institute of Technology, the University of California and NASA. The Observatory was made possible by the generous financial support of the W. M. Keck Foundation. The KMOS data were obtained at the Very Large Telescope of the European Southern Observatory, Paranal, Chile, and provided by the KGES survey team and the public release of KMOS$^\mathrm{3D}$. The SINFONI data were obtained at the same facility and provided in the public release of the SINS/zC-AO survey. \textit{HST} data were obtained from the data archive at the Space Telescope Science Institute. Simulations in this work were performed on the OzStar national facility at Swinburne University of Technology and the National Collaborative Research Infrastructure Strategy (NCRIS). We acknowledge the open-source software packages used throughout this work including \textsc{astropy} (\citealp{astropy}), \textsc{scipy} (\citealp{scipy}), \textsc{numpy} (\citealp{numpy}), CM\textsc{asher} (\citealp{cmasher}), \textsc{matplotlib} (\citealp{matplotlib}).

\section{Data availability}

The data underlying this article will be shared on reasonable request to the corresponding author.

%%%%%%%%%%%%%%%%%%%%%%%%%%%%%%%%%%%%%%%%%%%%%%%%%%



\bibliographystyle{mnras}
%\bibliography{references} % if your bibtex file is called references.bib

\begin{thebibliography}{}
\makeatletter
\relax
\def\mn@urlcharsother{\let\do\@makeother \do\$\do\&\do\#\do\^\do\_\do\%\do\~}
\def\mn@doi{\begingroup\mn@urlcharsother \@ifnextchar [ {\mn@doi@}
  {\mn@doi@[]}}
\def\mn@doi@[#1]#2{\def\@tempa{#1}\ifx\@tempa\@empty \href
  {http://dx.doi.org/#2} {doi:#2}\else \href {http://dx.doi.org/#2} {#1}\fi
  \endgroup}
\def\mn@eprint#1#2{\mn@eprint@#1:#2::\@nil}
\def\mn@eprint@arXiv#1{\href {http://arxiv.org/abs/#1} {{\tt arXiv:#1}}}
\def\mn@eprint@dblp#1{\href {http://dblp.uni-trier.de/rec/bibtex/#1.xml}
  {dblp:#1}}
\def\mn@eprint@#1:#2:#3:#4\@nil{\def\@tempa {#1}\def\@tempb {#2}\def\@tempc
  {#3}\ifx \@tempc \@empty \let \@tempc \@tempb \let \@tempb \@tempa \fi \ifx
  \@tempb \@empty \def\@tempb {arXiv}\fi \@ifundefined
  {mn@eprint@\@tempb}{\@tempb:\@tempc}{\expandafter \expandafter \csname
  mn@eprint@\@tempb\endcsname \expandafter{\@tempc}}}

\bibitem[\protect\citeauthoryear{{Airy}}{{Airy}}{1835}]{Airy}
{Airy} G.~B.,  1835, Transactions of the Cambridge Philosophical Society, \href
  {https://ui.adsabs.harvard.edu/abs/1835TCaPS...5..283A} {5, 283}

\bibitem[\protect\citeauthoryear{{Astropy Collaboration} et~al.,}{{Astropy
  Collaboration} et~al.}{2013}]{astropy}
{Astropy Collaboration} et~al., 2013, \mn@doi [\aap]
  {10.1051/0004-6361/201322068}, \href
  {https://ui.adsabs.harvard.edu/abs/2013A&A...558A..33A} {558, A33}

\bibitem[\protect\citeauthoryear{{Avila}}{{Avila}}{2017}]{ACS}
{Avila} R.~J.,  2017, {Advanced Camera for Surveys Instrument Handbook for
  Cycle 25 v. 16.0}

\bibitem[\protect\citeauthoryear{{Bacon} et~al.,}{{Bacon}
  et~al.}{2010}]{Bacon_2010_MUSE}
{Bacon} R.,  et~al., 2010, in {McLean} I.~S.,  {Ramsay} S.~K.,   {Takami} H.,
  eds,  Society of Photo-Optical Instrumentation Engineers (SPIE) Conference
  Series Vol. 7735, Ground-based and Airborne Instrumentation for Astronomy
  III. p. 773508, \mn@doi{10.1117/12.856027}

\bibitem[\protect\citeauthoryear{{Bassett} et~al.,}{{Bassett}
  et~al.}{2014}]{Bassett_2014}
{Bassett} R.,  et~al., 2014, \mn@doi [\mnras] {10.1093/mnras/stu1029}, \href
  {https://ui.adsabs.harvard.edu/abs/2014MNRAS.442.3206B} {442, 3206}

\bibitem[\protect\citeauthoryear{{Begeman}}{{Begeman}}{1989}]{Begeman}
{Begeman} K.~G.,  1989, \aap, \href
  {https://ui.adsabs.harvard.edu/abs/1989A&A...223...47B} {223, 47}

\bibitem[\protect\citeauthoryear{Bekiaris, Glazebrook, Fluke  \&
  Abraham}{Bekiaris et~al.}{2016}]{Bekiaris2016}
Bekiaris G.,  Glazebrook K.,  Fluke C.~J.,   Abraham R.,  2016, \mn@doi
  [Monthly Notices of the Royal Astronomical Society] {10.1093/mnras/stv2292},
  455, 754

\bibitem[\protect\citeauthoryear{{Bell} \& {de Jong}}{{Bell} \& {de
  Jong}}{2001}]{Bell_2001}
{Bell} E.~F.,  {de Jong} R.~S.,  2001, \mn@doi [\apj] {10.1086/319728}, \href
  {https://ui.adsabs.harvard.edu/abs/2001ApJ...550..212B} {550, 212}

\bibitem[\protect\citeauthoryear{{Bell}, {McIntosh}, {Katz}  \&
  {Weinberg}}{{Bell} et~al.}{2003}]{Bell_colors_M_L_ratio}
{Bell} E.~F.,  {McIntosh} D.~H.,  {Katz} N.,   {Weinberg} M.~D.,  2003, \mn@doi
  [\apjs] {10.1086/378847}, \href
  {https://ui.adsabs.harvard.edu/abs/2003ApJS..149..289B} {149, 289}

\bibitem[\protect\citeauthoryear{{Bellocchi}, {Arribas}  \&
  {Colina}}{{Bellocchi} et~al.}{2012}]{Bellocchi}
{Bellocchi} E.,  {Arribas} S.,   {Colina} L.,  2012, \mn@doi [\aap]
  {10.1051/0004-6361/201117894}, \href
  {https://ui.adsabs.harvard.edu/abs/2012A&A...542A..54B} {542, A54}

\bibitem[\protect\citeauthoryear{{Bezanson} et~al.,}{{Bezanson}
  et~al.}{2018}]{Bezanson_2018}
{Bezanson} R.,  et~al., 2018, \mn@doi [\apj] {10.3847/1538-4357/aabc55}, \href
  {https://ui.adsabs.harvard.edu/abs/2018ApJ...858...60B} {858, 60}

\bibitem[\protect\citeauthoryear{Blumenthal, Faber, Primack  \&
  Rees}{Blumenthal et~al.}{1984}]{Blumenthal:1984bp}
Blumenthal G.~R.,  Faber S.,  Primack J.~R.,   Rees M.~J.,  1984, \mn@doi
  [Nature] {10.1038/311517a0}, 311, 517

\bibitem[\protect\citeauthoryear{{Boissier}, {Prantzos}, {Boselli}  \&
  {Gavazzi}}{{Boissier} et~al.}{2003}]{Boissier_2003}
{Boissier} S.,  {Prantzos} N.,  {Boselli} A.,   {Gavazzi} G.,  2003, \mn@doi
  [\mnras] {10.1111/j.1365-2966.2003.07170.x}, \href
  {https://ui.adsabs.harvard.edu/abs/2003MNRAS.346.1215B} {346, 1215}

\bibitem[\protect\citeauthoryear{{Bouch{\'e}} et~al.,}{{Bouch{\'e}}
  et~al.}{2007}]{Bouche_2007}
{Bouch{\'e}} N.,  et~al., 2007, \mn@doi [\apj] {10.1086/522221}, \href
  {https://ui.adsabs.harvard.edu/abs/2007ApJ...671..303B} {671, 303}

\bibitem[\protect\citeauthoryear{{Bouch{\'e}}, {Carfantan}, {Schroetter},
  {Michel-Dansac}  \& {Contini}}{{Bouch{\'e}} et~al.}{2015}]{Galpak}
{Bouch{\'e}} N.,  {Carfantan} H.,  {Schroetter} I.,  {Michel-Dansac} L.,
  {Contini} T.,  2015, \mn@doi [\aj] {10.1088/0004-6256/150/3/92}, \href
  {https://ui.adsabs.harvard.edu/abs/2015AJ....150...92B} {150, 92}

\bibitem[\protect\citeauthoryear{{Bouch{\'e}} et~al.,}{{Bouch{\'e}}
  et~al.}{2021}]{Bouche_2021}
{Bouch{\'e}} N.~F.,  et~al., 2021, arXiv e-prints, \href
  {https://ui.adsabs.harvard.edu/abs/2021arXiv210112250B} {p. arXiv:2101.12250}

\bibitem[\protect\citeauthoryear{{Burkert} et~al.,}{{Burkert}
  et~al.}{2016}]{Burkert}
{Burkert} A.,  et~al., 2016, \mn@doi [\apj] {10.3847/0004-637X/826/2/214},
  \href {https://ui.adsabs.harvard.edu/abs/2016ApJ...826..214B} {826, 214}

\bibitem[\protect\citeauthoryear{{Carignan}, {Chemin}, {Huchtmeier}  \&
  {Lockman}}{{Carignan} et~al.}{2006}]{Caringan}
{Carignan} C.,  {Chemin} L.,  {Huchtmeier} W.~K.,   {Lockman} F.~J.,  2006,
  \mn@doi [\apjl] {10.1086/503869}, \href
  {https://ui.adsabs.harvard.edu/abs/2006ApJ...641L.109C} {641, L109}

\bibitem[\protect\citeauthoryear{Catelan \& Theuns}{Catelan \&
  Theuns}{1996a}]{Catelan_Theuns_b}
Catelan P.,  Theuns T.,  1996a, \mn@doi [Monthly Notices of the Royal
  Astronomical Society] {10.1093/mnras/282.2.436}, 282, 436

\bibitem[\protect\citeauthoryear{{Catelan} \& {Theuns}}{{Catelan} \&
  {Theuns}}{1996b}]{Catelan_Theuns}
{Catelan} P.,  {Theuns} T.,  1996b, \mn@doi [\mnras] {10.1093/mnras/282.2.455},
  \href {https://ui.adsabs.harvard.edu/abs/1996MNRAS.282..455C} {282, 455}

\bibitem[\protect\citeauthoryear{{Catinella}, {Giovanelli}  \&
  {Haynes}}{{Catinella} et~al.}{2006}]{Catinella}
{Catinella} B.,  {Giovanelli} R.,   {Haynes} M.~P.,  2006, \mn@doi [\apj]
  {10.1086/500171}, \href
  {https://ui.adsabs.harvard.edu/abs/2006ApJ...640..751C} {640, 751}

\bibitem[\protect\citeauthoryear{{Conselice}, {Bershady}  \&
  {Jangren}}{{Conselice} et~al.}{2000}]{Conselice}
{Conselice} C.~J.,  {Bershady} M.~A.,   {Jangren} A.,  2000, \mn@doi [\apj]
  {10.1086/308300}, \href
  {https://ui.adsabs.harvard.edu/abs/2000ApJ...529..886C} {529, 886}

\bibitem[\protect\citeauthoryear{{Contini} et~al.,}{{Contini}
  et~al.}{2016}]{Contini_2016}
{Contini} T.,  et~al., 2016, \mn@doi [\aap] {10.1051/0004-6361/201527866},
  \href {https://ui.adsabs.harvard.edu/abs/2016A&A...591A..49C} {591, A49}

\bibitem[\protect\citeauthoryear{Cortese et~al.,}{Cortese
  et~al.}{2016}]{Cortese2016a}
Cortese L.,  et~al., 2016, \mn@doi [Monthly Notices of the Royal Astronomical
  Society] {10.1093/mnras/stw1891}, 463, 170

\bibitem[\protect\citeauthoryear{Cresci et~al.,}{Cresci
  et~al.}{2009}]{Cresci2009_DYSMAL}
Cresci G.,  et~al., 2009, \mn@doi [Astrophysical Journal]
  {10.1088/0004-637X/697/1/115}, 697, 115

\bibitem[\protect\citeauthoryear{{Davies} \& {Kasper}}{{Davies} \&
  {Kasper}}{2012}]{Davies_AO}
{Davies} R.,  {Kasper} M.,  2012, \mn@doi [\araa]
  {10.1146/annurev-astro-081811-125447}, \href
  {https://ui.adsabs.harvard.edu/abs/2012ARA&A..50..305D} {50, 305}

\bibitem[\protect\citeauthoryear{{Decarli} et~al.,}{{Decarli}
  et~al.}{2019}]{Decarli}
{Decarli} R.,  et~al., 2019, \mn@doi [\apj] {10.3847/1538-4357/ab30fe}, \href
  {https://ui.adsabs.harvard.edu/abs/2019ApJ...882..138D} {882, 138}

\bibitem[\protect\citeauthoryear{{Dekel} \& {Burkert}}{{Dekel} \&
  {Burkert}}{2014}]{Dekel_Burkert}
{Dekel} A.,  {Burkert} A.,  2014, \mn@doi [\mnras] {10.1093/mnras/stt2331},
  \href {https://ui.adsabs.harvard.edu/abs/2014MNRAS.438.1870D} {438, 1870}

\bibitem[\protect\citeauthoryear{{Dekel}, {Sari}  \& {Ceverino}}{{Dekel}
  et~al.}{2009}]{Dekel_2009}
{Dekel} A.,  {Sari} R.,   {Ceverino} D.,  2009, \mn@doi [\apj]
  {10.1088/0004-637X/703/1/785}, \href
  {https://ui.adsabs.harvard.edu/abs/2009ApJ...703..785D} {703, 785}

\bibitem[\protect\citeauthoryear{{Eisenhauer} et~al.,}{{Eisenhauer}
  et~al.}{2003}]{SINFONI}
{Eisenhauer} F.,  et~al., 2003, in {Iye} M.,  {Moorwood} A. F.~M.,  eds,
  Society of Photo-Optical Instrumentation Engineers (SPIE) Conference Series
  Vol. 4841, Instrument Design and Performance for Optical/Infrared
  Ground-based Telescopes. pp 1548--1561 (\mn@eprint {arXiv}
  {astro-ph/0306191}), \mn@doi{10.1117/12.459468}

\bibitem[\protect\citeauthoryear{{Elmegreen}, {Elmegreen}, {Rubin}  \&
  {Schaffer}}{{Elmegreen} et~al.}{2005}]{Elmegreen}
{Elmegreen} D.~M.,  {Elmegreen} B.~G.,  {Rubin} D.~S.,   {Schaffer} M.~A.,
  2005, \mn@doi [\apj] {10.1086/432502}, \href
  {https://ui.adsabs.harvard.edu/abs/2005ApJ...631...85E} {631, 85}

\bibitem[\protect\citeauthoryear{Epinat, Amram, Balkowski  \& Marcelin}{Epinat
  et~al.}{2010}]{Epinat2010}
Epinat B.,  Amram P.,  Balkowski C.,   Marcelin M.,  2010, \mn@doi [Monthly
  Notices of the Royal Astronomical Society]
  {10.1111/j.1365-2966.2009.15688.x}, 401, 2113

\bibitem[\protect\citeauthoryear{{Epinat} et~al.,}{{Epinat}
  et~al.}{2012}]{Epinat_2012}
{Epinat} B.,  et~al., 2012, \mn@doi [\aap] {10.1051/0004-6361/201117711}, \href
  {https://ui.adsabs.harvard.edu/abs/2012A&A...539A..92E} {539, A92}

\bibitem[\protect\citeauthoryear{{Fall}}{{Fall}}{1983}]{fall83}
{Fall} S.~M.,  1983, in {Athanassoula} E.,  ed., ~ Vol. 100, Internal
  Kinematics and Dynamics of Galaxies. pp 391--398

\bibitem[\protect\citeauthoryear{{Fall} \& {Efstathiou}}{{Fall} \&
  {Efstathiou}}{1980}]{Fall_Efstathiou}
{Fall} S.~M.,  {Efstathiou} G.,  1980, \mn@doi [\mnras]
  {10.1093/mnras/193.2.189}, \href
  {https://ui.adsabs.harvard.edu/abs/1980MNRAS.193..189F} {193, 189}

\bibitem[\protect\citeauthoryear{{Fall} \& {Romanowsky}}{{Fall} \&
  {Romanowsky}}{2013}]{F_and_R_2013}
{Fall} S.~M.,  {Romanowsky} A.~J.,  2013, \mn@doi [\apjl]
  {10.1088/2041-8205/769/2/L26}, \href
  {https://ui.adsabs.harvard.edu/abs/2013ApJ...769L..26F} {769, L26}

\bibitem[\protect\citeauthoryear{{Fall} \& {Romanowsky}}{{Fall} \&
  {Romanowsky}}{2018}]{Fall_Romanowsky_2018}
{Fall} S.~M.,  {Romanowsky} A.~J.,  2018, \mn@doi [\apj]
  {10.3847/1538-4357/aaeb27}, \href
  {https://ui.adsabs.harvard.edu/abs/2018ApJ...868..133F} {868, 133}

\bibitem[\protect\citeauthoryear{{Fisher} \& {Drory}}{{Fisher} \&
  {Drory}}{2010}]{Fisher_Drory}
{Fisher} D.~B.,  {Drory} N.,  2010, \mn@doi [\apj]
  {10.1088/0004-637X/716/2/942}, \href
  {https://ui.adsabs.harvard.edu/abs/2010ApJ...716..942F} {716, 942}

\bibitem[\protect\citeauthoryear{{Fisher} et~al.,}{{Fisher}
  et~al.}{2017}]{Fisher}
{Fisher} D.~B.,  et~al., 2017, \mn@doi [\mnras] {10.1093/mnras/stw2281}, \href
  {https://ui.adsabs.harvard.edu/abs/2017MNRAS.464..491F} {464, 491}

\bibitem[\protect\citeauthoryear{{Foreman-Mackey}, {Hogg}, {Lang}  \&
  {Goodman}}{{Foreman-Mackey} et~al.}{2013}]{Foreman}
{Foreman-Mackey} D.,  {Hogg} D.~W.,  {Lang} D.,   {Goodman} J.,  2013, \mn@doi
  [\pasp] {10.1086/670067}, \href
  {https://ui.adsabs.harvard.edu/abs/2013PASP..125..306F} {125, 306}

\bibitem[\protect\citeauthoryear{{F{\"o}rster Schreiber} et~al.,}{{F{\"o}rster
  Schreiber} et~al.}{2009}]{Forster_2008_no_AO}
{F{\"o}rster Schreiber} N.~M.,  et~al., 2009, \mn@doi [\apj]
  {10.1088/0004-637X/706/2/1364}, \href
  {https://ui.adsabs.harvard.edu/abs/2009ApJ...706.1364F} {706, 1364}

\bibitem[\protect\citeauthoryear{{F{\"o}rster Schreiber} et~al.,}{{F{\"o}rster
  Schreiber} et~al.}{2018}]{SINS/zC-SINF}
{F{\"o}rster Schreiber} N.~M.,  et~al., 2018, \mn@doi [\apjs]
  {10.3847/1538-4365/aadd49}, \href
  {https://ui.adsabs.harvard.edu/abs/2018ApJS..238...21F} {238, 21}

\bibitem[\protect\citeauthoryear{{Fouque}, {Bottinelli}, {Gouguenheim}  \&
  {Paturel}}{{Fouque} et~al.}{1990}]{Fouque}
{Fouque} P.,  {Bottinelli} L.,  {Gouguenheim} L.,   {Paturel} G.,  1990,
  \mn@doi [\apj] {10.1086/168288}, \href
  {https://ui.adsabs.harvard.edu/abs/1990ApJ...349....1F} {349, 1}

\bibitem[\protect\citeauthoryear{{Freeman}}{{Freeman}}{1970}]{Freeman}
{Freeman} K.~C.,  1970, \mn@doi [\apj] {10.1086/150474}, \href
  {https://ui.adsabs.harvard.edu/abs/1970ApJ...160..811F} {160, 811}

\bibitem[\protect\citeauthoryear{{Genzel} et~al.,}{{Genzel}
  et~al.}{2006}]{Genzel_2006}
{Genzel} R.,  et~al., 2006, \mn@doi [\nat] {10.1038/nature05052}, \href
  {https://ui.adsabs.harvard.edu/abs/2006Natur.442..786G} {442, 786}

\bibitem[\protect\citeauthoryear{{Genzel} et~al.,}{{Genzel}
  et~al.}{2008}]{Genzel_2008}
{Genzel} R.,  et~al., 2008, \mn@doi [\apj] {10.1086/591840}, \href
  {https://ui.adsabs.harvard.edu/abs/2008ApJ...687...59G} {687, 59}

\bibitem[\protect\citeauthoryear{{Giavalisco} et~al.,}{{Giavalisco}
  et~al.}{2004}]{Goods}
{Giavalisco} M.,  et~al., 2004, \mn@doi [\apjl] {10.1086/379232}, \href
  {https://ui.adsabs.harvard.edu/abs/2004ApJ...600L..93G} {600, L93}

\bibitem[\protect\citeauthoryear{{Gillman} et~al.,}{{Gillman}
  et~al.}{2019}]{Gillman_Hizels}
{Gillman} S.,  et~al., 2019, \mn@doi [\mnras] {10.1093/mnras/stz765}, \href
  {https://ui.adsabs.harvard.edu/abs/2019MNRAS.486..175G} {486, 175}

\bibitem[\protect\citeauthoryear{{Gillman} et~al.,}{{Gillman}
  et~al.}{2020}]{Gillman}
{Gillman} S.,  et~al., 2020, \mn@doi [\mnras] {10.1093/mnras/stz3576}, \href
  {https://ui.adsabs.harvard.edu/abs/2020MNRAS.492.1492G} {492, 1492}

\bibitem[\protect\citeauthoryear{Glazebrook}{Glazebrook}{2013}]{Glazebrook_2013}
Glazebrook K.,  2013, \mn@doi [Publications of the Astronomical Society of
  Australia] {10.1017/pasa.2013.34}, 30, e056

\bibitem[\protect\citeauthoryear{{Glazebrook}, {Ellis}, {Santiago}  \&
  {Griffiths}}{{Glazebrook} et~al.}{1995}]{Glazebrook95}
{Glazebrook} K.,  {Ellis} R.,  {Santiago} B.,   {Griffiths} R.,  1995, \mn@doi
  [\mnras] {10.1093/mnras/275.1.L19}, \href
  {https://ui.adsabs.harvard.edu/abs/1995MNRAS.275L..19G} {275, L19}

\bibitem[\protect\citeauthoryear{{Gu{\'e}rou} et~al.,}{{Gu{\'e}rou}
  et~al.}{2017}]{Guerou_2017}
{Gu{\'e}rou} A.,  et~al., 2017, \mn@doi [\aap] {10.1051/0004-6361/201730905},
  \href {https://ui.adsabs.harvard.edu/abs/2017A&A...608A...5G} {608, A5}

\bibitem[\protect\citeauthoryear{{Harrison} et~al.,}{{Harrison}
  et~al.}{2017}]{Harrison_2017}
{Harrison} C.~M.,  et~al., 2017, \mn@doi [\mnras] {10.1093/mnras/stx217}, \href
  {https://ui.adsabs.harvard.edu/abs/2017MNRAS.467.1965H} {467, 1965}

\bibitem[\protect\citeauthoryear{{Heidmann}, {Heidmann}  \& {de
  Vaucouleurs}}{{Heidmann} et~al.}{1972}]{Heidmann}
{Heidmann} J.,  {Heidmann} N.,   {de Vaucouleurs} G.,  1972, \memras, \href
  {https://ui.adsabs.harvard.edu/abs/1972MmRAS..75...85H} {75, 85}

\bibitem[\protect\citeauthoryear{{Holmberg}}{{Holmberg}}{1946}]{Holmberg_inclination_axis_ratio}
{Holmberg} E.,  1946, Meddelanden fran Lunds Astronomiska Observatorium Serie
  II, \href {https://ui.adsabs.harvard.edu/abs/1946MeLuS.117....3H} {117, 3}

\bibitem[\protect\citeauthoryear{{Hubble}}{{Hubble}}{1926}]{Hubble_sequence}
{Hubble} E.,  1926, Contributions from the Mount Wilson Observatory / Carnegie
  Institution of Washington, \href
  {https://ui.adsabs.harvard.edu/abs/1926CMWCI.324....1H} {324, 1}

\bibitem[\protect\citeauthoryear{{Hunter}}{{Hunter}}{2007}]{matplotlib}
{Hunter} J.~D.,  2007, \mn@doi [Computing in Science \& Engineering]
  {10.1109/MCSE.2007.55}, 9, 90

\bibitem[\protect\citeauthoryear{{J{\'o}zsa}, {Kenn}, {Klein}  \&
  {Oosterloo}}{{J{\'o}zsa} et~al.}{2007}]{Tirific}
{J{\'o}zsa} G.~I.~G.,  {Kenn} F.,  {Klein} U.,   {Oosterloo} T.~A.,  2007,
  \mn@doi [\aap] {10.1051/0004-6361:20066164}, \href
  {https://ui.adsabs.harvard.edu/abs/2007A&A...468..731J} {468, 731}

\bibitem[\protect\citeauthoryear{{Kamphuis}, {J{\'o}zsa}, {Oh}, {Spekkens},
  {Urbancic}, {Serra}, {Koribalski}  \& {Dettmar}}{{Kamphuis}
  et~al.}{2015}]{Kamphuis}
{Kamphuis} P.,  {J{\'o}zsa} G.~I.~G.,  {Oh} S. .~H.,  {Spekkens} K.,
  {Urbancic} N.,  {Serra} P.,  {Koribalski} B.~S.,   {Dettmar} R.~J.,  2015,
  \mn@doi [\mnras] {10.1093/mnras/stv1480}, \href
  {https://ui.adsabs.harvard.edu/abs/2015MNRAS.452.3139K} {452, 3139}

\bibitem[\protect\citeauthoryear{{Koekemoer} et~al.,}{{Koekemoer}
  et~al.}{2011}]{Candels}
{Koekemoer} A.~M.,  et~al., 2011, \mn@doi [\apjs] {10.1088/0067-0049/197/2/36},
  \href {https://ui.adsabs.harvard.edu/abs/2011ApJS..197...36K} {197, 36}

\bibitem[\protect\citeauthoryear{{Krajnovi{\'c}}, {Cappellari}, {de Zeeuw}  \&
  {Copin}}{{Krajnovi{\'c}} et~al.}{2006}]{Kinemetry}
{Krajnovi{\'c}} D.,  {Cappellari} M.,  {de Zeeuw} P.~T.,   {Copin} Y.,  2006,
  \mn@doi [\mnras] {10.1111/j.1365-2966.2005.09902.x}, \href
  {https://ui.adsabs.harvard.edu/abs/2006MNRAS.366..787K} {366, 787}

\bibitem[\protect\citeauthoryear{{Larkin} et~al.,}{{Larkin}
  et~al.}{2006}]{OSIRIS}
{Larkin} J.,  et~al., 2006, in Society of Photo-Optical Instrumentation
  Engineers (SPIE) Conference Series. p. 62691A, \mn@doi{10.1117/12.672061}

\bibitem[\protect\citeauthoryear{{Law}, {Steidel}, {Erb}, {Larkin}, {Pettini},
  {Shapley}  \& {Wright}}{{Law} et~al.}{2007}]{Law}
{Law} D.~R.,  {Steidel} C.~C.,  {Erb} D.~K.,  {Larkin} J.~E.,  {Pettini} M.,
  {Shapley} A.~E.,   {Wright} S.~A.,  2007, \mn@doi [\apj] {10.1086/521786},
  \href {https://ui.adsabs.harvard.edu/abs/2007ApJ...669..929L} {669, 929}

\bibitem[\protect\citeauthoryear{{Law}, {Steidel}, {Erb}, {Larkin}, {Pettini},
  {Shapley}  \& {Wright}}{{Law} et~al.}{2009}]{Law_2}
{Law} D.~R.,  {Steidel} C.~C.,  {Erb} D.~K.,  {Larkin} J.~E.,  {Pettini} M.,
  {Shapley} A.~E.,   {Wright} S.~A.,  2009, \mn@doi [\apj]
  {10.1088/0004-637X/697/2/2057}, \href
  {https://ui.adsabs.harvard.edu/abs/2009ApJ...697.2057L} {697, 2057}

\bibitem[\protect\citeauthoryear{{Lawrence} et~al.,}{{Lawrence}
  et~al.}{2007}]{UDS}
{Lawrence} A.,  et~al., 2007, \mn@doi [\mnras]
  {10.1111/j.1365-2966.2007.12040.x}, \href
  {https://ui.adsabs.harvard.edu/abs/2007MNRAS.379.1599L} {379, 1599}

\bibitem[\protect\citeauthoryear{{Liao}, {Gao}, {Frenk}, {Guo}  \&
  {Wang}}{{Liao} et~al.}{2017}]{Liao}
{Liao} S.,  {Gao} L.,  {Frenk} C.~S.,  {Guo} Q.,   {Wang} J.,  2017, \mn@doi
  [\mnras] {10.1093/mnras/stx1391}, \href
  {https://ui.adsabs.harvard.edu/abs/2017MNRAS.470.2262L} {470, 2262}

\bibitem[\protect\citeauthoryear{{Livermore} et~al.,}{{Livermore}
  et~al.}{2015}]{Livermore}
{Livermore} R.~C.,  et~al., 2015, \mn@doi [\mnras] {10.1093/mnras/stv686},
  \href {https://ui.adsabs.harvard.edu/abs/2015MNRAS.450.1812L} {450, 1812}

\bibitem[\protect\citeauthoryear{{Madau} \& {Dickinson}}{{Madau} \&
  {Dickinson}}{2014}]{Madau_SFH}
{Madau} P.,  {Dickinson} M.,  2014, \mn@doi [\araa]
  {10.1146/annurev-astro-081811-125615}, \href
  {https://ui.adsabs.harvard.edu/abs/2014ARA&A..52..415M} {52, 415}

\bibitem[\protect\citeauthoryear{Marasco, Fraternali, Posti, Ijtsma, {Di
  Teodoro}  \& Oosterloo}{Marasco et~al.}{2019}]{Marasco2019}
Marasco A.,  Fraternali F.,  Posti L.,  Ijtsma M.,  {Di Teodoro} E.~M.,
  Oosterloo T.,  2019, \mn@doi [Astronomy and Astrophysics]
  {10.1051/0004-6361/201834456}

\bibitem[\protect\citeauthoryear{{Markwardt}}{{Markwardt}}{2009}]{MPFIT}
{Markwardt} C.~B.,  2009, in {Bohlender} D.~A.,  {Durand} D.,   {Dowler} P.,
  eds,  Astronomical Society of the Pacific Conference Series Vol. 411,
  Astronomical Data Analysis Software and Systems XVIII. p.~251 (\mn@eprint
  {arXiv} {0902.2850})

\bibitem[\protect\citeauthoryear{{Mieda}}{{Mieda}}{2015}]{Mieda}
{Mieda} E.,  2015, PhD thesis, University of Toronto

\bibitem[\protect\citeauthoryear{{Mo}, {Mao}  \& {White}}{{Mo}
  et~al.}{1998}]{Mo_galaxy_formation}
{Mo} H.~J.,  {Mao} S.,   {White} S. D.~M.,  1998, \mn@doi [\mnras]
  {10.1046/j.1365-8711.1998.01227.x}, \href
  {https://ui.adsabs.harvard.edu/abs/1998MNRAS.295..319M} {295, 319}

\bibitem[\protect\citeauthoryear{{Mobasher} et~al.,}{{Mobasher}
  et~al.}{2015}]{Mobasher}
{Mobasher} B.,  et~al., 2015, \mn@doi [\apj] {10.1088/0004-637X/808/1/101},
  \href {https://ui.adsabs.harvard.edu/abs/2015ApJ...808..101M} {808, 101}

\bibitem[\protect\citeauthoryear{{Moffat}}{{Moffat}}{1969}]{Moffat}
{Moffat} A.~F.~J.,  1969, \aap, \href
  {https://ui.adsabs.harvard.edu/abs/1969A&A.....3..455M} {3, 455}

\bibitem[\protect\citeauthoryear{{Mowla} et~al.,}{{Mowla}
  et~al.}{2019}]{DASH-HST}
{Mowla} L.~A.,  et~al., 2019, \mn@doi [\apj] {10.3847/1538-4357/ab290a}, \href
  {https://ui.adsabs.harvard.edu/abs/2019ApJ...880...57M} {880, 57}

\bibitem[\protect\citeauthoryear{{Naab} \& {Ostriker}}{{Naab} \&
  {Ostriker}}{2017}]{Naab}
{Naab} T.,  {Ostriker} J.~P.,  2017, \mn@doi [\araa]
  {10.1146/annurev-astro-081913-040019}, \href
  {https://ui.adsabs.harvard.edu/abs/2017ARA&A..55...59N} {55, 59}

\bibitem[\protect\citeauthoryear{{Obreschkow} \& {Glazebrook}}{{Obreschkow} \&
  {Glazebrook}}{2014a}]{KarlandDanail}
{Obreschkow} D.,  {Glazebrook} K.,  2014a, \mn@doi [\apj]
  {10.1088/0004-637X/784/1/26}, \href
  {https://ui.adsabs.harvard.edu/abs/2014ApJ...784...26O} {784, 26}

\bibitem[\protect\citeauthoryear{{Obreschkow} \& {Glazebrook}}{{Obreschkow} \&
  {Glazebrook}}{2014b}]{Obreschkow_2014}
{Obreschkow} D.,  {Glazebrook} K.,  2014b, \mn@doi [\apj]
  {10.1088/0004-637X/784/1/26}, \href
  {https://ui.adsabs.harvard.edu/abs/2014ApJ...784...26O} {784, 26}

\bibitem[\protect\citeauthoryear{{Obreschkow} et~al.,}{{Obreschkow}
  et~al.}{2015}]{Obreschkow_2015}
{Obreschkow} D.,  et~al., 2015, \mn@doi [\apj] {10.1088/0004-637X/815/2/97},
  \href {https://ui.adsabs.harvard.edu/abs/2015ApJ...815...97O} {815, 97}

\bibitem[\protect\citeauthoryear{{Palunas} \& {Williams}}{{Palunas} \&
  {Williams}}{2000}]{Palunas_Williams}
{Palunas} P.,  {Williams} T.~B.,  2000, \mn@doi [\aj] {10.1086/316878}, \href
  {https://ui.adsabs.harvard.edu/abs/2000AJ....120.2884P} {120, 2884}

\bibitem[\protect\citeauthoryear{{Peebles}}{{Peebles}}{1969}]{Peebles}
{Peebles} P.~J.~E.,  1969, \mn@doi [\apj] {10.1086/149876}, \href
  {https://ui.adsabs.harvard.edu/abs/1969ApJ...155..393P} {155, 393}

\bibitem[\protect\citeauthoryear{{Pichon}, {Pogosyan}, {Kimm}, {Slyz},
  {Devriendt}  \& {Dubois}}{{Pichon} et~al.}{2011}]{Pichon}
{Pichon} C.,  {Pogosyan} D.,  {Kimm} T.,  {Slyz} A.,  {Devriendt} J.,
  {Dubois} Y.,  2011, \mn@doi [\mnras] {10.1111/j.1365-2966.2011.19640.x},
  \href {https://ui.adsabs.harvard.edu/abs/2011MNRAS.418.2493P} {418, 2493}

\bibitem[\protect\citeauthoryear{{Posti}, {Fraternali}, {Di Teodoro}  \&
  {Pezzulli}}{{Posti} et~al.}{2018}]{Posti_2018}
{Posti} L.,  {Fraternali} F.,  {Di Teodoro} E.~M.,   {Pezzulli} G.,  2018,
  \mn@doi [\aap] {10.1051/0004-6361/201833091}, \href
  {https://ui.adsabs.harvard.edu/abs/2018A&A...612L...6P} {612, L6}

\bibitem[\protect\citeauthoryear{{Rodrigues}, {Hammer}, {Flores}, {Puech}  \&
  {Athanassoula}}{{Rodrigues} et~al.}{2017}]{Rodrigues}
{Rodrigues} M.,  {Hammer} F.,  {Flores} H.,  {Puech} M.,   {Athanassoula} E.,
  2017, \mn@doi [\mnras] {10.1093/mnras/stw2711}, \href
  {https://ui.adsabs.harvard.edu/abs/2017MNRAS.465.1157R} {465, 1157}

\bibitem[\protect\citeauthoryear{{Romanowsky} \& {Fall}}{{Romanowsky} \&
  {Fall}}{2012}]{Romanowsky}
{Romanowsky} A.~J.,  {Fall} S.~M.,  2012, \mn@doi [\apjs]
  {10.1088/0067-0049/203/2/17}, \href
  {https://ui.adsabs.harvard.edu/abs/2012ApJS..203...17R} {203, 17}

\bibitem[\protect\citeauthoryear{{Scoville} et~al.,}{{Scoville}
  et~al.}{2007}]{COSMOS}
{Scoville} N.,  et~al., 2007, \mn@doi [\apjs] {10.1086/516585}, \href
  {https://ui.adsabs.harvard.edu/abs/2007ApJS..172....1S} {172, 1}

\bibitem[\protect\citeauthoryear{{Sellwood} \& {Spekkens}}{{Sellwood} \&
  {Spekkens}}{2015}]{Diskfit}
{Sellwood} J.~A.,  {Spekkens} K.,  2015, arXiv e-prints, \href
  {https://ui.adsabs.harvard.edu/abs/2015arXiv150907120S} {p. arXiv:1509.07120}

\bibitem[\protect\citeauthoryear{{Sharples} et~al.,}{{Sharples}
  et~al.}{2013}]{KMOS}
{Sharples} R.,  et~al., 2013, The Messenger, \href
  {https://ui.adsabs.harvard.edu/abs/2013Msngr.151...21S} {151, 21}

\bibitem[\protect\citeauthoryear{{Simon}, {Bolatto}, {Leroy}  \&
  {Blitz}}{{Simon} et~al.}{2003}]{RINGFIT}
{Simon} J.~D.,  {Bolatto} A.~D.,  {Leroy} A.,   {Blitz} L.,  2003, \mn@doi
  [\apj] {10.1086/378200}, \href
  {https://ui.adsabs.harvard.edu/abs/2003ApJ...596..957S} {596, 957}

\bibitem[\protect\citeauthoryear{{Simons} et~al.,}{{Simons}
  et~al.}{2017}]{Kassin_disk_assembly}
{Simons} R.~C.,  et~al., 2017, \mn@doi [\apj] {10.3847/1538-4357/aa740c}, \href
  {https://ui.adsabs.harvard.edu/abs/2017ApJ...843...46S} {843, 46}

\bibitem[\protect\citeauthoryear{Simons et~al.,}{Simons
  et~al.}{2019}]{Simons2019}
Simons R.~C.,  et~al., 2019, \mn@doi [The Astrophysical Journal]
  {10.3847/1538-4357/ab07c9}, 874, 59

\bibitem[\protect\citeauthoryear{{Stott} et~al.,}{{Stott}
  et~al.}{2016}]{Stott_2016}
{Stott} J.~P.,  et~al., 2016, \mn@doi [\mnras] {10.1093/mnras/stw129}, \href
  {https://ui.adsabs.harvard.edu/abs/2016MNRAS.457.1888S} {457, 1888}

\bibitem[\protect\citeauthoryear{{Sweet} et~al.,}{{Sweet} et~al.}{2019}]{Sweet}
{Sweet} S.~M.,  et~al., 2019, \mn@doi [\mnras] {10.1093/mnras/stz750}, \href
  {https://ui.adsabs.harvard.edu/abs/2019MNRAS.485.5700S} {485, 5700}

\bibitem[\protect\citeauthoryear{{Swinbank} et~al.,}{{Swinbank}
  et~al.}{2017}]{Swinbank}
{Swinbank} A.~M.,  et~al., 2017, \mn@doi [\mnras] {10.1093/mnras/stx201}, \href
  {https://ui.adsabs.harvard.edu/abs/2017MNRAS.467.3140S} {467, 3140}

\bibitem[\protect\citeauthoryear{{Szomoru}, {Franx}, {Bouwens}, {van Dokkum},
  {Labb{\'e}}, {Illingworth}  \& {Trenti}}{{Szomoru} et~al.}{2011}]{Szomoru}
{Szomoru} D.,  {Franx} M.,  {Bouwens} R.~J.,  {van Dokkum} P.~G.,  {Labb{\'e}}
  I.,  {Illingworth} G.~D.,   {Trenti} M.,  2011, \mn@doi [\apjl]
  {10.1088/2041-8205/735/1/L22}, \href
  {https://ui.adsabs.harvard.edu/abs/2011ApJ...735L..22S} {735, L22}

\bibitem[\protect\citeauthoryear{{Tacchella} et~al.,}{{Tacchella}
  et~al.}{2015}]{Tacchella_sins_sizes}
{Tacchella} S.,  et~al., 2015, \mn@doi [\apj] {10.1088/0004-637X/802/2/101},
  \href {https://ui.adsabs.harvard.edu/abs/2015ApJ...802..101T} {802, 101}

\bibitem[\protect\citeauthoryear{{Tacchella}, {Dekel}, {Carollo}, {Ceverino},
  {DeGraf}, {Lapiner}, {Mand elker}  \& {Primack Joel}}{{Tacchella}
  et~al.}{2016}]{Tacchella}
{Tacchella} S.,  {Dekel} A.,  {Carollo} C.~M.,  {Ceverino} D.,  {DeGraf} C.,
  {Lapiner} S.,  {Mand elker} N.,   {Primack Joel} R.,  2016, \mn@doi [\mnras]
  {10.1093/mnras/stw131}, \href
  {https://ui.adsabs.harvard.edu/abs/2016MNRAS.457.2790T} {457, 2790}

\bibitem[\protect\citeauthoryear{{Tacconi} et~al.,}{{Tacconi}
  et~al.}{2013}]{Tacconi}
{Tacconi} L.~J.,  et~al., 2013, \mn@doi [\apj] {10.1088/0004-637X/768/1/74},
  \href {https://ui.adsabs.harvard.edu/abs/2013ApJ...768...74T} {768, 74}

\bibitem[\protect\citeauthoryear{Teklu, Remus, Dolag, Beck, Burkert, Schmidt,
  Schulze  \& Steinborn}{Teklu et~al.}{2015}]{Teklu}
Teklu A.,  Remus R.,  Dolag K.,  Beck A.,  Burkert A.,  Schmidt A.,  Schulze
  F.,   Steinborn L.,  2015, \mn@doi [The Astrophysical Journal]
  {10.1088/0004-637X/812/1/29}, 812

\bibitem[\protect\citeauthoryear{{Teuben}}{{Teuben}}{2004}]{NEMO}
{Teuben} P.~J.,  2004, in {Ochsenbein} F.,  {Allen} M.~G.,   {Egret} D.,  eds,
  Astronomical Society of the Pacific Conference Series Vol. 314, Astronomical
  Data Analysis Software and Systems (ADASS) XIII. p.~621

\bibitem[\protect\citeauthoryear{{Tiley} et~al.,}{{Tiley}
  et~al.}{2019a}]{Tiley_2019}
{Tiley} A.~L.,  et~al., 2019a, \mn@doi [\mnras] {10.1093/mnras/sty2794}, \href
  {https://ui.adsabs.harvard.edu/abs/2019MNRAS.482.2166T} {482, 2166}

\bibitem[\protect\citeauthoryear{{Tiley} et~al.,}{{Tiley}
  et~al.}{2019b}]{Tiley}
{Tiley} A.~L.,  et~al., 2019b, \mn@doi [\mnras] {10.1093/mnras/stz428}, \href
  {https://ui.adsabs.harvard.edu/abs/2019MNRAS.485..934T} {485, 934}

\bibitem[\protect\citeauthoryear{{Tiley} et~al.,}{{Tiley}
  et~al.}{2021}]{KGES_alfie}
{Tiley} A.~L.,  et~al., 2021, \mn@doi [\mnras] {10.1093/mnras/stab1692}, \href
  {https://ui.adsabs.harvard.edu/abs/2021MNRAS.506..323T} {506, 323}

\bibitem[\protect\citeauthoryear{{Tully} \& {Fisher}}{{Tully} \&
  {Fisher}}{1977}]{Tully_fisher}
{Tully} R.~B.,  {Fisher} J.~R.,  1977, \aap, \href
  {https://ui.adsabs.harvard.edu/abs/1977A&A....54..661T} {500, 105}

\bibitem[\protect\citeauthoryear{{Ventou} et~al.,}{{Ventou}
  et~al.}{2017}]{Ventou17}
{Ventou} E.,  et~al., 2017, \mn@doi [\aap] {10.1051/0004-6361/201731586}, \href
  {https://ui.adsabs.harvard.edu/abs/2017A&A...608A...9V} {608, A9}

\bibitem[\protect\citeauthoryear{{Ventou} et~al.,}{{Ventou}
  et~al.}{2019}]{Ventou19}
{Ventou} E.,  et~al., 2019, \mn@doi [\aap] {10.1051/0004-6361/201935597}, \href
  {https://ui.adsabs.harvard.edu/abs/2019A&A...631A..87V} {631, A87}

\bibitem[\protect\citeauthoryear{{Virtanen} et~al.,}{{Virtanen}
  et~al.}{2019}]{scipy}
{Virtanen} P.,  et~al., 2019, arXiv e-prints, \href
  {https://ui.adsabs.harvard.edu/abs/2019arXiv190710121V} {p. arXiv:1907.10121}

\bibitem[\protect\citeauthoryear{{Wechsler} \& {Tinker}}{{Wechsler} \&
  {Tinker}}{2018}]{Wechsler}
{Wechsler} R.~H.,  {Tinker} J.~L.,  2018, \mn@doi [\araa]
  {10.1146/annurev-astro-081817-051756}, \href
  {https://ui.adsabs.harvard.edu/abs/2018ARA&A..56..435W} {56, 435}

\bibitem[\protect\citeauthoryear{{Whitaker} et~al.,}{{Whitaker}
  et~al.}{2019}]{HLF}
{Whitaker} K.~E.,  et~al., 2019, \mn@doi [\apjs] {10.3847/1538-4365/ab3853},
  \href {https://ui.adsabs.harvard.edu/abs/2019ApJS..244...16W} {244, 16}

\bibitem[\protect\citeauthoryear{Wisnioski}{Wisnioski}{2012}]{Emily}
Wisnioski E.,  2012, PhD thesis, Swinburne University of Technology

\bibitem[\protect\citeauthoryear{{Wisnioski} et~al.,}{{Wisnioski}
  et~al.}{2015}]{KMOS3D}
{Wisnioski} E.,  et~al., 2015, \mn@doi [\apj] {10.1088/0004-637X/799/2/209},
  \href {https://ui.adsabs.harvard.edu/abs/2015ApJ...799..209W} {799, 209}

\bibitem[\protect\citeauthoryear{{Wisnioski} et~al.,}{{Wisnioski}
  et~al.}{2019}]{Wisnioski_2019}
{Wisnioski} E.,  et~al., 2019, \mn@doi [\apj] {10.3847/1538-4357/ab4db8}, \href
  {https://ui.adsabs.harvard.edu/abs/2019ApJ...886..124W} {886, 124}

\bibitem[\protect\citeauthoryear{{Zasov}, {Saburova}, {Khoperskov}  \&
  {Khoperskov}}{{Zasov} et~al.}{2017}]{Zasov}
{Zasov} A.~V.,  {Saburova} A.~S.,  {Khoperskov} A.~V.,   {Khoperskov} S.~A.,
  2017, \mn@doi [Physics Uspekhi] {10.3367/UFNe.2016.03.037751}, \href
  {https://ui.adsabs.harvard.edu/abs/2017PhyU...60....3Z} {60, 3}

\bibitem[\protect\citeauthoryear{{de Blok}, {Walter}, {Brinks}, {Trachternach},
  {Oh}  \& {Kennicutt}}{{de Blok} et~al.}{2008}]{de_Blok}
{de Blok} W.~J.~G.,  {Walter} F.,  {Brinks} E.,  {Trachternach} C.,  {Oh}
  S.~H.,   {Kennicutt} R.~C. J.,  2008, \mn@doi [\aj]
  {10.1088/0004-6256/136/6/2648}, \href
  {https://ui.adsabs.harvard.edu/abs/2008AJ....136.2648D} {136, 2648}

\bibitem[\protect\citeauthoryear{di Teodoro \& Fraternali}{di~Teodoro \&
  Fraternali}{2015}]{DiTeodoro2015}
di Teodoro E.~M.,  Fraternali F.,  2015, \mn@doi [Monthly Notices of the Royal
  Astronomical Society] {10.1093/mnras/stv1213}, 451, 3021

\bibitem[\protect\citeauthoryear{{van de Hulst}, {Raimond}  \& {van
  Woerden}}{{van de Hulst} et~al.}{1957}]{Van_de_Hulst}
{van de Hulst} H.~C.,  {Raimond} E.,   {van Woerden} H.,  1957, \bain, \href
  {https://ui.adsabs.harvard.edu/abs/1957BAN....14....1V} {14, 1}

\bibitem[\protect\citeauthoryear{{van den Bosch}}{{van den
  Bosch}}{1998}]{van_den_Bosch}
{van den Bosch} F.~C.,  1998, \mn@doi [\apj] {10.1086/306354}, \href
  {https://ui.adsabs.harvard.edu/abs/1998ApJ...507..601V} {507, 601}

\bibitem[\protect\citeauthoryear{{van der Hulst}, {Terlouw}, {Begeman},
  {Zwitser}  \& {Roelfsema}}{{van der Hulst} et~al.}{1992}]{GIPSY}
{van der Hulst} J.~M.,  {Terlouw} J.~P.,  {Begeman} K.~G.,  {Zwitser} W.,
  {Roelfsema} P.~R.,  1992, in {Worrall} D.~M.,  {Biemesderfer} C.,   {Barnes}
  J.,  eds,  Astronomical Society of the Pacific Conference Series Vol. 25,
  Astronomical Data Analysis Software and Systems I. p.~131

\bibitem[\protect\citeauthoryear{{van der Velden}}{{van der
  Velden}}{2020}]{cmasher}
{van der Velden} E.,  2020, \mn@doi [The Journal of Open Source Software]
  {10.21105/joss.02004}, \href
  {https://ui.adsabs.harvard.edu/abs/2020JOSS....5.2004V} {5, 2004}

\bibitem[\protect\citeauthoryear{{van der Walt}, {Colbert}  \&
  {Varoquaux}}{{van der Walt} et~al.}{2011}]{numpy}
{van der Walt} S.,  {Colbert} S.~C.,   {Varoquaux} G.,  2011, \mn@doi
  [Computing in Science and Engineering] {10.1109/MCSE.2011.37}, \href
  {https://ui.adsabs.harvard.edu/abs/2011CSE....13b..22V} {13, 22}

\bibitem[\protect\citeauthoryear{{van der Wel} et~al.,}{{van der Wel}
  et~al.}{2014}]{van_der_Wel}
{van der Wel} A.,  et~al., 2014, \mn@doi [\apj] {10.1088/0004-637X/788/1/28},
  \href {https://ui.adsabs.harvard.edu/abs/2014ApJ...788...28V} {788, 28}

\makeatother
\end{thebibliography}


%\begin{thebibliography}{99}
%\end{thebibliography}

%%%%%%%%%%%%%%%%%%%%%%%%%%%%%%%%%%%%%%%%%%%%%%%%%%

%%%%%%%%%%%%%%%%% APPENDICES %%%%%%%%%%%%%%%%%%%%%

\appendix

\section{Kinematic fitting code}
\label{section:appendix_code}

We developed a fitting code based on a maximum likelihood estimation with \textsc{emcee}\footnote{\url{https://emcee.readthedocs.io/en/stable/}} (\citealp{Foreman}) that takes the velocity fields from two different datasets of the same galaxy and finds the best kinematic model. As mentioned in the main text, the two datasets can have different spatial resolutions, radial extents, alignments and pixel scales. Additionally, the code can also find the best fit from individual data.

The deprojected velocity models are built following \cite{KarlandDanail} (Appendix B) where the parameters to fit are position angle $\theta_\mathrm{PA}$, inclination $i$, kinematic centre coordinates ($x_0$, $y_0$), asymptotic velocity $v_{\textrm{flat}}$ and characteristic radius of the rotation curve $r_{\textrm{flat}}$. The deprojected velocity model is obtained by multiplying the circular velocity model $v_c(x,y)$ by a deprojection function in each pixel as

\begin{ceqn}
\begin{align}
   v(x,y) = \mathcal{C}(x,y) v_c(x,y),
\label{eq:v deprojected}
\end{align}
\end{ceqn}

where $\mathcal{C}(x,y)$ is called the deprojection factor and is calculated as:

\begin{ceqn}
\begin{align}
   \mathcal{C}(x,y) = \frac{\sin i}{\sqrt{1+(y'/x')^2}}  \sign (x'),
\label{eq:deprojection factor}
\end{align}
\end{ceqn}

with $x'$ and $y'$ the face-on coordinates that are computed from the parameters position angle $\theta_\mathrm{PA}$, inclination $i$ and kinematic centre ($x_0,y_0$) as

\begin{ceqn}
\begin{align}
   x' = \cos \theta_\mathrm{PA}(x-x_0) + \sin \theta_\mathrm{PA}(y-y_0),
\label{eq:x face on}
\end{align}
\end{ceqn}

\begin{ceqn}
\begin{align}
   y' = \frac{ \cos \theta_\mathrm{PA}(y-y_0) - \sin\theta_\mathrm{PA}(x-x_0)}{\cos i}.
\label{eq:y face on}
\end{align}
\end{ceqn}

After the deprojected velocity model is created, it is masked so that it matches the input velocity field in each pixel and it is then used to create a model datacube. The position of the model emission line is drawn from the velocity model and the line intensity is given by an input surface mass density model characterized by the size of the galaxy $r_d$. The width of the line is modelled according to the instrument spectral resolution. The model datacube is then convolved with the data PSF accordingly (Airy Disk + Moffat kernels in the adaptive optics case and a Gaussian kernel for natural seeing) and the velocity maps are then extracted from the convolved cubes. 

The best fit model is obtained from minimizing the negative log-likelihood function (which corresponds to maximizing the log-likelihood) using the two input velocity fields and the two model velocity fields. The initial guess for the position angle is estimated using \textsc{pafit}\footnote{ \url{http://www-astro.physics.ox.ac.uk/~mxc/software/\#pafit}} and the initial guess of the kinematic centre is estimated from a visual inspection of the peak in the H$\alpha$ intensity maps. The best-fit parameters from the probability distribution are calculated as the median of the marginalized parameters in the last 100 steps of the MCMC run which corresponds to the maximum of the likelihood function in the hyperspace of parameters.

In the case of fitting to a single resolution, the $\chi^2$ used in the likelihood estimation is drawn from the individual fit. In the case of the combined analysis, where the data at both resolutions is used, the $\chi^2$ is the sum of the two different contributions from the two datasets

\begin{ceqn}
\begin{align}
   \chi^2 =  \underbrace{\sum_{i,j} \frac{(v_{i,j}- \tilde{v}_{i,j})^2}{\sigma_{i,j}^2}}_{\chi_\mathrm{NS}^2} + 
   \underbrace{\sum_{k,l} \frac{(v_{k,l}- \tilde{v}_{k,l})^2}{\sigma_{k,l}^2}}_{\chi_\mathrm{AO}^2}.
\label{eq:chi squared combined}
\end{align}
\end{ceqn}

where $i,j$ represent the coordinates in the natural seeing data and $k,l$ represent the coordinates in the adaptive optics data, which do not need to be correlated since each model is fit separately. 

\begin{figure}
  \includegraphics[width=8.5cm]{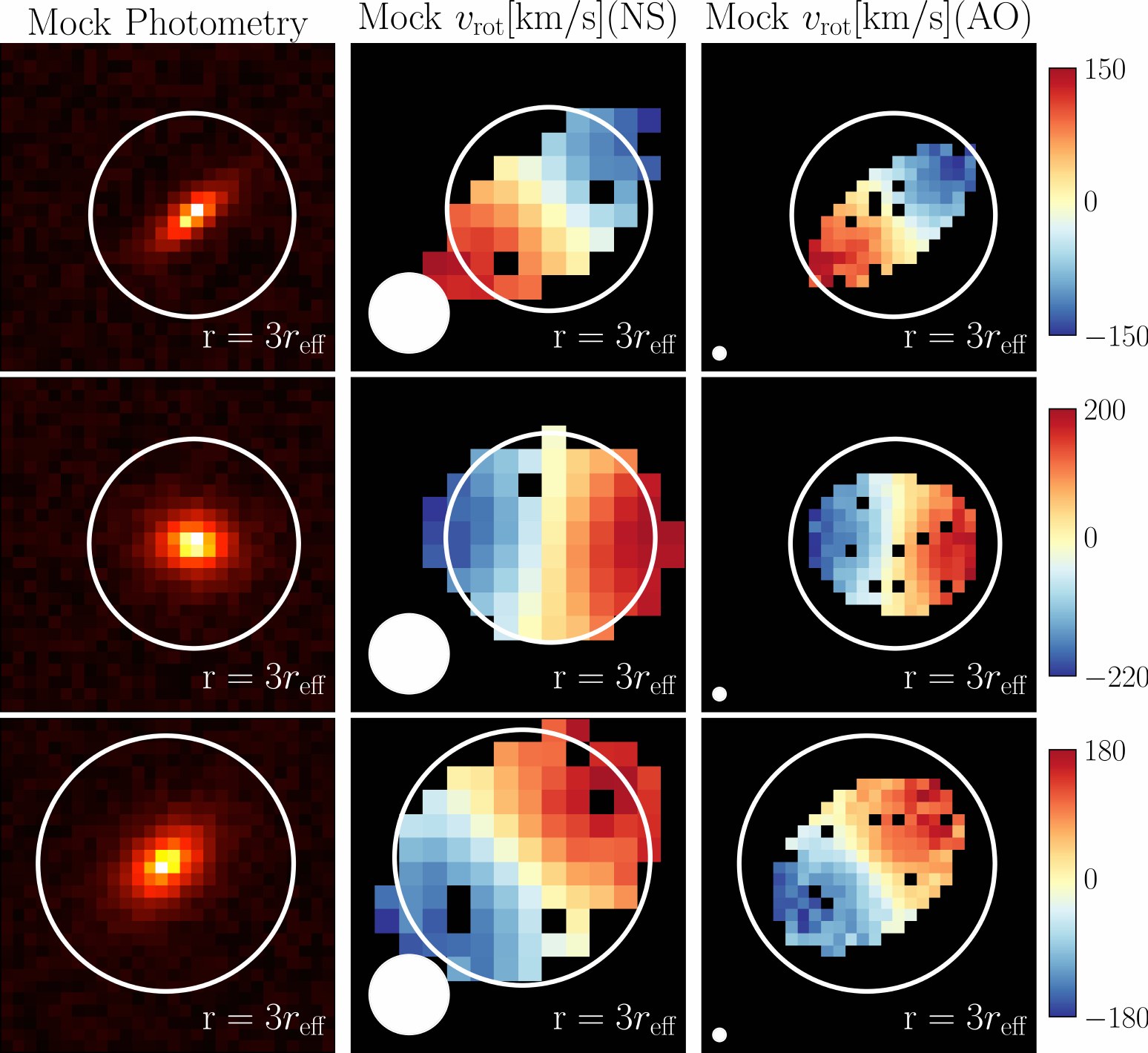}
    \caption{Example of three of the 10$^3$ mock galaxies used to test the kinematic code. Each row shows a simulated mock galaxy in photometry and the kinematics from the natural seeing and adaptive optics resolutions respectively. White empty circles represent the $r=3r_\mathrm{eff}$ boundary to show the extent of the data while the filled circles represent the size of the PSF.}
    \label{fig:Mock galaxies}
\end{figure}

In order to verify the efficacy of the code and quantify the improvement when doing the combined analysis, we simulated a sample of $10^3$ mock galaxies that resemble our real sample. The mock galaxies are built by creating a perfect intensity cube and photometric map with different values of position angle ($0^\circ\leq \theta_\mathrm{PA} \leq 360^\circ$), inclination ($20^\circ \leq i \leq 90^\circ$), galaxy centre $[x_0,y_0]$ at a random distance of $d\leq 3$ pixels from the centre, asymptotic velocity ($80 \leq v_{\textrm{flat}} \leq 320$ km/s), characteristic radius of the rotation curve ($1\leq r_{\textrm{flat}} \leq 2.52$ kpc) and disk scale length ($1.5 \leq r_d \leq 2.52$ kpc). After extracting the velocity field we introduced Gaussian noise at the level of the real data but also instrumental limitations and loss of random pixels due to signal to noise. Figure \ref{fig:Mock galaxies} shows an example of three of these mock galaxies.

For each mock galaxy, we ran the fitting code in the three cases (natural seeing only, adaptive optics only and combined resolutions). In each case, we obtain a best fit value of specific angular momentum $j_\mathrm{*,fit}$ which can be compared to the input value of the model $j_\mathrm{*,real}$.

\begin{figure}
    \includegraphics[width=8.5cm]{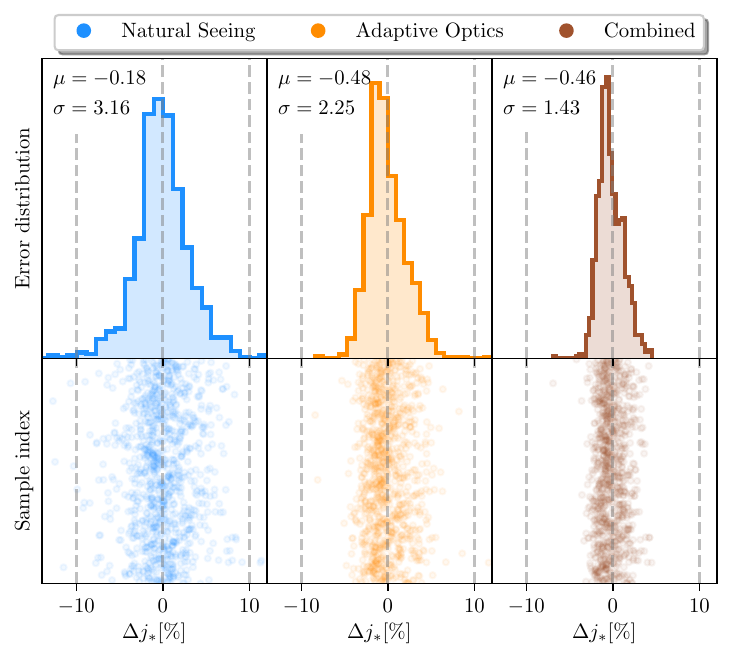}
    \caption{Results from the test of the kinematic code when applied to the 10$^3$ mock galaxies. \textbf{Bottom panels:} Fractional error of the obtained value of $j_*$ with respect to the real input value $j_\mathrm{real}$ for each of the mock galaxies. \textbf{Top panels:} Histograms representing the spread in errors where $\mu$ is the systematic fractional error and $\sigma$ the spread in the random errors.}
    \label{fig:simulations}
\end{figure}

The first notable result is that the spread in the percentage error $\Delta j_* = 100\times(j_\mathrm{*,fit} - j_\mathrm{*,real})/j_\mathrm{*,real}$ is largest in the natural seeing only case with $\sigma=3.2\%$, followed by the adaptive optics only case with $\sigma=2.3\%$. In the case of the combined analysis, the spread in the percentage error is the lowest with $\sigma=1.4\%$, as shown by the histograms in Figure \ref{fig:simulations}. Additionally, the mean systematic percentage error $\mu$ in each case is low (-0.2\%, -0.5\% and -0.5\% respectively) which illustrates the robustness of the code.

\begin{figure*}
    \centering
    \includegraphics[width=0.98\textwidth]{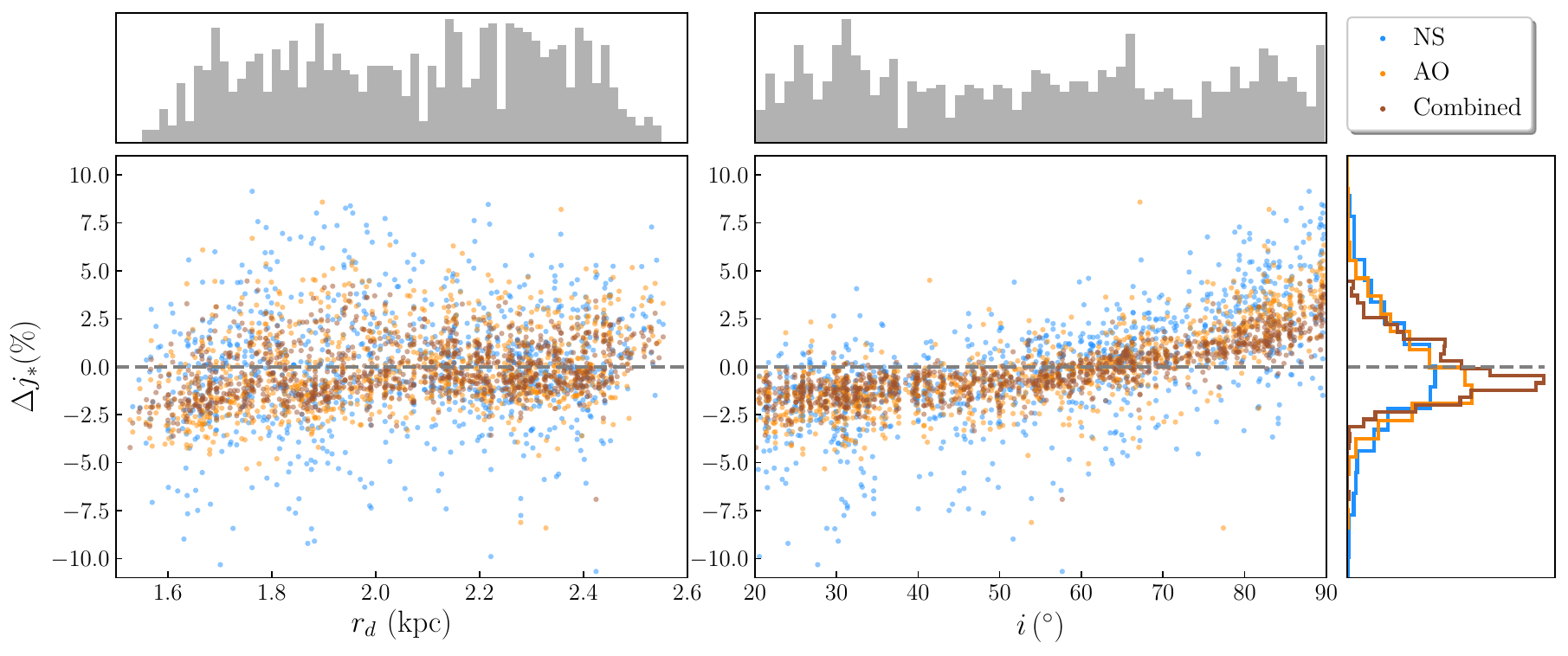}
    \caption{Distribution of errors (\%) from the simulations against the input disk scale length $r_d$ and inclination $i$. Top panels show the near-uniform distribution of simulated values of $r_d$ and $i$ which were chosen to be in the range $1.5\leq r_d \leq 2.52$ kpc and $20^\circ \leq i \leq 90^\circ$ respectively. Right panel shows the histograms of the individual and combined analysis showing only a small negative systematic offset ($\sim 1\%$) towards lower values of $j_*$.}
    \label{fig:sims_residuals}
\end{figure*}

Additionally, we test the accuracy of the angular momentum measurement as a function of the input size and inclination of the mock galaxies by comparing the percentage errors $\Delta j_* = 100\times(j_\mathrm{*,fit} - j_\mathrm{*,real})/j_\mathrm{*,real}$ with the input values of $r_d$ and $i$ as shown in Figure \ref{fig:sims_residuals}. We find no significant dependence on disk scale length as the mean trend is within 2.5\% and the RMS is of a similar order for all three cases. As expected, the uncertainty decreases slightly towards larger radii since larger disks have a higher number of useful pixels in the fit, thus the error is smaller. The effect of disk inclination is noticeable for high and low values of $i$ corresponding to near edge-on and near face-on projections of the disk. In the former, the small number of useful pixels in the deprojection affects the accuracy of the fit and in the latter, the line of sight velocities are low and closer to the noise values. These effects are particularly noticeable in the NS case given the large pixel scales where $\Delta j_*$ can be up to $\sim10\%$ in some cases. However, given the conservative range that we used for the inclinations ($i>20^\circ$) and the lack of additional features such as bulges or non-symmetrical clumps in the mock photometry maps, the inclinations found by the code are in very good agreement with the input values so these effects are small and merit special attention only when individual galaxies have very low inclinations at low resolution.

\clearpage

%%%%%%%%%% Merge with supplemental materials %%%%%%%%%%
\pagebreak
%\widetext
\onecolumn
\begin{center}
\textbf{\large Supplemental Materials:}
\end{center}
%%%%%%%%%% Merge with supplemental materials %%%%%%%%%%
%%%%%%%%%% Prefix a "S" to all equations, figures, tables and reset the counter
\setcounter{equation}{0}
\setcounter{figure}{0}
\setcounter{table}{0}
\setcounter{page}{1}
\makeatletter
\renewcommand{\theequation}{S\arabic{equation}}
\renewcommand{\thefigure}{S\arabic{figure}}
\renewcommand{\bibnumfmt}[1]{[S#1]}
\renewcommand{\citenumfont}[1]{S#1}
%%%%%%%%%% Prefix a "S" to all equations, figures, tables and reset the counter

\begin{center}
\section{Case by case discussion}
\label{section:appendix_case_by_case}
\end{center}

\begin{figure*}

  \begin{subfigure}{15.8cm}
    \centering\includegraphics[width=15.8cm]{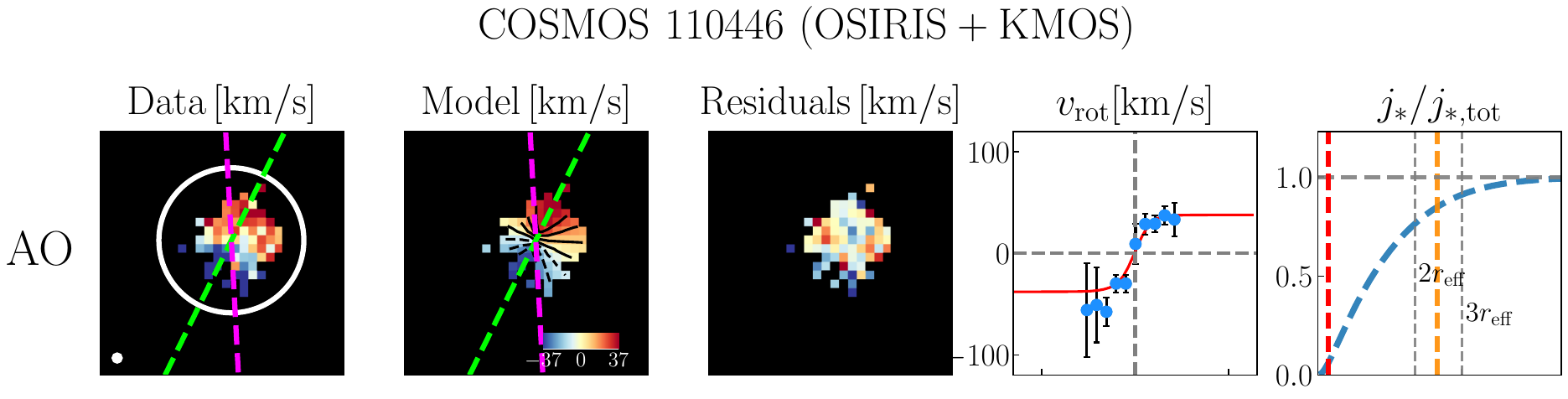}
  \end{subfigure}\hfill
  \begin{subfigure}{15.8cm}
    \centering\includegraphics[width=15.8cm]{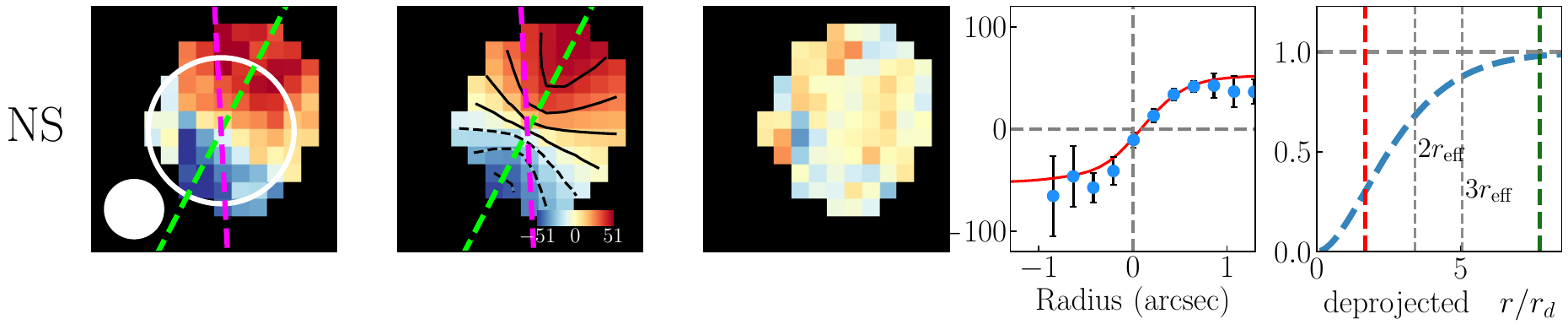}
  \end{subfigure}\hfill
  \begin{subfigure}{15.8cm}
    \centering\includegraphics[width=15.8cm]{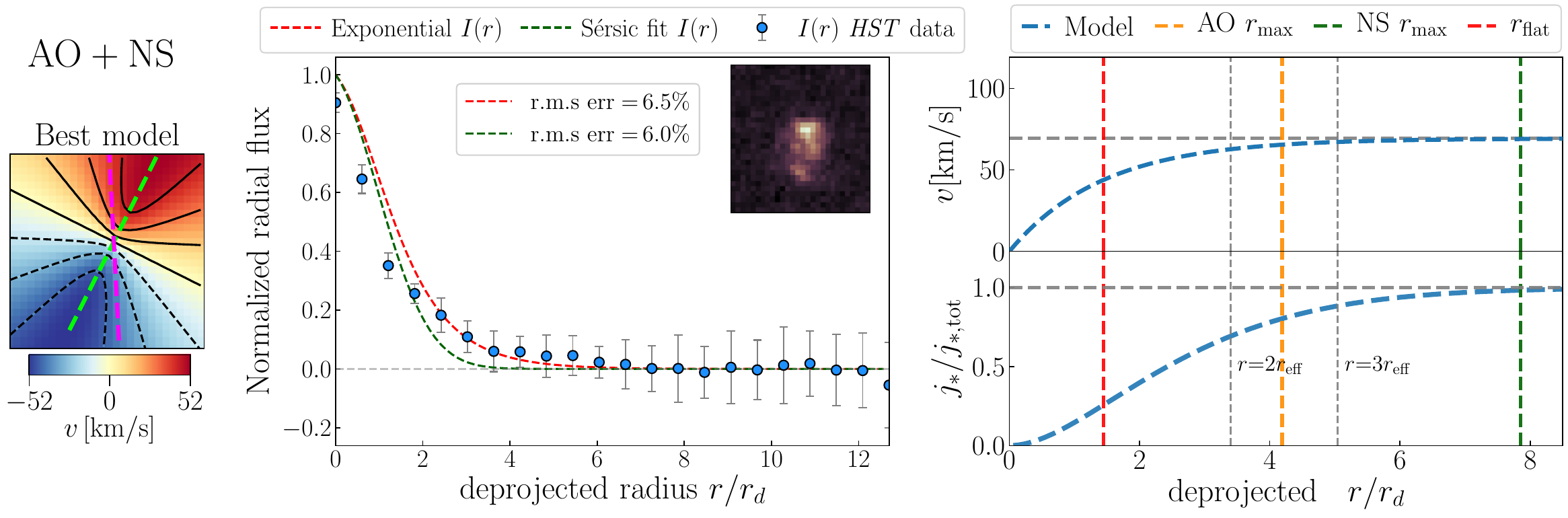}
  \end{subfigure}
  
  \caption{Main results of galaxy COSMOS 110446 at the individual and combined resolutions. \textbf{Top row:} Analysis of the adaptive optics assisted data. From left to right: 1) Velocity field from the data where the solid white circle at the centre denotes the $r=3r_\mathrm{eff}$ boundary to illustrate the spatial scale and white filled circles at the bottom left indicate the PSF size. Green and purple dashed lines represent the kinematic and morphological main axes respectively. 2) Model velocity field from the best fit with the corresponding contours. 3) Residuals (sharing the same color bar as the data and model). 4) Position-velocity diagram along the kinematic axis (blue points) with the corresponding best fit model in red. 5) Cumulative stellar specific angular momentum profile where grey dashed lines represent the radius at 2 and 3 times $r_\mathrm{eff}$ and orange line indicates the maximum radial extent of the data. \textbf{Middle row:} Similar to the top row but for the natural seeing data. Green dashed line in the last column indicates the maximum radial extent of the natural seeing data. \textbf{Bottom row:} Results from the combined analysis with 1) the best kinematic model, 2) the one-dimensional surface brightness profiles from the exponential and S\'ersic fits with the average radial value represented by the blue dots and 3) shape of the rotation curve (top) as well as cumulative angular momentum profile (bottom) with the corresponding relevant radial values given by the dashed lines.}
  \label{fig:COSMOS 110446}
\end{figure*}

\begin{figure*}
  \begin{subfigure}{14.1cm}
    \centering\includegraphics[width=14.1cm]{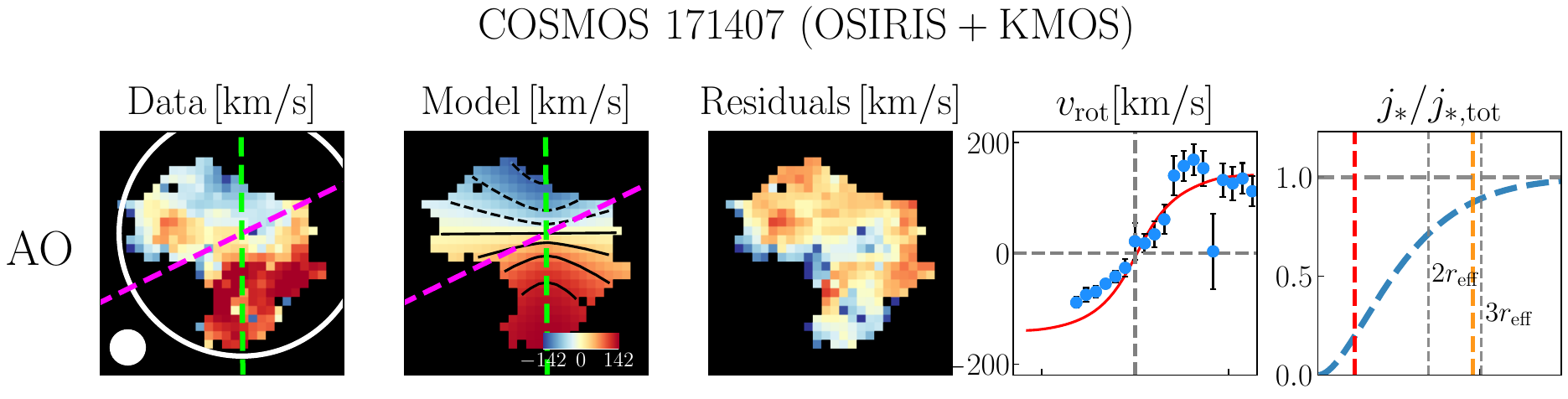}
  \end{subfigure}\hfill
  \begin{subfigure}{14.1cm}
    \centering\includegraphics[width=14.1cm]{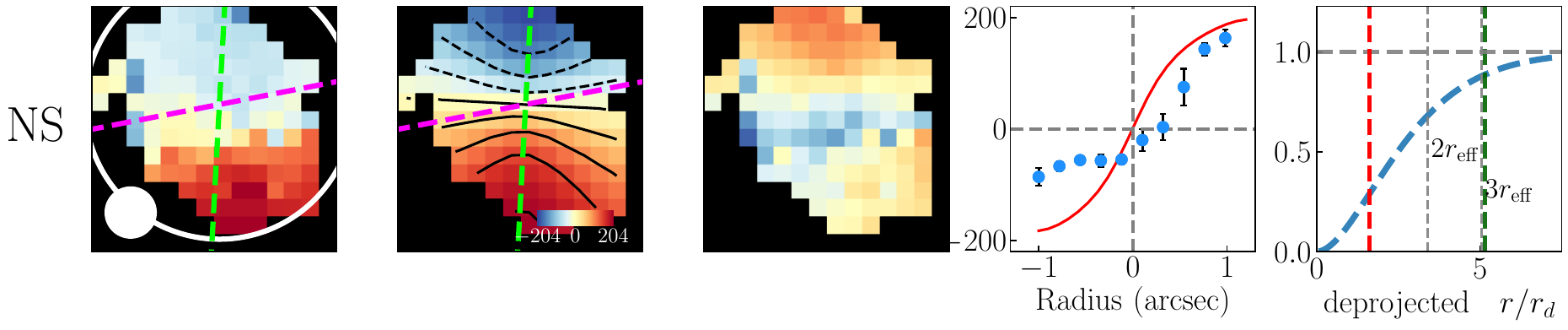}
  \end{subfigure}\hfill
  \begin{subfigure}{14.3cm}
    \centering\includegraphics[width=14.3cm]{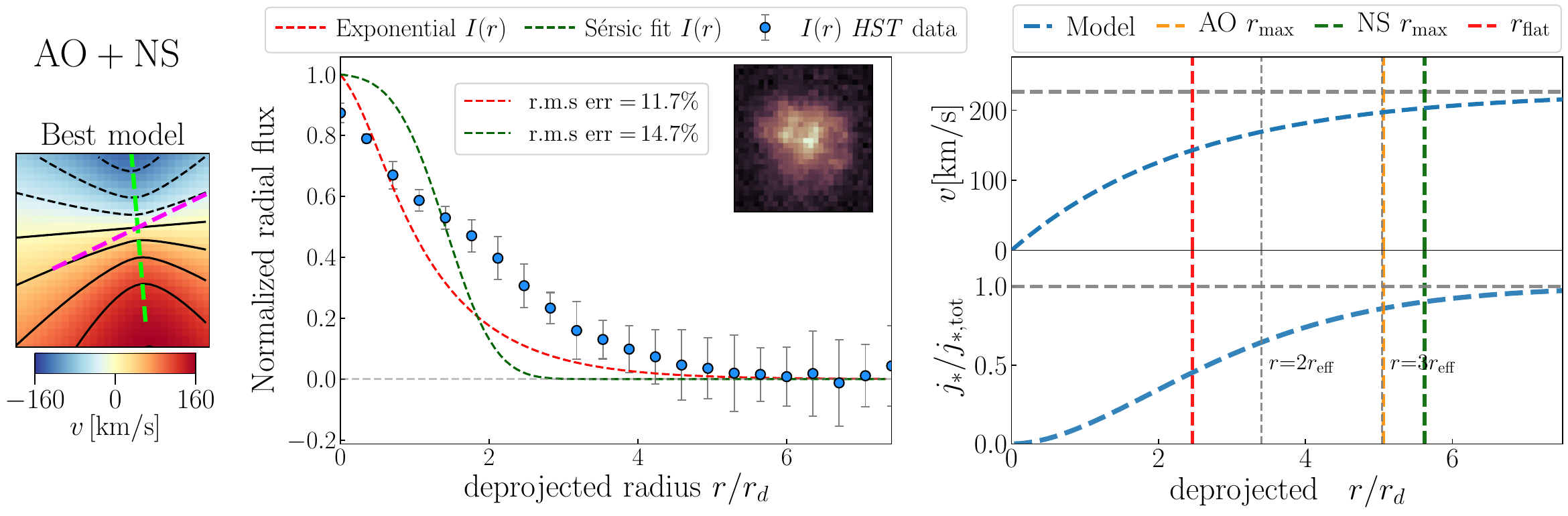}
  \end{subfigure}
  
  \par\bigskip
  \begin{subfigure}{14.1cm}
    \centering\includegraphics[width=14.1cm]{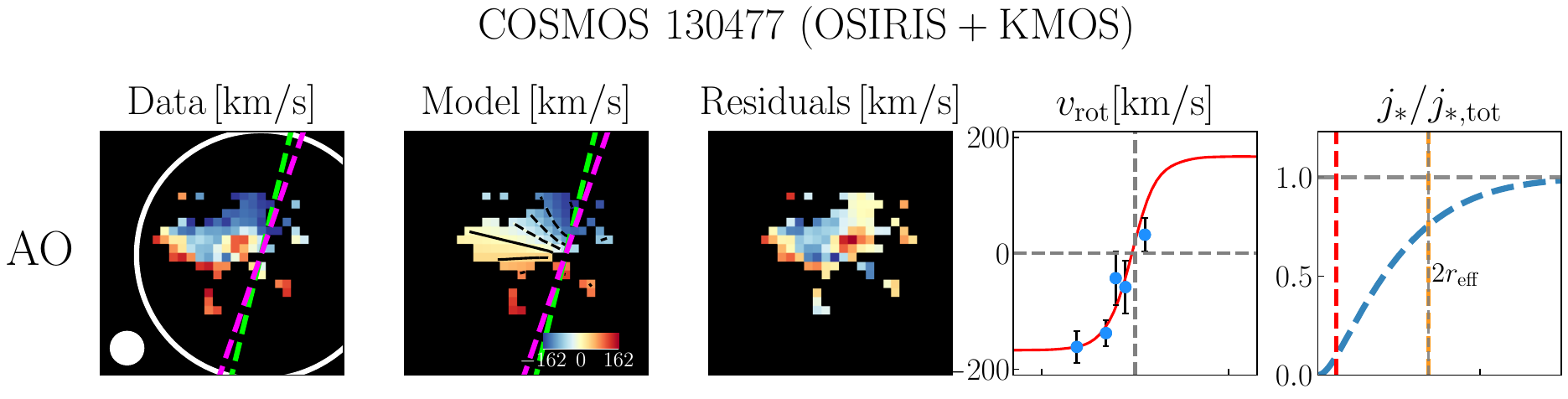}
  \end{subfigure}\hfill
  \begin{subfigure}{14.1cm}
    \centering\includegraphics[width=14.1cm]{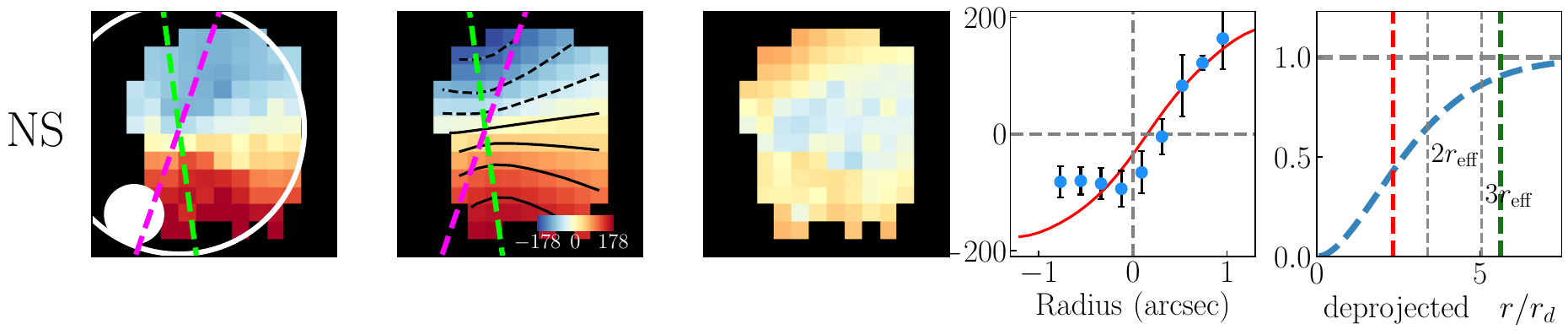}
  \end{subfigure}\hfill
  \begin{subfigure}{14.3cm}
    \centering\includegraphics[width=14.3cm]{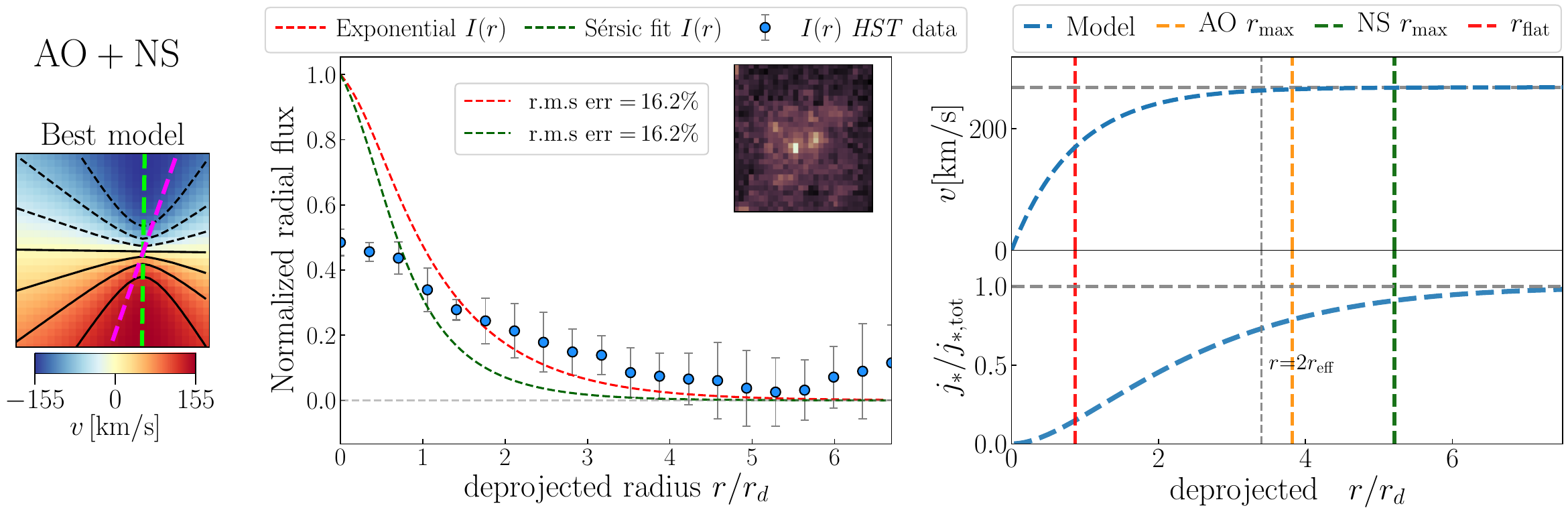}
  \end{subfigure}
  
  \caption{Analysis of COSMOS 171407 (top) and COSMOS 130477 (bottom). Similar description to that of Figure \ref{fig:COSMOS 110446}.}
  \label{fig:COSMOS_171407 and COSMOS_130477}
\end{figure*}

\begin{figure*}
  \begin{subfigure}{14.1cm}
    \centering\includegraphics[width=14.1cm]{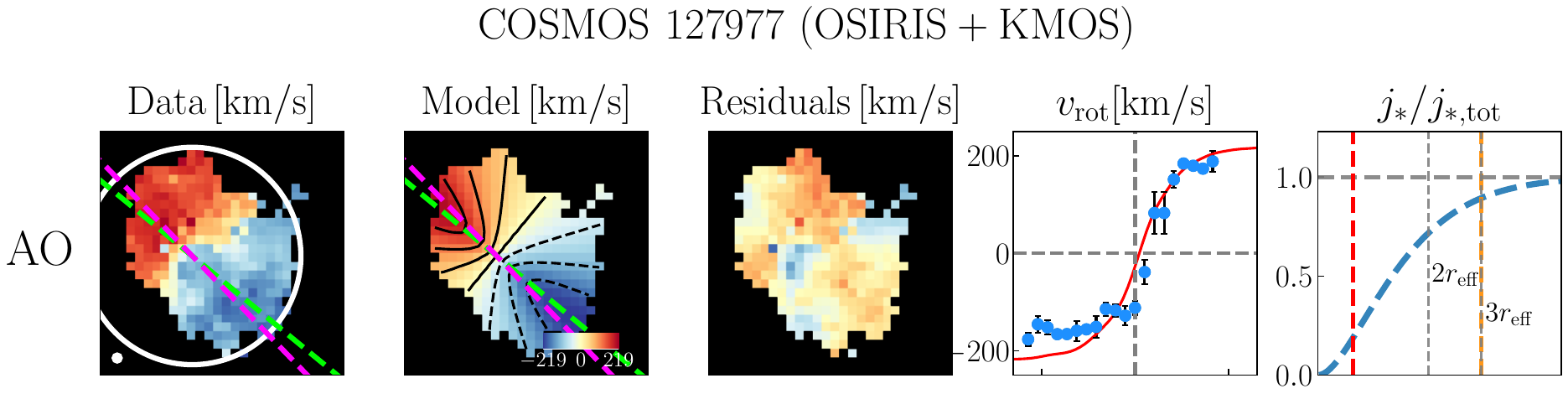}
  \end{subfigure}\hfill
  \begin{subfigure}{14.1cm}
    \centering\includegraphics[width=14.1cm]{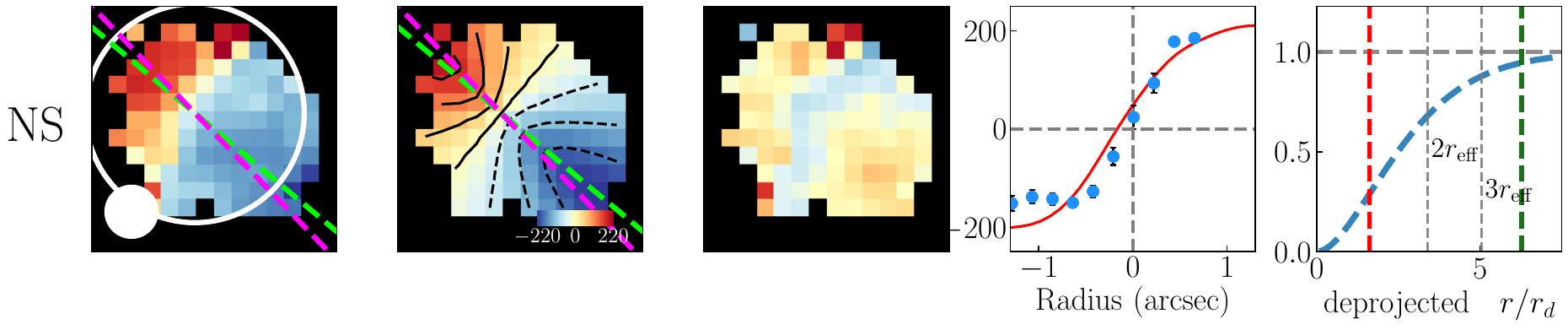}
  \end{subfigure}\hfill
  \begin{subfigure}{14.3cm}
    \centering\includegraphics[width=14.3cm]{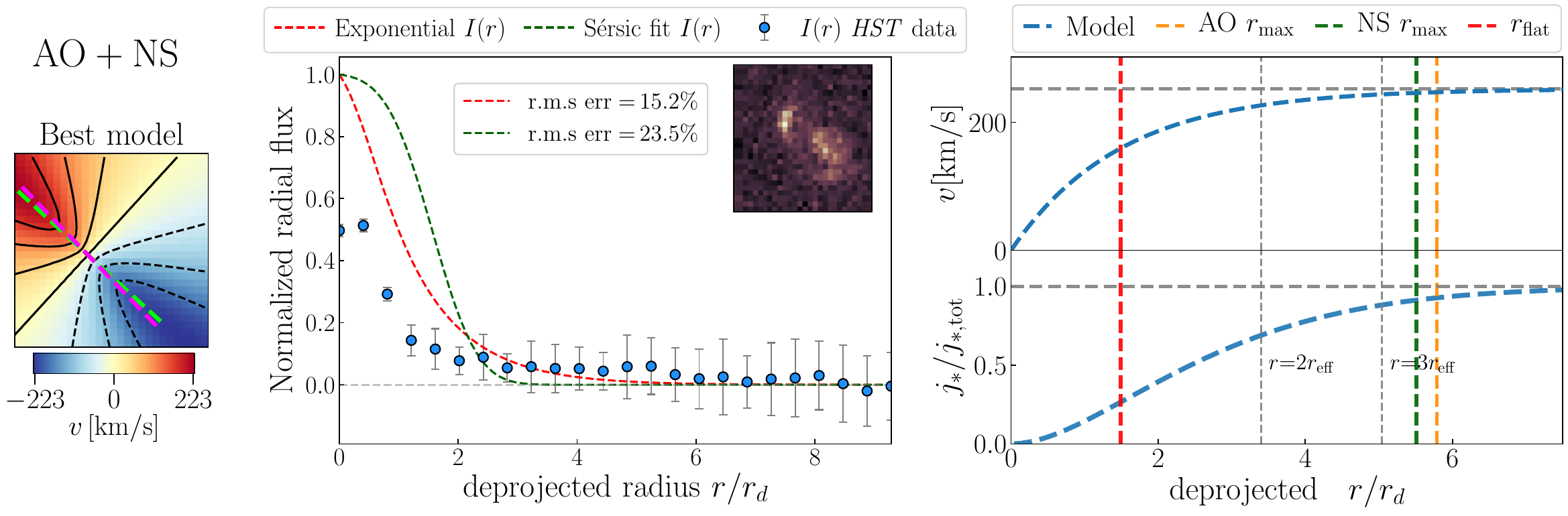}
  \end{subfigure}
  
  \par\bigskip
  \begin{subfigure}{14.1cm}
    \centering\includegraphics[width=14.1cm]{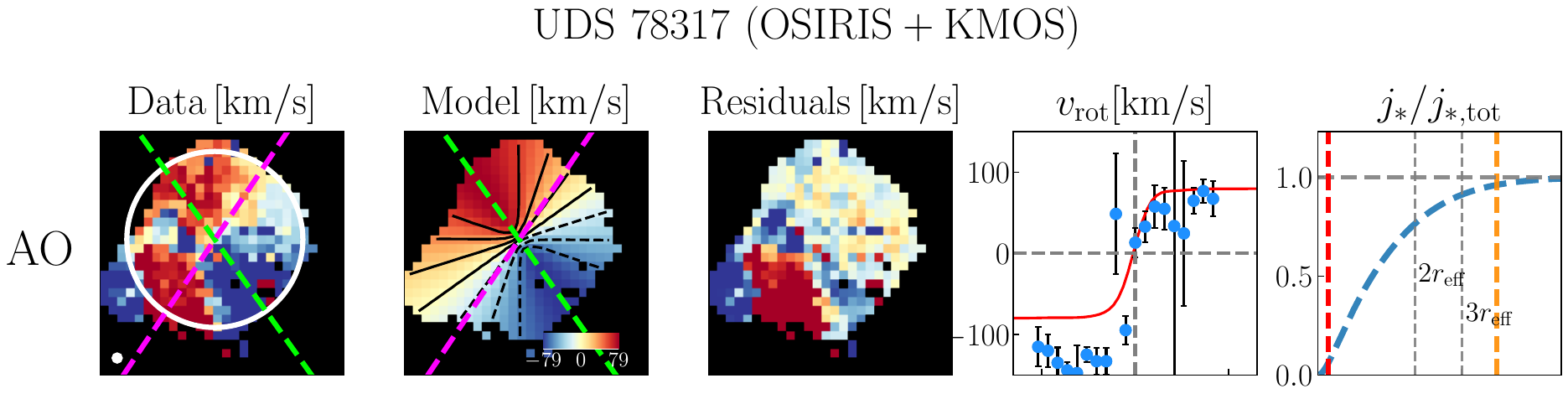}
  \end{subfigure}\hfill
  \begin{subfigure}{14.1cm}
    \centering\includegraphics[width=14.1cm]{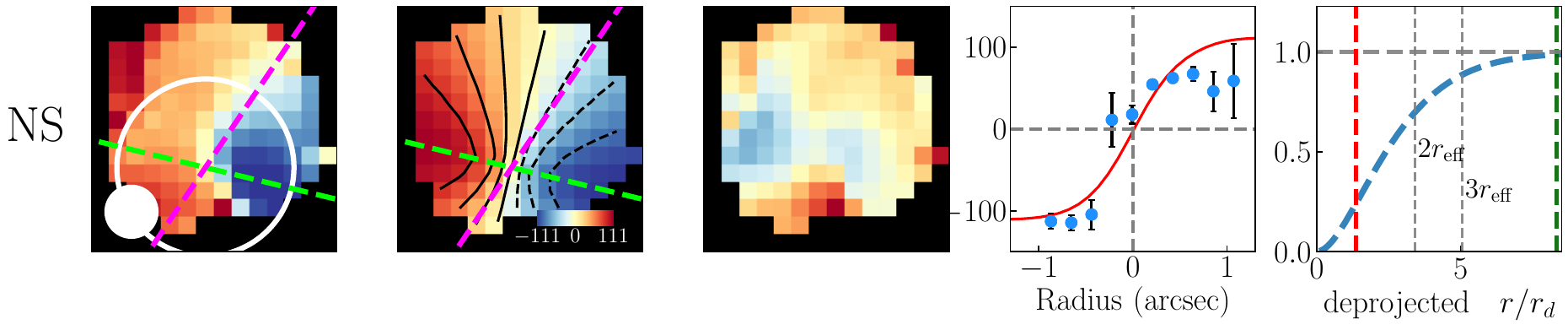}
  \end{subfigure}\hfill
  \begin{subfigure}{14.3cm}
    \centering\includegraphics[width=14.3cm]{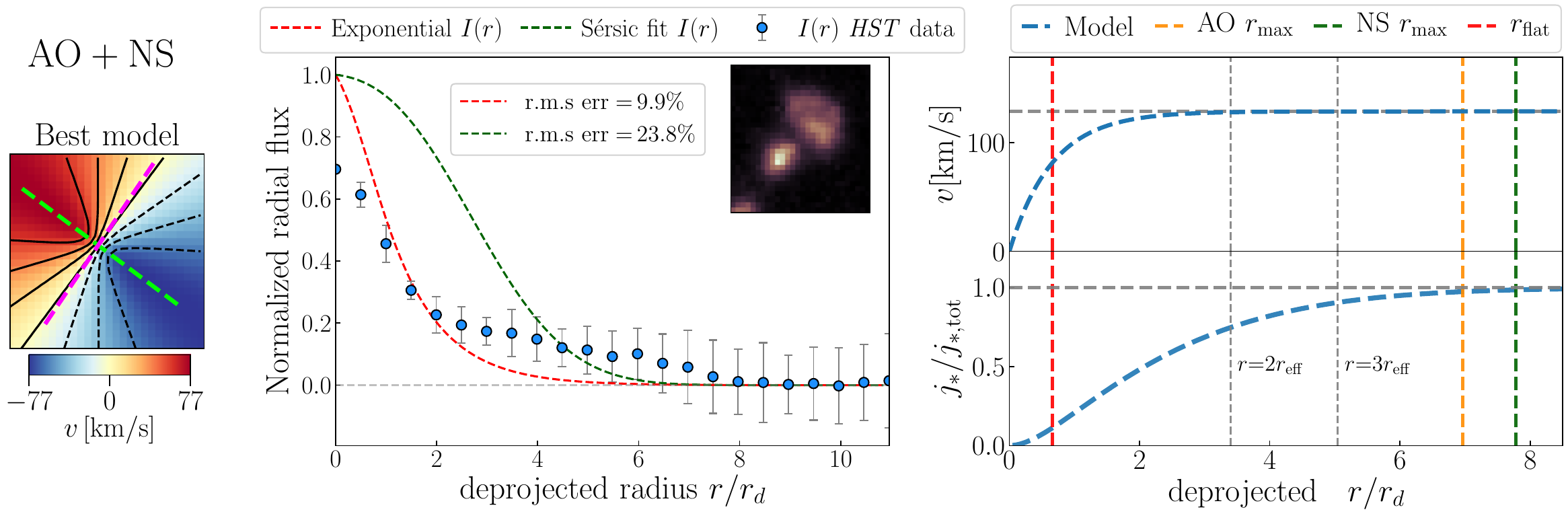}
  \end{subfigure}
  
  \caption{Analysis of COSMOS 127977 (top) and \textit{UDS 78317} (bottom). Similar description to that of Figure \ref{fig:COSMOS 110446}.}
  \label{fig:COSMOS_127977 and UDS_78317}
\end{figure*}

\begin{figure*}
  \begin{subfigure}{14.1cm}
    \centering\includegraphics[width=14.1cm]{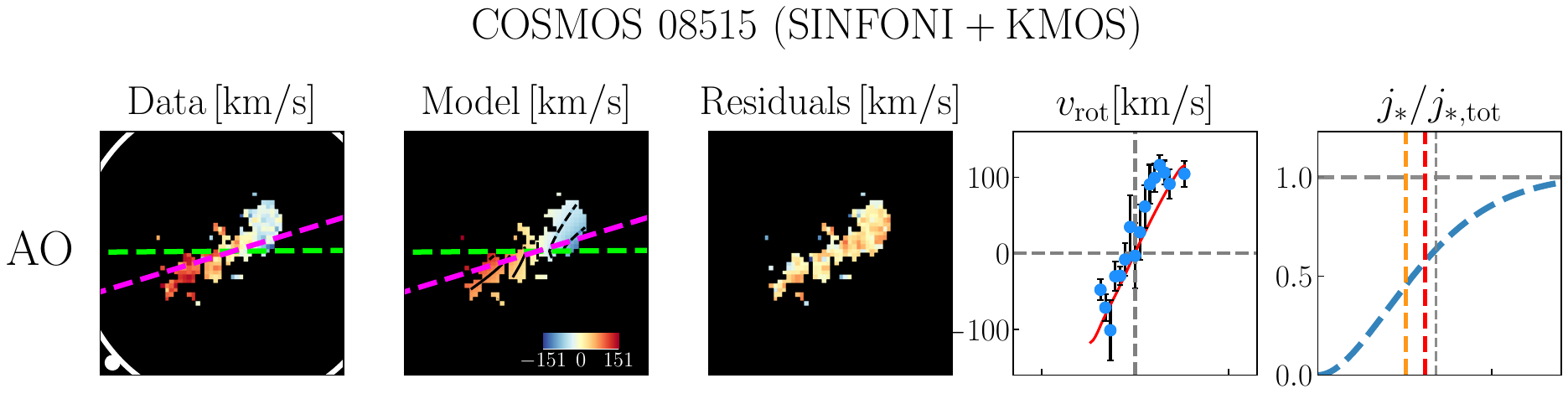}
  \end{subfigure}\hfill
  \begin{subfigure}{14.1cm}
    \centering\includegraphics[width=14.1cm]{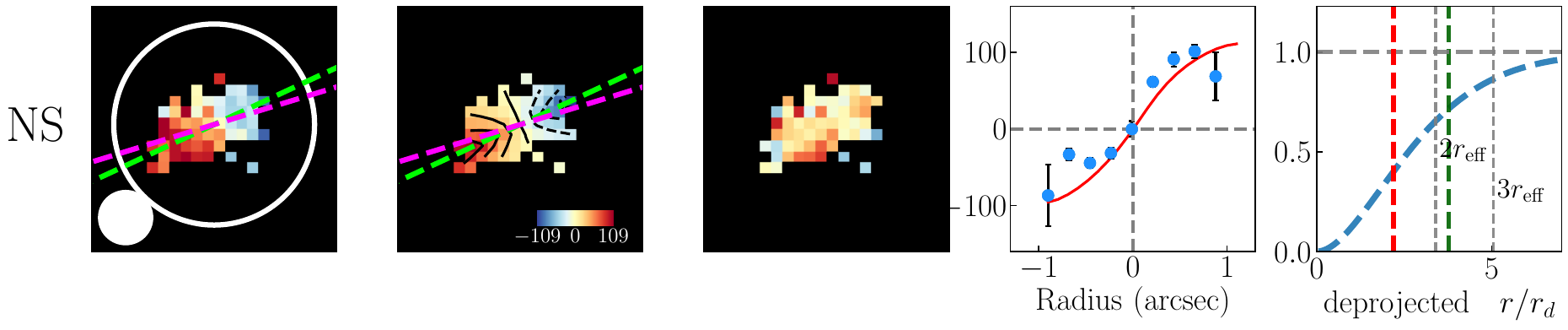}
  \end{subfigure}\hfill
  \begin{subfigure}{14.3cm}
    \centering\includegraphics[width=14.3cm]{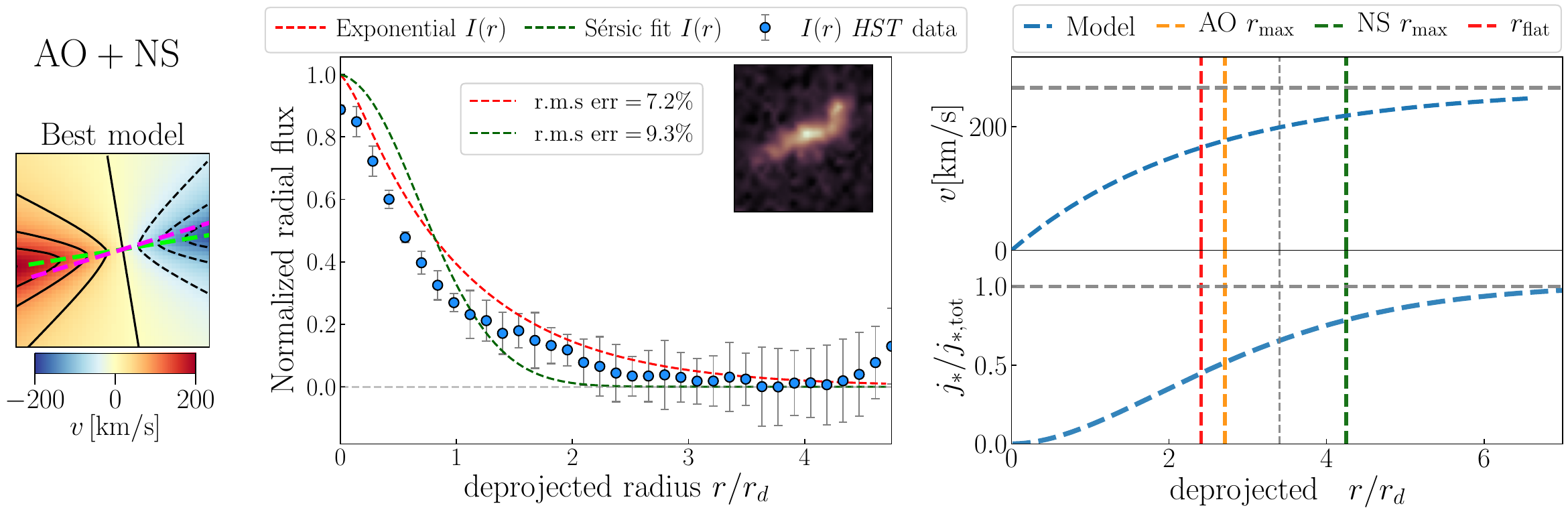}
  \end{subfigure}
  
  \par\bigskip
  \begin{subfigure}{14.1cm}
    \centering\includegraphics[width=14.1cm]{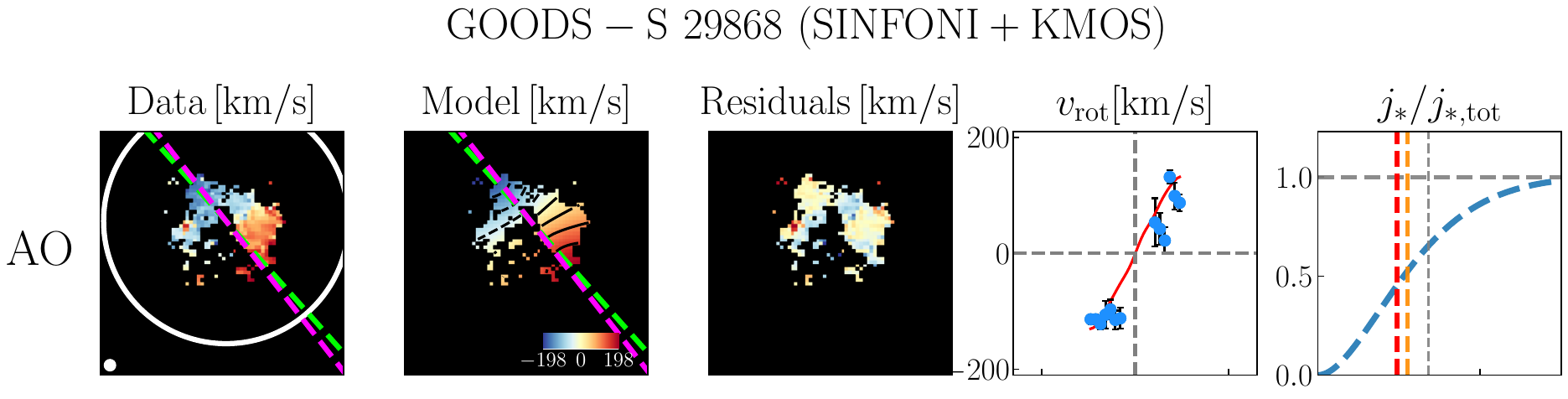}
  \end{subfigure}\hfill
  \begin{subfigure}{14.1cm}
    \centering\includegraphics[width=14.1cm]{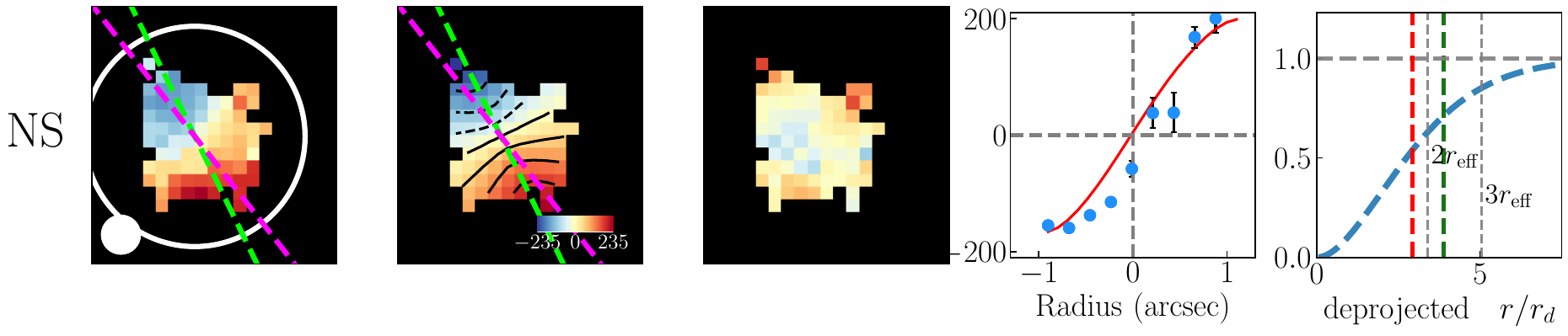}
  \end{subfigure}\hfill
  \begin{subfigure}{14.3cm}
    \centering\includegraphics[width=14.3cm]{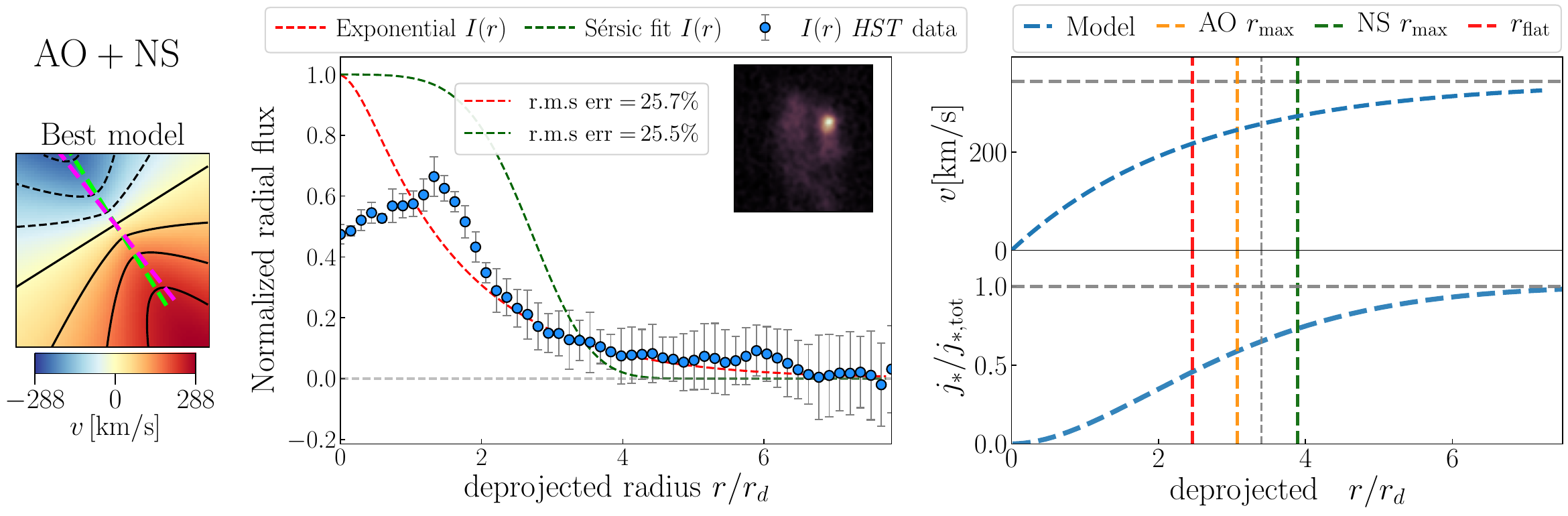}
  \end{subfigure}
  
  \caption{Analysis of COS4 08515 (top) and GOODS-S 29868 (bottom). Similar description to that of Figure \ref{fig:COSMOS 110446}.}
  \label{fig:COS4_08515 and GS4_29868}
\end{figure*}

\begin{figure*}
  \begin{subfigure}{14.1cm}
    \centering\includegraphics[width=14.1cm]{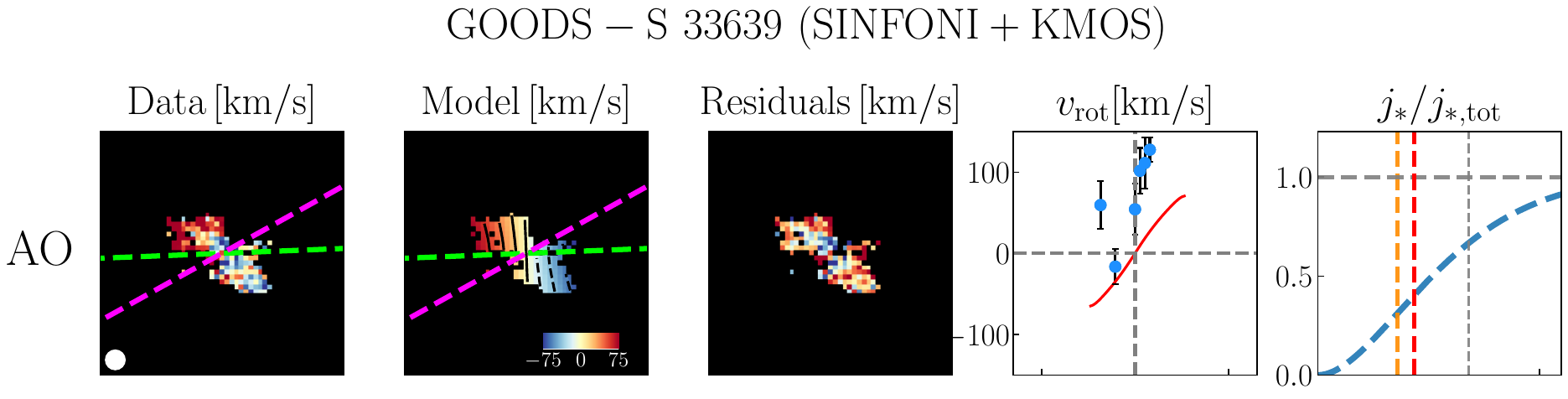}
  \end{subfigure}\hfill
  \begin{subfigure}{14.1cm}
    \centering\includegraphics[width=14.1cm]{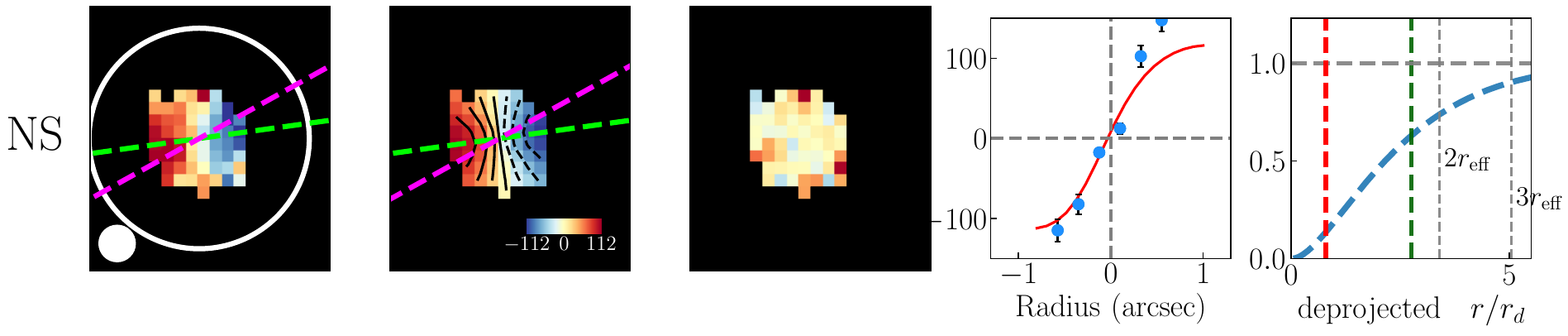}
  \end{subfigure}\hfill
  \begin{subfigure}{14.3cm}
    \centering\includegraphics[width=14.3cm]{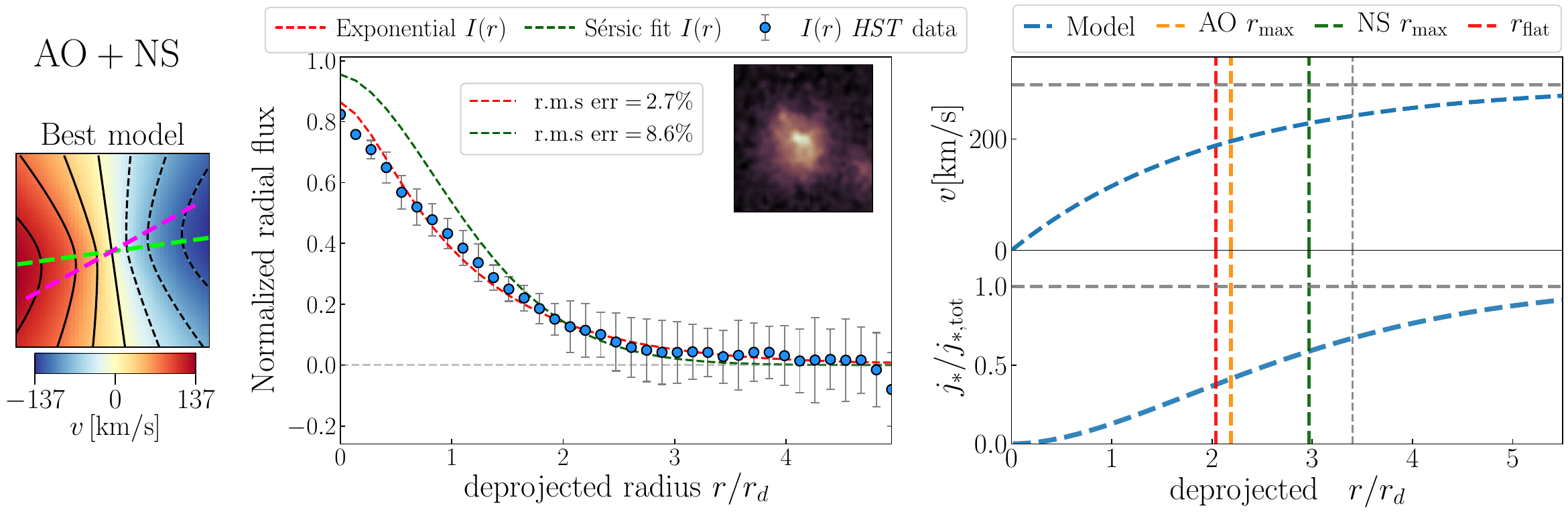}
  \end{subfigure}
  
  \par\bigskip
  \begin{subfigure}{14.1cm}
    \centering\includegraphics[width=14.1cm]{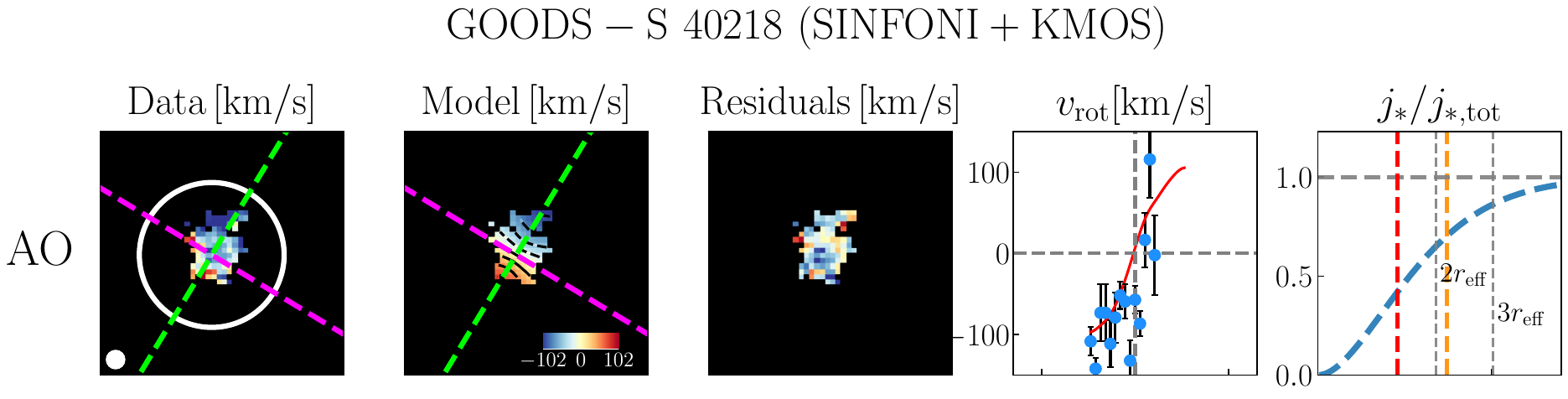}
  \end{subfigure}\hfill
  \begin{subfigure}{14.1cm}
    \centering\includegraphics[width=14.1cm]{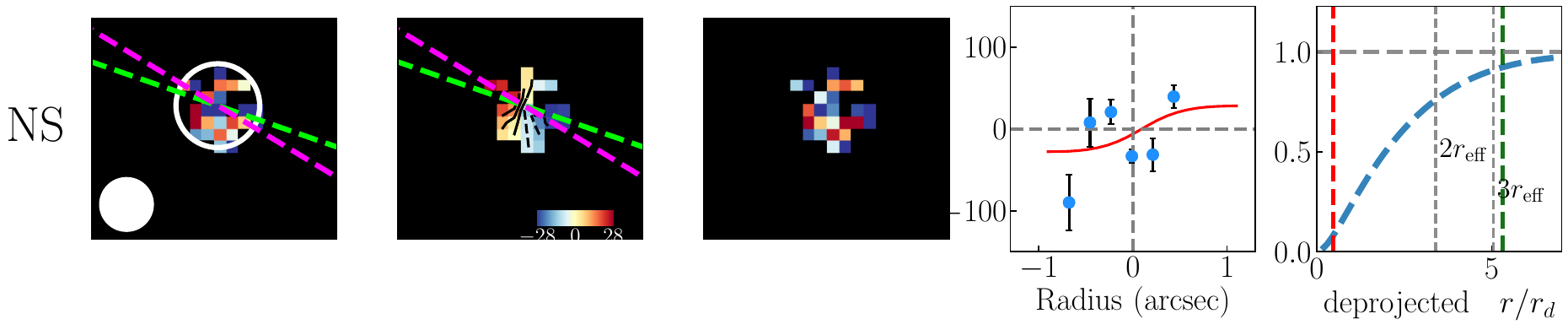}
  \end{subfigure}\hfill
  \begin{subfigure}{14.3cm}
    \centering\includegraphics[width=14.3cm]{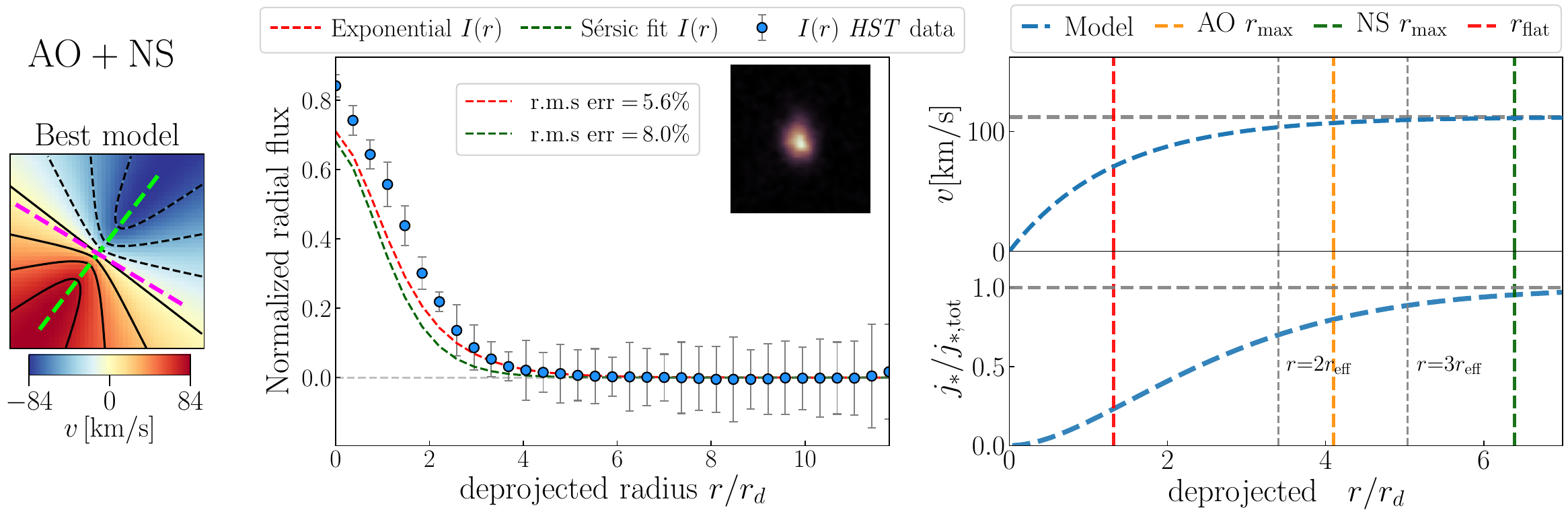}
  \end{subfigure}
  
  \caption{Analysis of GOODS-S 33639 (top) and \textit{GOODS-S 40218} (bottom). Similar description to that of Figure \ref{fig:COSMOS 110446}.}
  \label{fig:GS4_33639 and GS4_40218}
\end{figure*}

\begin{figure*}
  \begin{subfigure}{14.5cm}
    \centering\includegraphics[width=14.5cm]{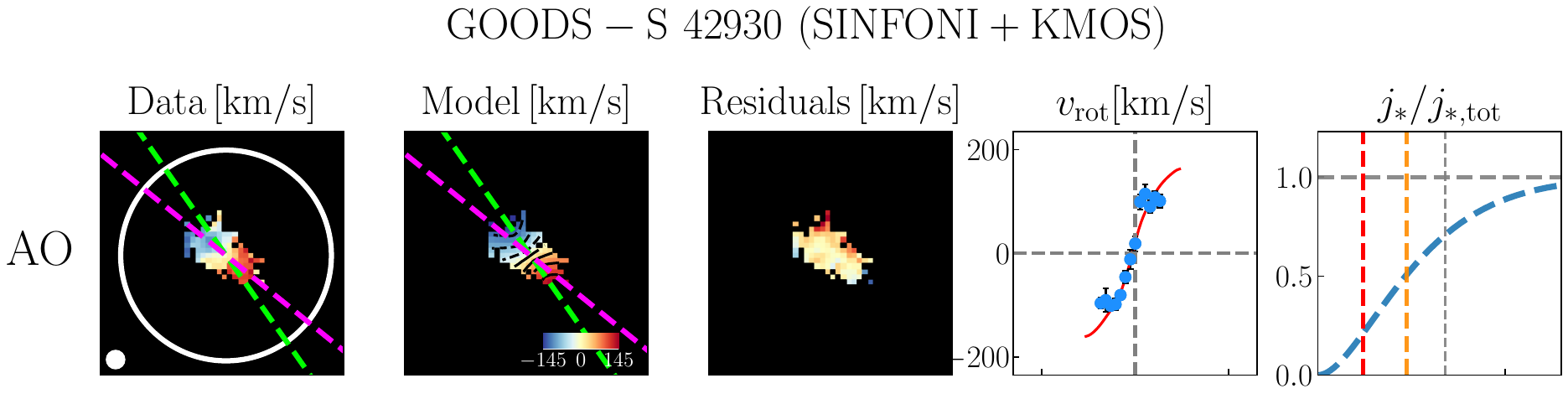}
  \end{subfigure}\hfill
  \begin{subfigure}{14.5cm}
    \centering\includegraphics[width=14.5cm]{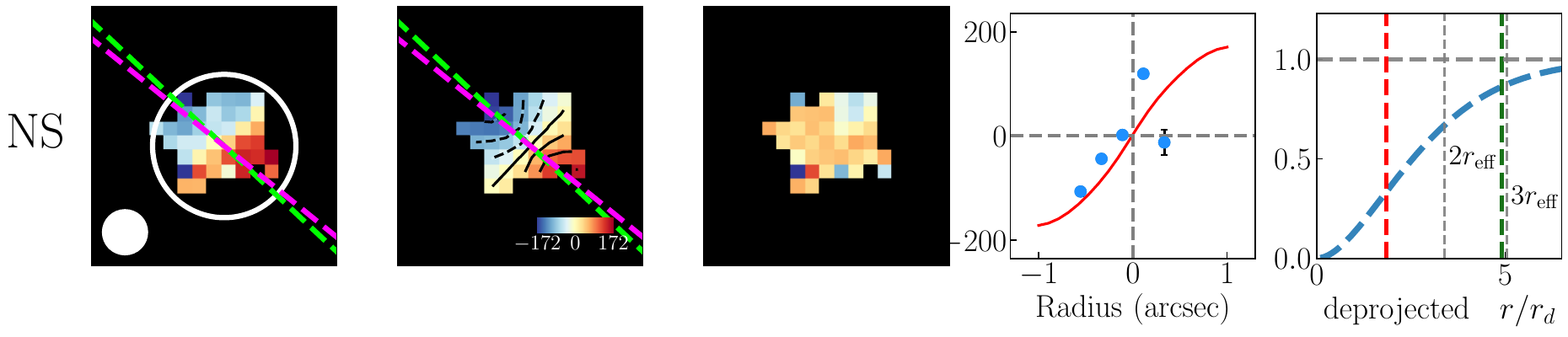}
  \end{subfigure}\hfill
  \begin{subfigure}{14.5cm}
    \centering\includegraphics[width=14.5cm]{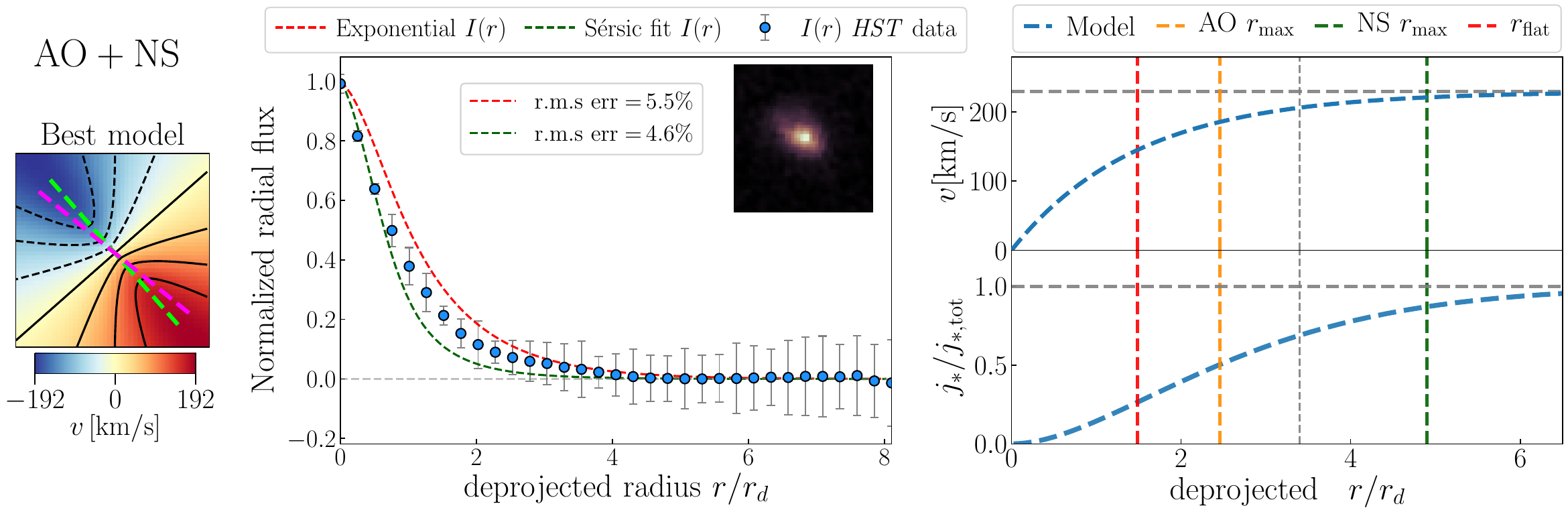}
  \end{subfigure}
  
  \caption{Analysis of GOODS-S 42930. Similar description to that of Figure \ref{fig:COSMOS 110446}.}
  \label{fig:GS4_42930}
\end{figure*}

In this supplementary section we discuss the results and individual characteristics of each galaxy. The supporting figures for the whole sample are shown bellow with the main results of each galaxy, including the velocity fields and angular momentum measurements at the single and combined resolutions (Figures \ref{fig:COSMOS 110446} to \ref{fig:GS4_42930}).

\begin{comment}
COSMOS110446      NS=1.8       AO=2.2
COSMOS171407      NS=2.3       AO=2.8
COSMOS130477      NS=3.0       AO=4.8
COSMOS127977      NS=2.4       AO=3.1
UDS 78317      NS=1.3       AO=1.2
COS4 08515      NS=3.5       AO=3.0
GS4 29868      NS=3.9       AO=3.4
GS4 33639      NS=3.8       AO=3.2
GS4 40218      NS=1.0       AO=0.9
GS4 42930      NS=3.3       AO=3.4
\end{comment}

\begin{itemize}

\item \textbf{COSMOS 110446 (disk):} The data shows a smooth monotonic velocity gradient with consistent position angles in the two resolutions and a $v/\sigma=1.8$ in the case of NS and $v/\sigma=2.2$ in the case of the AO data. Despite a $\sim30^\circ$ misalignment between the kinematic and photometric position angles, we classify this galaxy as a rotating disk as it fulfils the rest of the requirements listed in Section \ref{subsection:Kinematic state of the galaxies}.

\item \textbf{COSMOS 171407 (minor merger treated as a disk):} After smoothing the original adaptive optics datacube with a median filter, the gain in resolution decreases to about $\times$2. The adaptive optics data shows the presence of a large redshifted clump on the blueshifted side of the disk and a more complex H$\alpha$ intensity distribution. These features are also visible in the \textit{HST} data which differs significantly from the exponential disk approach. We classify this galaxy as a minor merger but treat it as a disk given its ordered rotation (with $v/\sigma=2.8$ in the AO data) and matching kinematic and morphological centres.

\item \textbf{COSMOS 130477 (disk):} This galaxy fulfils all the conditions to be classified as a disk, with kinematics clearly dominated by rotation with $v/\sigma=3$ and $v/\sigma=4.8$ in NS and AO respectively. Despite the spatial smoothing on the original high-resolution datacube, there are very few pixels with enough SNR to retrieve a value in the velocity field and thus the kinematic centre is less reliable. For that reason, the natural seeing map has a higher weight in the fit. However, the adaptive optics data contribution is still evident in the rapid rise of the rotation curve. We emphasize that for an exponential rotation velocity model of disk galaxies, the rotation curve rises rapidly in the innermost regions of the galaxies where the adaptive optics data is a better tracer. This galaxy is a good example of that behavior.

\item \textbf{COSMOS 127977 (disk):} The data quality is good in both resolutions and in this particular case, the extent of the high-resolution data is similar to the seeing limited one. It fulfils all conditions to be classified as a disk with $v/\sigma=2.4$ (NS) and $v/\sigma=3.1$ (AO). The model of the NS data shows a slowly rising rotation curve as compared to the high-resolution data which is consistent with the expected effects of beam smearing.

\item \textbf{\textit{UDS 78317} (merger):} This galaxy is a very clear example of how low resolution can affect the correct classification of galaxies at high redshift. The seeing limited maps suggest a smooth rotating disk with a clear rotation axis despite a low $v/\sigma$ ratio of 1.3. On the other hand, the adaptive optics data clearly shows a complex kinematic state in both the kinematic and H$\alpha$ intensity maps which resemble that of a merging system or a highly disrupted disk with $v/\sigma=1.2$. In the analysis of this galaxy at both resolutions conducted by \cite{Sweet}, they also concluded that this galaxy is a merger.

\item \textbf{COSMOS 08515 (disk):} The adaptive optics gain in resolution is $\times$6 but the radial extent of adaptive optics data is much lower than the seeing limited one so the combination is ideal for a good determination of the velocity field. The two datasets clearly show an ordered rotation with $v/\sigma=3.5$ (NS) and $v/\sigma=3.0$ (AO), and fulfil all criteria to be classified as a disk.

\item \textbf{GOODS-S 29868 (disk):} This is an extended galaxy with a ring-like shape from the \textit{HST} and adaptive optics H$\alpha$ intensity maps (not visible in the natural seeing maps) that shows an ordered rotation, with only one small additional kinematic component arising from the high-resolution data. The complexity in the \textit{HST} map makes the assumption of an exponential disk problematic since the kinematic and photometric centre disagree significantly. This galaxy clearly shows the effects of beam smearing in the low-resolution maps since they resemble a smooth disk with ordered rotation with considerable flux in the centre. However, in the high-resolution maps, the real shape of the galaxy (the ring-like structure as seen in the \textit{HST} map) becomes more clear and the low flux at the centre becomes evident. This system is clearly dominated by rotation with $v/\sigma=3.9$ (NS) and $v/\sigma=3.4$ (AO) and despite the mismatch between the photometric and kinematic centres, we classify it as a disk.

\item \textbf{GOODS-S 33639 (disk):} The high-resolution data contains many pixels where the direction of the velocity differs from that of the best velocity field, however the error bar from the Gaussian fit in some of these points is large so they are likely to be dominated by noise rather than signatures from the H$\alpha$ emission. Overall the galaxy shows a moderately ordered rotation in both resolutions with $v/\sigma=3.8$ (NS) and $v/\sigma=3.2$ (AO), similar kinematic position angles at both resolutions, and similar photometric and kinematic centres so we classify it as a disk. There is also a significant misalignment with the photometric axis in both cases because the continuum map does not show a clear major axis. The measurement of the angular momentum content of the galaxy is largely uncertain as evidenced by the large errors quoted in \ref{tab:results}.

\item \textbf{\textit{GOODS-S 40218} (merger):} The results from this galaxy are the most defective from our sample since very few pixels have enough SNR for a reliable Gaussian fit and due to its compact size. We classify this galaxy as a merger given the lack of ordered rotation and low $v/\sigma$ ratios (1.0 for NS and 0.9 for AO). The angular momentum measurement is unreliable and should only be taken for reference. However, by measuring the uncertainty on this target (by MCMC resampling), we have an indication of the limits of the code for future work where we have similar data.

\item \textbf{GOODS-S 42930 (disk):} This galaxy has a coherent ordered rotation in both datasets and shows clearly the difference in the shape of rotation curves from low- and high- resolution (in the latter, the rotation curve rises more rapidly). The combination of the datasets improves the measurement since it yields a more realistic model where the rotation curve rises fast but reaches a consistent velocity at large radii. The kinematic and photometric centres and position angles coincide and the $v/\sigma$ ratios are large (3.3 for NS and 3.4 for AO) so we confidently classify this galaxy as a disk.
\end{itemize}

%%%%%%%%%%%%%%%%%%%%%%%%%%%%%%%%%%%%%%%%%%%%%%%%%%

% Don't change these lines
\bsp	% typesetting comment
\label{lastpage}
\end{document}